\documentclass[pre,superscriptaddress,floatfix,nofootinbib,twocolumn]{revtex4}

\usepackage[T1]{fontenc}
\usepackage{times}
\usepackage{textcomp}
\usepackage[english]{babel}
\usepackage[pdftex]{graphicx}
\usepackage{amssymb,amsfonts,amsmath}
\usepackage{xcolor}

\begin{document}

\title{Rationality, irrationality and escalating behavior in lowest unique bid auctions}

\author{Filippo Radicchi}

\affiliation{Howard Hughes Medical Institute (HHMI), Northwestern University, Evanston, Illinois 60208 USA}

\affiliation{Department of Chemical and Biological Engineering, Northwestern University, Evanston, Illinois 60208 USA}

\affiliation{Departament d'Enginyeria Quimica, Universitat Rovira i Virgili, Av. Paisos Catalans 26, 43007 Tarragona, Catalunya, Spain}

\author{Andrea Baronchelli}
\affiliation{Departament de F\'\i sica i Enginyeria Nuclear, Universitat Polit\`ecnica de Catalunya, Campus Nord B4, 08034 Barcelona, Spain} 

\author{Lu\'{\i}s A. N. Amaral}

\affiliation{Howard Hughes Medical Institute (HHMI), Northwestern University, Evanston, Illinois 60208 USA}

\affiliation{Department of Chemical and Biological Engineering, Northwestern University, Evanston, Illinois 60208 USA}

\affiliation{Northwestern Institute on Complex Systems (NICO), Northwestern University, Evanston, Illinois 60208 USA}



 %

\begin{abstract}
Information technology has revolutionized the traditional 
structure of markets.
The removal of  geographical and time constraints has fostered
the growth of online auction markets,
which now include millions of economic agents worldwide and
annual transaction volumes in the billions of dollars.
Here, we analyze bid histories
of  a little studied type of online auctions
--- lowest unique bid auctions. 
Similarly to what has been reported for foraging
animals searching for scarce food, 
we find that agents  adopt L\'evy flight search strategies
in their exploration of ``bid space''.
The L\'evy regime, which is characterized
by a  power-law  decaying probability
distribution of step lengths, holds over
nearly three orders of magnitude. 
We develop a quantitative model for 
lowest unique bid online auctions that
reveals that agents use nearly optimal  
bidding strategies.
However, 
agents participating in 
these auctions do not optimize their financial gain.
Indeed, as long as there are many auction
participants, a rational profit optimizing agent would choose not
to participate in these auction markets.
\end{abstract}

\maketitle



\section*{Introduction}
\noindent Animals searching for scarce food
resources display movement patterns that can be statistically classified 
as L\'evy flights~\cite{viswanathan96,bartumeus03,ramos04,bertrand05,
sims08,viswanathan08,humpries10,viswanathan10}.
L\'evy flights~\cite{shlesinger93} represent 
the best strategy that can 
be adopted by a searcher looking for a scarce resource in an 
unknown environment~\cite{viswanathan99}, and
foraging animals seem therefore to have learned
the best strategy for survival.
L\'evy flights describe also
the movement patterns of humans in real space~\cite{brockmann06} 
and the variability of economic indices~\cite{mantegna95},
but these observations do not correspond to
search processes as in the case of foraging animals.
Surprisingly, there is no indication
of whether humans also use L\'evy flight
strategies when searching for scarce
resources.
Analyzing apparently unrelated data regarding
online auctions, we address here this question and show that, 
when searching for scarce resources,
humans explore the relevant space in the same 
class of strategies as foraging animals do.

\

\noindent  Lowest unique bid  auctions are
a new generation of online markets~\cite{malone87,bakos98,grieger03,vanheck98,lucking00,lucking07}.
Agents winning lowest unique bid auctions
may purchase expensive goods for absurdly low
prices; cars, boats and even houses
can be bought for only hundreds
of dollars.
The idea of the auction is strikingly simple.
A good, typically with a market value $V$ of at least
a thousand dollars, is put up for
auction. The auction duration
is fixed {\it a priori}. A bid can be
any amount from one cent to a pre-determined
maximum value $M$, generally lower than one hundred dollars.
 Each time
an agent makes a bid on a value $1 \leq b \leq M$, 
she pays a fee $c$, which ranges
from one to ten dollars depending
on the auction. During the bidding period,
an agent knows only the status of her new bid, that is,
whether it is winning or not.
None of the agents knows on what values the other
agents have bid until the end of the auction. 
When the bidding period expires,
the agent who made the lowest unmatched bid
can purchase the good
for the value of the winning bid 
(see Fig.~1 for an illustration of the determination of the winning bid).

\

\begin{figure}[!ht]
\begin{center}
\includegraphics[width=8.7cm]{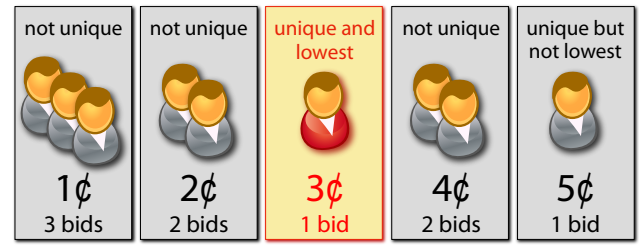}
\caption{{\bf Unique bid auctions.} 
Illustration of 
the rules of a lowest unique bid auction. 
At the end of the auction, the winner results 
to be the agent who has bid $3$\textcent, which 
represents the lowest unique bid.
All other bids are not unique apart from the one of
$5$\textcent, which is not the
lowest one. In highest unique bid auctions
the mechanism is reversed, and the winner
is the agent making the highest unmatched bid.
Illustration of 
the rules of a lowest unique bid auction. 
At the end of the auction, the winner results 
to be the agent who has bid $3$\textcent, which 
represents the lowest unique bid.
All other bids are not unique apart from the one of
$5$\textcent, which is not the
lowest one. In highest unique bid auctions
the mechanism is reversed, and the winner
is the agent making the highest unmatched bid.}
\end{center}
\end{figure}

\begin{figure}[!ht]
\begin{center}
\includegraphics[width=8.7cm]{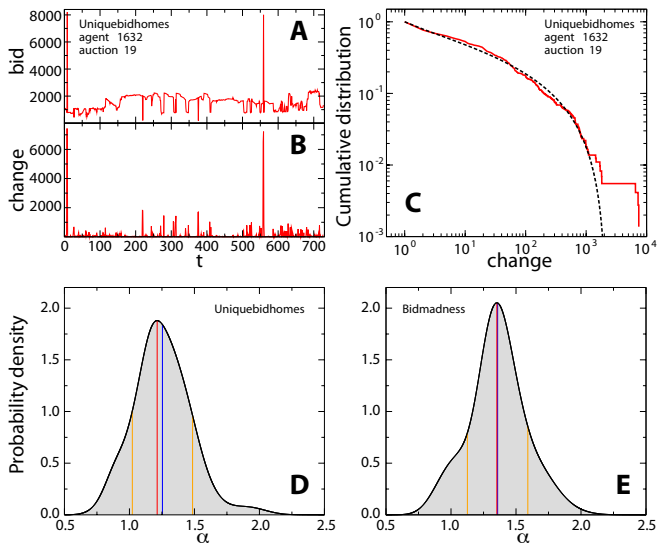}
\caption{{\bf Individual activity.}
{\bf (A)}  Bid values explored
by agent $1632$ on auction $19$ 
in the data set
{\tt www.uniquebidhomes.com}.
Bids are sorted chronologically, and the figure 
reports the value $b_t$ of the $t$-th bid.
The unit of the bid amount is one hundredth of an Australian dollar.
{\bf (B)}  Absolute value
of the difference between two consecutive bids. 
The exploration
of the bid space is characterized by a bursty behavior, where
many small movements are occasionally followed by large
jumps. 
{\bf (C)}  
Cumulative distribution function  of the change in bid value. 
The distribution is well fitted
by a power-law, with decay exponent consistent with $\alpha = 1.3 \pm 0.1$
 (dashed line). The agent  therefore explores the bid space using 
a L\'evy flight strategy.
Notice that the curve bends down because
of the finiteness of the bid space.
{\bf (D)} Probability density function
of the L\'evy-flight exponents
adopted by agents in lowest
unique bid auctions ({\tt www.uniquebidhomes.com}).
The blue line indicates the average value $\langle \alpha \rangle \simeq 1.26$ 
of the distribution,
the red line identifies the mode $\alpha_b \simeq 1.21$
of the distribution, the orange lines
bound the region within one standard deviation $\sigma \simeq 0.23$ from the average.
{\bf (E)} Probability density function
of the L\'evy-flight exponents
adopted by agents in highest
unique bid auctions ({\tt www.bidmadness.com.au}).
In this case we find   
$\langle \alpha \rangle \simeq 1.36$, $\alpha_b\simeq 1.35$
and $\sigma \simeq 0.23$.}
\end{center}
\end{figure}

\noindent  Lowest unique bid auction markets are
competitive arenas. Each agent performs
a search for a single target whose position changes
from auction to auction, as it
is determined by the bid history of the whole population
of agents.  Since the cost of each bid
is as much as $100$ times larger than the natural unit of the bid,
the number of bids that can be made by a single agent
is limited and allows only a partial exploration of
the bid space. Successful agents need to identify good strategies 
in order to maximize their winning chances and thus
limit their risk.

\

\noindent 
Lowest unique bid auctions are just a particular variant
of online pay-to-bid auctions, but other
types of pay-to-bid auctions
are regularly hosted on the web. For example, in highest
unique bid auction the mechanism of lowest unique bid auction is inverted,
and the winning bid is determined by the highest value
closest to a pre-determined upper bound value.
Since these auctions  still
involve
 a blind search of the winning value, highest
unique bid auctions are
equivalent to lowest unique bid auctions. Indeed, in this
paper we analyze data taken from both types of auctions.
\\
Other online pay-to-bid auctions,
however, can be very different from lowest unique bid auctions.   
For example, the so-called penny auctions, which
have acquired a great popularity in recent years,
appear quite similar to but are not.
As in the case of lowest unique bid auctions, 
the cost of the fee is at least $100$ times larger
of the bid increment, and as a consequence,
the final value of the winning bid is much lower
than the real value of the good up for auction.
However, in penny auctions
the value of the winning bid is publicly known and can only grow 
during the auction (i.e., the word ``penny''
is used because, in penny auctions, bid increments are equal to one cent).
While escalation plays a very important role in penny auctions,
in this type of auctions agents do not need to
explore the bid space because the value
of the winning bid is known.
Penny auctions have been the focus of some theoretical and empirical 
studies~\cite{augenblick09,hinnosaar10,byers10,mittal10,platt10}.


\section*{Results}

\noindent We collected data from three distinct web sites hosting
lowest unique bid auctions. We automatically downloaded
and parsed the content of the tables
reporting the bid history of closed auctions. 
These data sets contain all the information
on individual auctions, including
the details of each bid: its value, when it was made and
who placed it. These data allow us to
keep track of all the movements performed on bid space 
by a given agent bidding in a specific auction.

\noindent We show in 
Figure~2A a typical exploration of the bid 
space performed by a single agent.
The exploration of the bid space is bursty:
consecutive bid values are generally close to each other,
but from time to time the agent performs ``long jumps''
in bid space.
We first compute the jump lengths (Fig.~2B)  
and estimate their probability
distribution function (Fig.~2C).
We find a strikingly robust power-law scaling consistent
with the exploration of the bid space using a 
L\'evy flight search strategy~\cite{shlesinger93}.
Note that here we use the notion of discrete L\'evy flights.
Time and space are in fact discrete, and the exploration
of the bid space is modeled as a discrete
time Markov chain [with transition
probability defined in Eq.~\ref{eq:transition}].
Our discrete model converges to a 
standard L\'evy flight only in the continuum limit of space 
and time~\cite{hughes81}.
The power-law scaling can 
be observed both at the level of
single agents (whenever the number of
bids is sufficiently large for estimating the distribution; 
c.f. Figs.~2C and Supporting Information) and globally, by
aggregating the length of the jumps made by all agents in all auctions 
(Figs.~3A and Supporting Information).
The density distribution of the exponents calculated over single
agents is  peaked
around a mean value $\langle \alpha \rangle \simeq 1.3$ 
(Figs.~2D,~2E and Supporting Information),
the same exponent value we estimate for the aggregated data.
Significant variations around the average value are anyway present,
and reflect the heterogeneity of the agent strategies.
The density distributions of Figs.~2D and~2E are in fact calculated 
by considering different agents bidding in different auctions.

\

\noindent
The power-law scaling and its measured exponent
are very stable. Exponent estimates do not depend on the direction
of the jumps 
(Figs.~3B and Supporting Information) or the level of activity 
of the agent (Figs.~3C and Supporting Information). Surprisingly,
performing L\'evy flights
does not appear to be a learned strategy. Instead
it appears to be
an intrinsic feature of the mental search process: the jump lengths 
in the bid space follow the same power-law at any stage of the auction
(Figs.~3D and Supporting Information). 

\

\noindent Our results represent
the strongest  empirical evidence for the
use of L\'evy flight strategies in the search
of scarce resources reported in literature up to now.
Differently from previous studies  where 
``two orders of magnitude of scaling can represent a luxury''~\cite{viswanathan08}, here the power-law decay can be clearly observed even over four orders
of magnitude.
It is unlikely, though, that adopting 
L\'evy flight strategies is
a deliberate choice of the agents, just
as it is not likely
that animals searching for food consciously follow
a L\'evy flight strategy.
Nevertheless, the data demonstrate that the changes
in bid value
are statistically consistent with a power-law
decaying distribution over several orders of magnitude 
(see and Supporting Information)~\cite{clauset09}.
Simple correlation measurements  show also that the lengths of
consecutive jumps are independent of each other 
(see and Supporting Information).
We believe that
the power-law is valid over such a broad regime
because the space is not
strictly physical. That is,
movements of tens of thousands of cents can be performed
for the same cost of those of only one cent. Agents
thus explore the bid space in an effectively super-diffusive fashion,
and steps are made with infinite velocity.

\begin{figure}[!ht]
\begin{center}
\includegraphics[width=8.7cm]{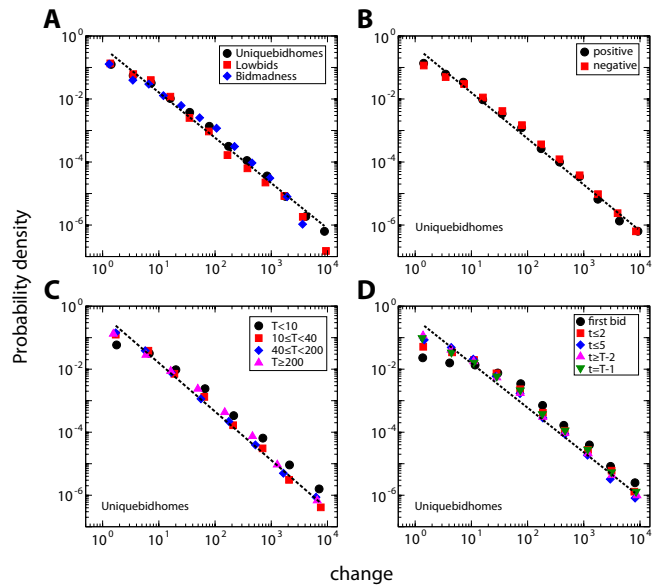}
\caption{{\bf Bidding strategies of agents are L\'evy flights.}
{\bf (A)} Probability density function 
of the bid change for all agents in all auctions.
We analyze data sets from three
different web sites hosting auctions:
{\tt www.uniquebidhomes.com} (black circles), 
{\tt www.lowbids.com.au} (red squares ) and
{\tt www.bidmadness.com.au} (blue diamonds). 
{\bf (B)} Probability density function  
of positive (black circles) and negative (red squares) bid changes. 
{\bf (C)}  Probability density function of the change amount
for data aggregated over agents with different levels
of activity ($T$ indicates the total number of bids made 
by an agent in a single auction). 
{\bf (D)} Probability density function of the change amount
at different stages of the auctions ($t$ stands for order
of the bid change in the bid history of an agent). 
In (A), (B) and (D) results have been obtained for lowest unique bid auctions
({\tt www.uniquebidhomes.com}).  All
dashed lines stand for best power-law fits (least square) 
and all exponent values are consistent with $\alpha=1.4 \pm 0.1$.
The unit of the bid value change amount 
is one hundredth of an Australian dollar.}
\end{center}
\end{figure}

\section*{Model}
\noindent Next, we model the lowest unique bid auction process.
Consider $N$ agents competing in a lowest unique bid auction. 
We model the successive bids of
these agents as L\'evy flight searches on bid space. 
Each agent moves in a bounded one-dimensional 
lattice with an {\it a priori} chosen exponent value,
which may be regarded as the agent's strategy in the auction.
In our formulation,
every agent performs the same number $T$ of bids and
may return to already visited sites. 
At the beginning of the auction, every agent sits at 
the leftmost site on the lattice and
then performs $T$ movements by changing, at each step, her
actual position by an amount randomly drawn from a power-law 
distribution. If at stage $t-1$ the agent with strategy $\alpha$ is 
sitting at position $j$,
then at stage $t$ she jumps to position $i$ with probability 
proportional to $\left| i - j \right|^{-\alpha}$.
This model provides us with an independent way
to determine the exponent values of the L\'evy flights
and offers a strikingly good statistical description
of the data  (Fig.~2B and Supporting Information).

\

\begin{figure}[!ht]
\begin{center}
\includegraphics[width=8.7cm]{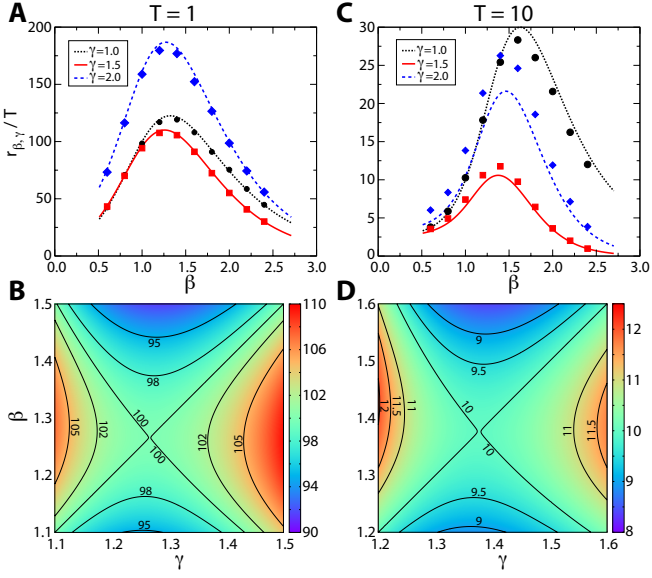}
\caption{{\bf Model predictions.}
Economic return $r_{\beta, \gamma}$ [Eq.~(\ref{eq:income})], divided
by the number of bids $T$, of an agent bidding with strategy $\beta$ when
competing, in a lowest unique bid auction with
upper-bound $M=1,000$ and for a good of value $V=10,000$,
against $N=100$ opponents bidding with strategy $\gamma$.
Unless specified, the quantity $r_{\beta, \gamma}$
reported in this plots
is computed by numerically solving the equations of the model.
{\bf (A)} Case where each agent performs a single bid
in the auction, for three values of $\gamma$. 
Theoretical predictions (lines) are compared with the results
of numerical simulations (symbols). In each simulation of the
auction, we randomly extracted $N$ bid values $j$ with probability 
proportional to $j^{-\gamma}$, and a single bid value $v$ with
probability proportional to $v^{-\beta}$. For
a given set of parameters, we repeated the same
simulation $G=10,000$ times, and  calculate the
number of times $g$ in which the bid value
extracted from the power-law distribution
with exponent $\beta$ was the winning bid, and
the sum $I$ of these winning bid values.
The economic return has been finally calculated 
as $r_{\beta, \gamma} = \left(g V - I \right)/G$.
{\bf (B)} Exploration of
parameter space reveals the existence of a saddle point
at $\beta_s=\gamma_s\simeq 1.27$.
{\bf (C)} Case where each agent performs $10$ bids
in the auction, for three values of $\gamma$.
Numerical simulations have been carried
out as in the former case, but considering agents
moving in the bid space according to Eq.~(\ref{eq:transition}).
{\bf (D)}  Exploration of
parameter space reveals the existence of a saddle point
at $\beta_s=\gamma_s \simeq 1.38$.}
\end{center}
\end{figure}

\noindent  We focus our attention
on a generic agent bidding with strategy
$\beta$ and on her chances to win
auctions in which the rest of the population
is bidding with strategy $\gamma$.
More complicated situations may in principle be 
studied with the same formalism.


\noindent



\subsection*{Single bid}
\noindent Consider first the case in which agents
make a single bid. The probability that a generic 
opponent, using bidding strategy $\gamma$, bids on value 
$i$ is
\begin{equation}
p_\gamma \left(i\right) = i^{-\gamma} / m\left(\gamma\right) \;\; ,
\label{eq:single_gen} 
\end{equation}
with $m\left(\gamma\right)=\sum_{j=1}^M j^{-\gamma}$
proper normalization constant.
Here we consider
the simple case in which all agents adopt the same 
bidding strategy $\gamma$. The probability of Eq.~(\ref{eq:single_gen})
can be anyway made more general by assuming
that agents chose strategies from a density distribution
$g\left(\alpha\right)$ and calculating the probability
of Eq.~\ref{eq:single_gen} as $p\left(i\right) = \int \, d \alpha \; i^{-\alpha} / m\left(\alpha\right) \; g\left(\alpha\right)$.
After all agents have bid, there will be $n_k$ bids 
on the $k$-th bid value. Such variables clearly obey the constraint 
$ N = \sum_{k=1}^M \, n_k$. The probability to observe a 
particular configuration 
$\left\{ n \right\} = \left(n_1, n_2, \ldots, n_k, \ldots , n_M\right)$
is given by
\begin{equation}
P_\gamma \left( \left\{ n \right\} \right) =  N! \; \prod_{k=1}^M \, \frac{\left[p_\gamma \left(k\right)\right]^{n_k}}{n_k!} \;\;,
\label{eq:multinomial} 
\end{equation}
which is a multinomial distribution with weights given 
by Eq.~(\ref{eq:single_gen}).
In particular, the probability that only one bid (i.e., a unique bid)
is made on value $i$ is 
\begin{equation}
\begin{array}{l}
u_\gamma \left(i\right) = P_\gamma \left( n_i=1 \right)
\; = \; \sum_{\sum_{k \neq i} n_k = N-1} P_\gamma \left( \left\{ n \right\} \right) 
\\
\;
= \; N p_\gamma \left(i\right) \left[ 1 - p_\gamma \left(i\right) \right]^{N-1}  
\end{array}
\;\;. 
\label{eq:unic}
\end{equation}

\noindent Focus now on the agent with bidding strategy $\beta$. 
The probability that, making a bid
on value $v$, she makes
a lowest unique bid can be calculated exactly
by summing the multinomial distribution of Eq.~(\ref{eq:multinomial})
over all configurations for which there are no bids on the value $v$ and
there is not a unique bid on a value smaller than $v$, and
finally multiplying this factor by the probability that
the agent with bidding strategy $\beta$ bids on the value $v$.
Such exact calculation is however unfeasible due
to the extremely high number of possible combinations, and therefore
we approximate the probability that, making a bid
on value $v$, the agent with bidding strategy $\beta$ makes
a lowest unique bid as
\begin{equation}
l_{\beta, \gamma}\left(v\right)=p_\beta\left(v\right)\,
\, \left[1-p_\gamma\left(v\right)\right]^N \, \prod_{k<v} \, \left[ 1 - u_\gamma\left(k\right)  \right]\;.
\label{eq:unic_low} 
\end{equation}
The r.h.s. of Eq.~(\ref{eq:unic_low}) is the product of 
three terms: $p_\beta\left(v\right)$
is the probability that the agent bids on value $v$; 
$\left[1-p_\gamma\left(v\right)\right]^N$ is the probability
that none of the opponents have bid on value $v$;
$\prod_{k<v} \, \left[ 1 - u_\gamma \left(k\right)\right]$ is 
the probability  that none of the bid values smaller than $v$ are
occupied by a single bid made by one of the opponents.
In spite of the fact that Eq.~(\ref{eq:unic_low}) is just
an approximation
of the real $l_{\beta, \gamma}\left(v\right)$, the approximation
can be considered good because able to reproduce
the results obtained from the direct simulation
of the process (see the section Results). Moreover
in the simplest case in which $N=1$, it correctly reduces
to the exact value $l_{\beta, \gamma}\left(v\right) =p_\beta\left(v\right)\, 
\prod_{k\leq v} \, \left[ 1 - p_\gamma\left(k\right)  \right]$.

\

\noindent Finally, the probability that the
agent with bidding strategy $\beta$ wins the auction is
\begin{equation}
w_{\beta, \gamma} = \sum_{v=1}^M \, l_{\beta, \gamma}\left(v\right)
\label{eq:prob_win}
\end{equation}
and, on average, the value of her winning bid is
\begin{equation}
\langle v \rangle_{\beta, \gamma} = \sum_{v=1}^M \, v \; l_{\beta, \gamma}\left(v\right) \;\; .
\label{eq:av_win}
\end{equation}

\subsection*{Repeated auctions}
\noindent \noindent Imagine now to repeat
the same auction $G$ independent times. 
The probability that the agent bidding with strategy $\beta$
wins $g$ times out of $G$ total auctions is given by a binomial distribution 
\[
P_{\beta, \gamma} \left(g\right)= {G \choose g} \, \left(w_{\beta, \gamma} \right)^g \, \left( 1- w_{\beta, \gamma} \right)^{G-g}
\; .
\]
If the agent with bidding strategy
$\beta$ wins $g$ auctions, the 
sum of her winning bids is a random variable $I$ whose
probability is determined by 
\[
R_{\beta, \gamma}\left(I\left|g\right.\right) = 
\sum_{v_1 + v_2 + \ldots + v_g =I} 
\, l_{\beta, \gamma}\left(v_1\right) \, l_{\beta, \gamma} \left(v_2\right) \, \cdots \, l_{\beta, \gamma}\left(v_g\right) \;\; ,
\]
where the sum runs over the integer indices $v_1$, $v_2$, \ldots, $v_g$
with the constraint that their sum should equal $I$.
Excluding bidding costs, the average return of the agent in $g$ victories is
\[
r_{\beta, \gamma} \left(g\right) = \left(g V - I \right)/G \;\; .
\] 
In general, the probability
that the sum of the winning bids is equal 
to $I$ in an arbitrary number
of  auctions won by the player with bidding strategy $\beta$ 
can be calculated as 
\[
R_{\beta, \gamma}\left(I\right)
 = \sum_g P_{\beta, \gamma}\left(g\right) \, R_{\beta, \gamma} \left(I\left|g\right.\right) \;\; ,
\] 
and a similar expression can be derived for the
distribution of $r_{\beta, \gamma}\left(g\right)$. However, we are interested
in the case in which the number
of auctions diverges ($G \gg 1$). In this limit, 
we can
approximate the number of victories with
its average $\langle g \rangle = G \,  w_{\beta, \gamma}$
as well as the sum of the winning bids as
 $I = \langle g \rangle \langle v \rangle_{\beta, \gamma} =  G \,  w_{\beta, \gamma} \,  \langle v \rangle_{\beta, \gamma}$. The return of
the agent with bidding strategy $\beta$
is therefore 
\begin{equation}
r_{\beta, \gamma} = w_{\beta, \gamma}\, \left(V-\langle v \rangle_{\beta, \gamma}\right) \;\; .
\label{eq:income}
\end{equation}
For $r_{\beta, \gamma}> c$,  the agent has a positive return for
participating in the auction, whereas,
for $r_{\beta, \gamma}< c$, her return is negative.

\subsection*{Multiple bids}
\noindent Given a generic agent with bidding strategy $\alpha$, her first 
bid is placed on value $i$ with probability 
$q^{(1)}_\alpha\left(i\right) = i^{-\alpha}/m\left(\alpha\right)$. 
For the subsequent bids, we need to define a transition matrix $Q_\alpha$,
whose generic element $\left(Q_\alpha\right)_{ji}$ gives the probability
that the agent bids on value $i$ when her previous bid has been made 
on value $j$. In our model, we have
\begin{equation}
\left(Q_\alpha \right)_{ji} = \frac{\left| i - j \right|^{-\alpha} \, \left[1-\delta \left(i-j\right)\right]}{m_j\left(\alpha\right)} \;\; ,
\label{eq:transition}
\end{equation} 
for all $i$ and $j$ in the interval $[1,M]$. $\delta\left(\cdot\right)$ 
is the Kronecker delta, equal to one if
its argument is equal to zero, and equal to zero otherwise. 
The normalization constant
$m_j\left( \alpha \right) = \sum_{i=1, i \neq j}^M \left| i - j \right|^{-\alpha} $ 
ensures the proper definition of the transition matrix. The matrix $Q$
describes a random walker performing 
uncorrelated L\'evy flights with exponent $\alpha$.
Notice that the agent has no memory
of her previous bid values and therefore she may
place more than a bid on the same value.
At the generic step $t$, the probability that the agent with bidding strategy
$\alpha$  bids on the value $i$ is
\[
q^{(t)}_\alpha \left( i \right) = \sum_{j=1}^M \; \left( Q_\alpha \right)_{ji} \; q^{(t-1)}_\alpha \left( j \right) \;\; .
\]
The probability that this agent has bid, during her $T$ bids,
on value $i$ is then
\[
s^{(T)}_\alpha\left( i  \right) = 1 - \prod_{t=1}^T \; \left[ \,1 - \, q^{(t)}_\alpha \left( i \right) \right] \;\;.
\]
The term $1-q^{(t)}_\alpha \left( i \right)$ counts the probability
that the agent has not bid on value $i$ at stage $t$. The probability
that the agent has not bid on value $i$ at any stage is therefore
the product of this single step probabilities. Finally, the probability
that the agent has bid on value $i$ at least once
is calculated as the probability to have bid on value $i$ an arbitrary 
number of times minus the probability to have never bid on value $i$.

\

\noindent Now go back to the situation in which an agent
with bidding strategy $\beta$ is opposed to a population
of $N$ agents with bidding strategy $\gamma$.
The probability that the agent with bidding strategy
$\beta$ has bid, in $T$ steps, at least once on value $i$ is
$s^{(T)}_\beta\left( i  \right)$. The probability
that one of the $N$ opponents, bidding
with strategy $\gamma$, makes
a unique bid on value $i$ is  given by 
\begin{equation} 
u^{(T)}_{\beta, \gamma}\left(i\right) = N \, s^{(T)}_\gamma\left(i\right) \, \left[ 1- s^{(T)}_\gamma\left(i\right)\right]^{N-1} \,  \left[ 1 - s^{(T)}_\beta\left(i\right)\right] \;\; .
\label{eq:multi_unique}  
\end{equation}
$u^{(T)}_{\beta, \gamma}\left(i\right)$ is the product of two terms: $N \, s^{(T)}_\gamma\left(i\right) \, \left[ 1- s^{(T)}_\gamma\left(i\right)\right]^{N-1}$ is 
the probability that a bid on value $i$ is unmatched by any of the other
$N-1$ opponents, while $ 1 - s^{(T)}_\beta\left(i\right) $ is the
probability that also the agent, with bidding strategy $\beta$, does not bid on 
value $i$. The probability that the agent 
with strategy $\beta$ wins the auction with a bid on value $v$ is
\begin{equation} 
l^{(T)}_{\beta, \gamma}\left(v\right) = s^{(T)}_\beta\left(v\right) \, \left[ 1- s^{(T)}_\gamma\left(v\right)\right]^{N} \, \prod_{k<v} \, \left[ 1 - u^{(T)}_{\beta, \gamma}\left(v\right)\right] \; \; ,
\label{eq:multi_lub}  
\end{equation}
respectively standing for the product 
of the probabilities that: she bids on value $v$; none
of the other agents bids on value $v$; none of the
bids with value smaller than $v$ is unique.
Eqs.~(\ref{eq:multi_unique}) and~(\ref{eq:multi_lub})
represent the generalization of Eqs.~(\ref{eq:unic}) and~(\ref{eq:unic_low}),
respectively.
In Eq.~(\ref{eq:multi_lub})
we made the same type of approximation  as the one used
for writing Eq.~(\ref{eq:unic_low}).
The probability $w^{(T)}_{\beta, \gamma}$ that the agent with
bidding strategy $\beta$ wins the auction  and
the average value $\langle v \rangle^{(T)}_{\beta, \gamma}$
 of her winning bids can be respectively calculated 
using Eqs.~(\ref{eq:prob_win}) and~(\ref{eq:av_win}). 
Finally, excluding bidding costs, the return
$r_{\beta, \gamma}$  of the agent with strategy $\beta$ over an infinite 
number of auctions is again given by Eq.~(\ref{eq:income}).
For $r_{\beta, \gamma}>T\, c$,  the agent has a positive return for
participating in the auction, whereas,
for $r_{\beta, \gamma}<T\, c$, her return is negative.

\begin{figure}[!ht]
\begin{center}
\includegraphics[width=8.7cm]{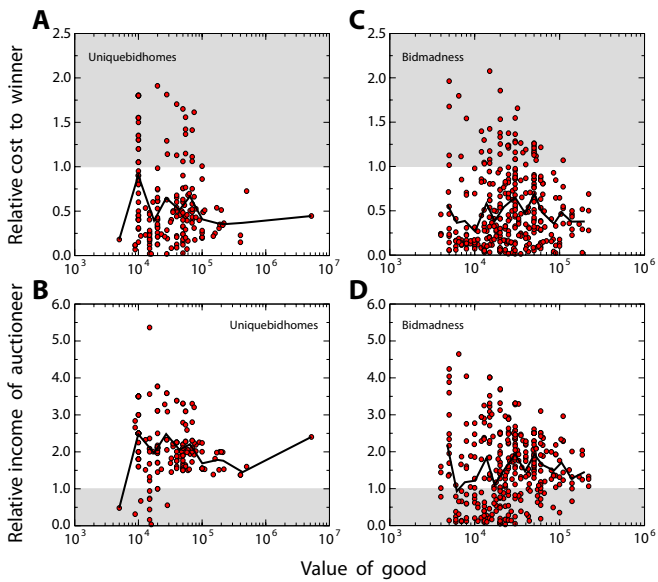}
\caption{{\bf Economic return of agents.}
Auction winners tend to pay half of the value
of the good, while auctioneers tend to earn twice
the value of the good. We consider two different  data sets, one
regarding lowest unique bid auctions [{\tt www.uniquebidhomes.com}, 
(A) and (C)]
and the other highest unique bid auctions 
[{\tt www.bidmadness.com.au}, (B) and (D)]. The unit is one hundredth of an Australian dollar.
{\bf (A)} Relation between the relative cost 
$\left(T_w \, c + b^*\right)/V$ to the winner
of an auction and her income $V$. $T_w$ indicates the number of bids
made by the winner of the auction, $c$ is the cost of the fee, $b^*$ indicates
the value of the winning bid, and $V$ is the value of the good
put up for auction. Each point represents an auction.
The gray area corresponds to the region of negative return
for winners of the auctions. The black line indicates the average value
of the relative cost to the winners.
Auctions are, on average, very profitable for the 
agent winning the auction, but
the probability of a specific agent winning 
the auction is
very low. 
Data refer to the data set {\tt www.uniquebidhomes.com}.
{\bf (B)}
Relative income 
$\left(B \, c + b^*\right)/V$ of the auctioneers
as a function of their investment $V$.
$B$ is the number of bids placed by all agents.
The gray area denotes the region of negative
return for the auctioneers.
The black line indicates the average value
of the relative profit of the auctioneers. On average, organizing
auctions is very profitable. 
Data refer to the data set {\tt www.uniquebidhomes.com}.
{\bf (C)} Relation between the relative cost to the winner
of an auction and her income for the data set 
{\tt  www.bidmadness.com.au}. 
{\bf (D)} Relative income of the auctioneers
as a function of their investment for the data set 
{\tt  www.bidmadness.com.au}.
}
\end{center}
\end{figure}

\subsection*{Model predictions}
\noindent We show in Fig.~4 the results obtained
with our analytical model. The presence of a saddle point at $\gamma_s=\beta_s$
indicates that $\beta_s$ is an optimal 
strategy or Nash equilibrium~\cite{nash50, rapoport70, basar95}.  
When the opponents do not bid rationally 
(i.e., $\gamma \neq \gamma_s$), it is more
convenient to use a strategy  $\beta \neq \beta_s$.
On the other hand, when the other agents bid rationally 
(i.e., $\gamma=\gamma_s$),
there is no better strategy than $\beta_s$.
The value of $\beta_s$ depends on
the parameters $N$ and $T$ , but for realistic
choices (see and Supporting Information
and Fig.~4), $\beta_s$ is in the range $1.2$ to $1.5$, 
the same range of the exponent values we estimated from the data.
Thus, despite its simplicity, our model captures the 
main features of the real auctions.
Performing L\'evy flights
with small exponents (ballistic motion) yields 
unique bids that are unlikely to be the lowest. 
On the other hand, performing preferentially short 
jumps (high exponents, diffusive motion) guarantees to always bid on 
small values which are
unlikely to be unique. Intermediate values of the exponent 
(super-diffusive motion) represent a compromise
between staying low and being unique, and therefore lead 
to maximal winning chances. These considerations are valid only
for finite values of $N$ and $T$, which is the realistic
case. Because the available positions in the lattice are finite,
when either $N$ or $T$ grow,
the probability to observe a unique bid 
progressively approaches zero~\cite{berkolaiko97}.
Notice that at 
the saddle point $\gamma_s=\beta_s$,
all $N+1$ agents are using the same bidding strategy
and therefore they all have the same
chances to win the auction.
In particular, the probability that a generic
agent wins the auction
is $w_{\beta_s, \gamma_s} \leq 1/\left(N+1\right)$, where
the inequality may arise because a unique
and lowest bid may not exist.


\

\noindent
The value of the exponent, corresponding
to the optimal  L\'evy flight strategy 
in lowest unique bid auctions,  
is distinct from the one found in the case
of purely random searches~\cite{viswanathan99},
and empirically
observed in the movement patterns of 
foraging animals~\cite{viswanathan96,bartumeus03,ramos04,bertrand05,
sims08,viswanathan08,humpries10,viswanathan10}.
The quantitative difference arises, we believe, as a consequence
of the anisotropy of  the bid space (low values
are favored), the role of competition, and, more 
importantly, the fact that the target is not ``static'' but moving 
according to the actions of the whole population of agents.

\section*{Discussion}

\noindent In lowest unique bid auctions, agents
have the possibility to win goods of high value
for impossibly low prices 
(Figs.~5A and~5C),
However, these  all-pay
auction markets are designed to be 
very profitable for the auctioneers~\cite{meyerson81,milgrom82,mcafee86,krishna02}, 
who, on average, double their investment 
(Figs.~5B,~5D and Supporting Information). 
For auctioneers, the profitability of lowest unique bid auctions is 
in fact guaranteed by the validity of the inequality $V < B \, c$,
where $B$ stands for the total number of bids and equals
$\left(N+1\right) \, T$ in our model. Under this constraint 
however, the payoff of a generic agent in a perfectly rational 
population is always negative since
\[
r_{\beta_s, \gamma_s} < w_{\beta_s, \gamma_s} \, V \leq  \frac{V}{N+1} < T\, c\;\;,
\]
and there 
is no expected economic gain to be obtained for
participating as a bidder in the auction markets.
The rationality of the economic agents in adopting
optimal strategies seems, therefore,
in contrast with the ultimate irrationality that induces
agents to take part in these auction markets. 

\

\noindent
Competitive 
irrationality, based  on rational choices, has been investigated 
in economic theories~\cite{staw76, staw89, bazerman92, wald08},
such as the dollar auction game~\cite{martin71}.
The decision to participate or not participate 
in lowest unique bid auctions 
presents a paradox
for potential bidders. If the number of agents participating in 
the auction is not too high, then the auction would bring a 
positive economic return to the agents, but not to the auctioneers. 
For example, in the case in which only one bidder participates
in the auction, this bidder would have  the maximal economic 
return by placing a single bid on 
the lowest value allowed. 
But by this token, every agent will
feel  that participating is profitable
as long as not many other agents have bid yet.
However, no agent can know how many other agents
will actually bid on the good.

\

\noindent
Our results raise a number of important research
questions. First, which brain regions
are responsible for implementing the search strategies
used by agents? Since agents use similar
search strategies to bees or birds, it is likely that there
is no frontal cortex
involvement.  Using neuroimaging techniques such as
fMRI it should be possible to answer this
question. Second, does the economic paradox
that the agents face reveal itself in  brain activity patterns?
Specifically, do some of the changes in brain activity
observed for preference reversal~\cite{grether79,tversky90} occur also in
this case? Additionally, our results
suggest that controlled
lowest unique bid auction markets would offer the possibility
to run large-scale experiments at relatively
low cost~\cite{salganik06}. These experiments
could be used for monitoring the behavior
of agents in auction markets
with tunable optimal search strategies,
and see if (and how fast) agents are able to adapt their
behavior to optimality.

\section*{materials}
\noindent Data have been collected from 
three publicly accessible web sites:
{\tt www.uniquebidhomes.com}, {\tt www.lowbids.com.au} and
{\tt www.bidmadness.com.au}. Also, we make available a 
version of these data at the\\
web page {\tt filrad.homelinux.org/resources}.

\begin{acknowledgments}
\noindent We thank A.~Arenas, A.~Flammini, S.~Fortunato, A.~Lancichinetti, and J.J.~Ramasco for useful discussions. T.~Rietz is gratefully acknowledged  for fundamental comments on the manuscript.
\end{acknowledgments}


\clearpage

\newpage

\renewcommand{\thesection}{S \arabic{section}}
\renewcommand{\theequation}{S\arabic{equation}}
\renewcommand{\thefigure}{S\arabic{figure}}
\renewcommand{\thetable}{S\arabic{table}}
\setcounter{figure}{0}
\setcounter{table}{0}
\setcounter{equation}{0}
\setcounter{section}{0}

\onecolumngrid

\section{``Lowest Unique Bid'' and ``Highest Unique Bid'' Auctions}

\subsection{Description of the auction}

\begin{figure}[!ht]
\begin{center}
\includegraphics[width=0.45\textwidth]{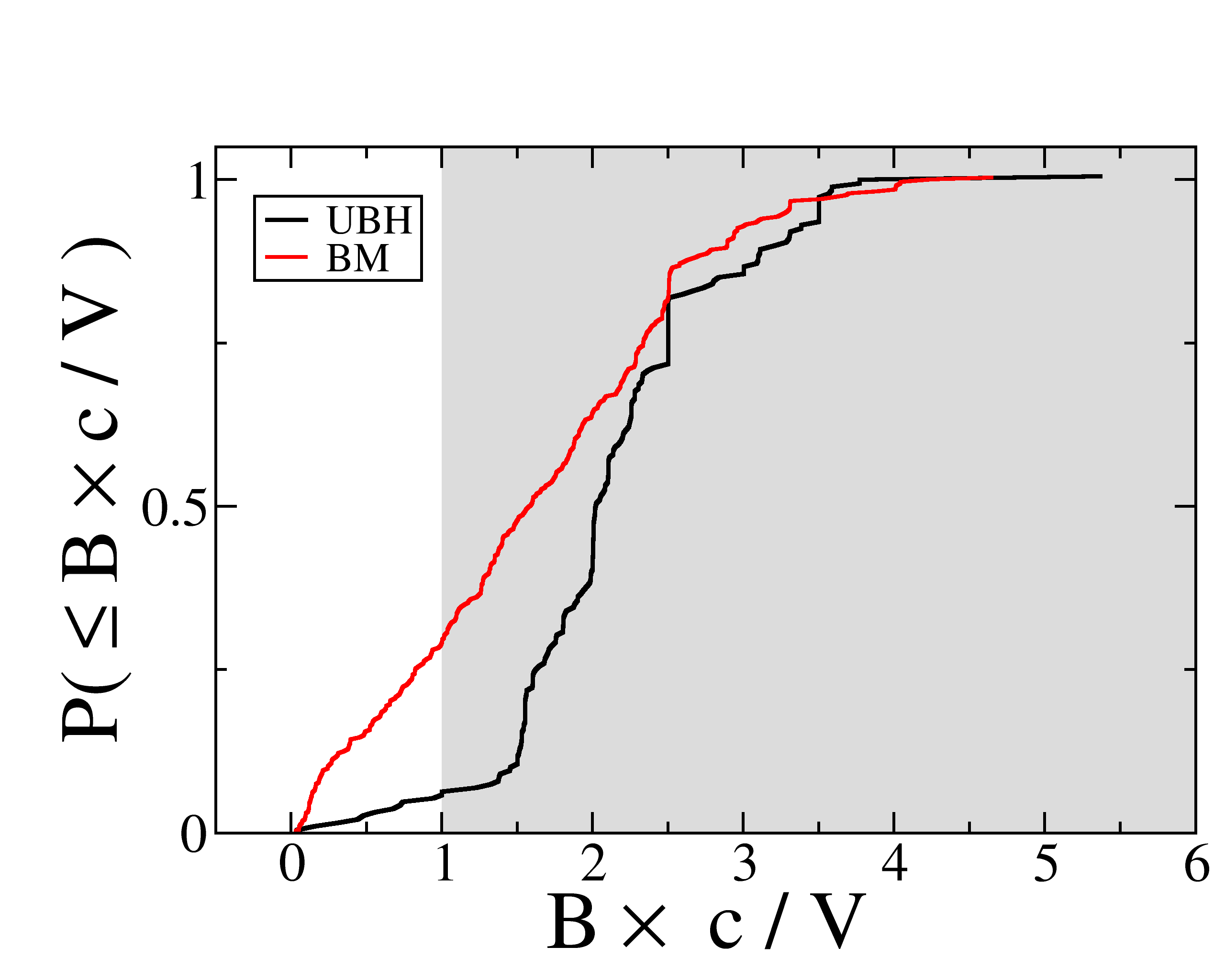}
\caption{Cumulative distribution
of the relative return $\left(B \times c\right) / V$
for the auctioneers. For {\tt www.uniquebidhomes.com} 
data set (black line)
the $95\%$ of the auctions are in the gray region of
positive return, where the relative return $\left(B \times c\right) / V > 1$.
On average, the relative return of the auctioneers is $2.2$.
For {\tt www.bidmadness.com.au}   data set (red line)
the $72\%$ of the auctions have produced a positive relative return
and on average the relative return is $1.6$.}
\label{fig:profit}
\end{center}
\end{figure}

\noindent Lowest Unique Bid (LUB) auctions are
special on-line auctions which have reached
a considerable success during last years.
Their peculiarity consists in the fact
that they are reverse auctions:
rather than the bidder with the highest bid 
(as in the case of traditional auctions), the winner is 
the person who makes the LUB (see Figure~1a in the main text).
The rules of a LUB auction are very simple.
At the beginning of each auction, the auctioneers
put up for auction a good of value $V$.
After the beginning of the auction and for a 
certain period of time
(in general of order of weeks),
agents participate to the auction by making bids.
The natural unit of the auction is one hundredth 
of dollars, euros, etc (i.e., the 
currency depends 
on the country where the auction is hosted).
Bids may be any amount (in cents) between 
one cent  and a maximal bid amount $M$ (generally
lower than ten hundreds of cents).
Sometimes, the value of $M$ is not fixed,
but the bid space is anyway naturally
bounded since none of the agents wants to bid more than $V$. 
The value of $V$ depends on the auction, but generally
its order of magnitude is of thousands of hundreds of cents.
Making a bid costs a fee $c$ (typically
from one hundred to ten hundreds cents). After each bid $b$, the agent
receives an automatic message, from the web site
hosting the auction,
saying whether that bid was the winning bid (i.e., the bid
is the LUB) or not.
The agent is constantly informed about the status of her bids
(i.e., whether one of them becomes the LUB or is not longer the LUB).
However, each agent knows only what she has bid, without
any information on which values the other agents have bid.
In general, there is no restriction for the number of bids that 
the same agent may place. When the
time dedicated to the auction expires, the winner is the agent
who made the LUB and can therefore purchase the good 
for the value of her winning bid.
If at the close of the auction a single lowest unique bid 
does not exist, the successful bid becomes the lowest one
made by only two agents and the winner is the one
 who has bid first on such value. In the case in which also a bid
made by only two agents does not exists, then the winning bid becomes
the lowest one made by only three agents and so on. Again
in these situations, the winner is the agent who has first placed
a bid on the winning value. In our data sets however, 
we always observe that the winning bid is an unmatched bid.
\\
There are several slight variations of this kind of auctions.
Very often, the end of the auction is not determined by an expiration time,
but by a minimum required number of bids, {\it a priori}
fixed by the auctioneers.
In other variations called Highest Unique Bid (HUB) auctions,
the winning bid is the unique one closest to $M$.
\\
Independently on the type of auctions,
this kind of auctions are particularly
profitable for both the auctioneers and the winners
of the auctions.
Figure~5 of the main text and Figure~\ref{fig:profit} 
clearly show that there are only few exceptions 
in which auctioneers or winners
have lost money, but in the majority of the auctions their returns
are positive.

\clearpage

\subsection{Description of the data sets}

\begin{table}
\begin{center}
\begin{tabular}{| c | c | c | c | c | c | c | c |}
\hline
Data set & Tot. Auctions &  Tot. Agents & Tot. Bids & $\langle M \rangle$ & $\langle c \rangle$ & $\langle  N \rangle$ & $\langle B \rangle$
\\ 
\hline
UBH & $189$ &  $3\, 740$ & $55\,041$ & $362$ & $437$ & $50$ & $6$
\\ 
\hline
LB & $55$ & $445$ & $3\, 740$ & $1\, 284$ & $478$ & $13$ & $6$
\\
\hline
BM & $336$ & $3\,719$& $127\,275$ &   $504$ & $174$ & $40$ & $14$
\\ 
\hline
\end{tabular}
\vskip .1cm
\caption{Summary table of the data sets analyzed in this paper.
We report, from left to right, the name of the data set, the total number
of auctions, the total number of different agents, the total number
of bids, the average value of the maximal bid value, the average
amount of the fee, the average number of agents involved in an auction and
the average number of bids made by a single agent in a single auction.
The unit of the bid values is one hundredth of an Australian dollar.}
\label{tab:summary}
\end{center}
\end{table}

\noindent We collected data from the web sites 
{\tt www.uniquebidhomes.com} (UBH),\\ 
{\tt www.lowbids.com.au} (LB)
both hosting LUB auctions and 
\\ 
from {\tt www.bidmadness.com.au} (BM)
organizing HUB auctions.
Data regard all auctions organized
during $2007$, $2008$, $2009$ and part of $2010$ by these web sites.
We collected detailed information concerning the auctions:
the value of the goods, the cost of the fee, 
the maximum bid amount, the duration of the auction
or eventually the required number of bids.
We report in Table~\ref{tab:summary} some of these quantities
calculated for our data sets.
We were able also to keep track, for all data sets,
of each single bid, getting information about its value, the time
when it was made and the agent who made it. 
Data sets were anonymised 
and can be downloaded at the page {\tt filrad.homelinux.org}.
In the following, we focus our analysis
mainly on the data sets UBH and BM since, given
their size, allow to perform much better statistics.

\clearpage

\subsection{Analysis}

\begin{figure}[!ht]
\begin{center}
\vskip .7cm
\includegraphics[width=7cm]{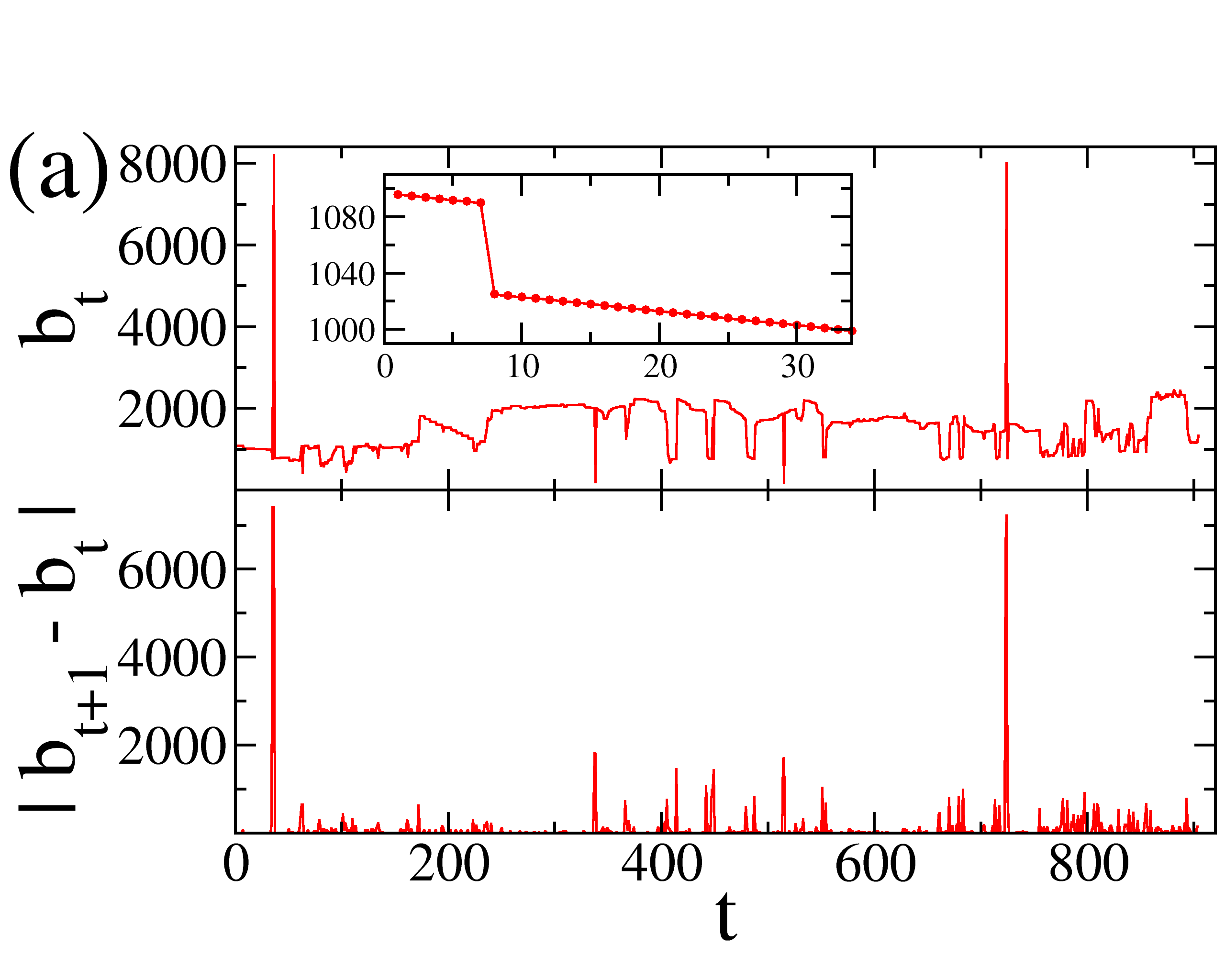}
\qquad
\includegraphics[width=7cm]{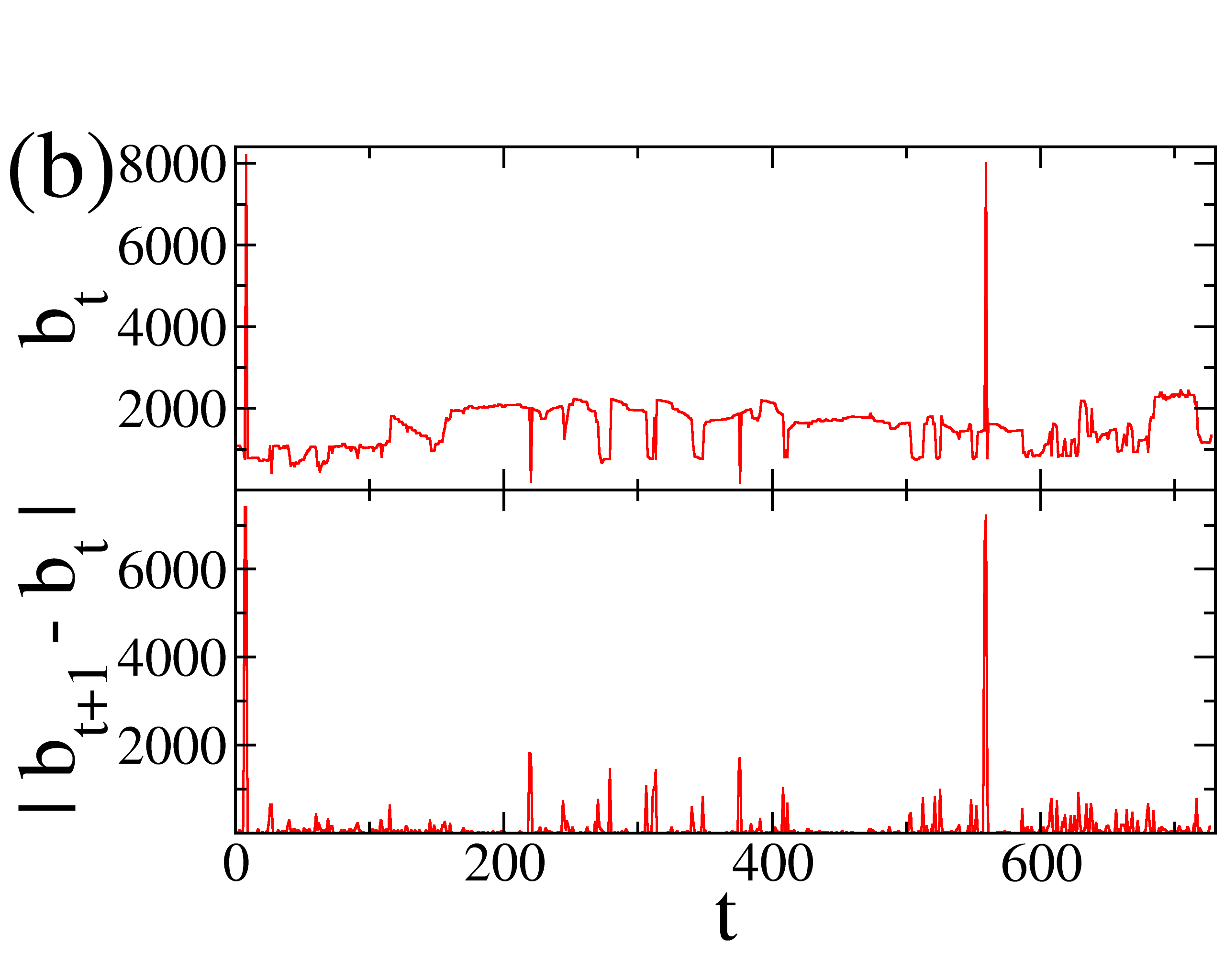}
\vskip .1cm
\caption{UBH data set. 
(a) In the top panel, the time series of bid values performed
by agent $u=1632$ in auction $a=19$ is shown.
In particular we zoom into a region where ``abnormal''
movements are present. In the bottom panel,
we plot the time series of the
length of  her jumps $d_t = \left| b_{t+1} - b_t \right|$. (b)
Same plots as those appearing in panel a, 
but for the cleaned version of the time series.
The indices $a$ and $u$ refer to the anonymised version
of the UBH data set.}
\label{fig:example_user}
\end{center}
\end{figure}

\noindent Fixed an auction $a$ and an agent $u$, we consider
the temporal series of her bids, whose total
number is denoted by $T_{a,u}$. This series
is basically a list of integers $b_1$, $b_2$, \dots, $b_T$,
where we have suppressed the indices $a$ and $u$ for shortness
of notation.
Their value is defined over the interval $[1,M]$. All $b$s are
different each other since no agent bids on a certain value
more than once. 
A typical example of these time series is reported in 
the top panel of Figure~\ref{fig:example_user}a. We calculate
the gap or difference between subsequent bid values and indicate
it with $d_t = \left| b_{t+1} - b_t \right|$. Given a list
of $T$ bid values we can in fact extract a list of
$T-1$ differences between consecutive bids. In 
the bottom panel of Figure~\ref{fig:example_user}a, we plot
the time series  $d_t$ for the same agent whose bid time series 
is plotted in the upper panel.

\clearpage

\subsection{Cleaning the data sets}
\noindent Generally, more ``professional'' agents perform
random searches followed by systematic coverages
of intervals. A typical example is
shown in the inset of the upper panel of Figure~\ref{fig:example_user}a.
Here a zoom of the time series appearing in the main plot is
reported. Systematic coverages are performed
by selecting a range of bid values and then placing a bid
on each single value in that interval. This is an opportunity
offered by the web site hosting auctions. Each bid
in this case is characterized by the same time stamp.
Such occurrence is likely for agents who make
a significant number of bids, but becomes less relevant
for agents who invest relatively small amount of money.
We cleaned data by removing all pieces of
the time series corresponding to this ``abnormal''
behavior. It should be remarked that the gaps 
between consecutive bids which can be measured
in these regions are in the majority
of the cases equal to one. Including
these regions will influence only gaps equal to unity
by overestimating their presence. We decided
to remove such systematic coverages in order to
focus our attention only on  ``normal'' bidding strategies.
The result, after the cleaning procedure of time series
shown in Figure~\ref{fig:example_user}a, is reported in
the upper panel of Figure~\ref{fig:example_user}b. The gaps
between consecutive bids in the cleaned time series
are reported in the bottom panel of Figure~\ref{fig:example_user}b.
\\ 
\begin{figure}[!ht]
\begin{center}
\vskip .7cm
\includegraphics[width=7cm]{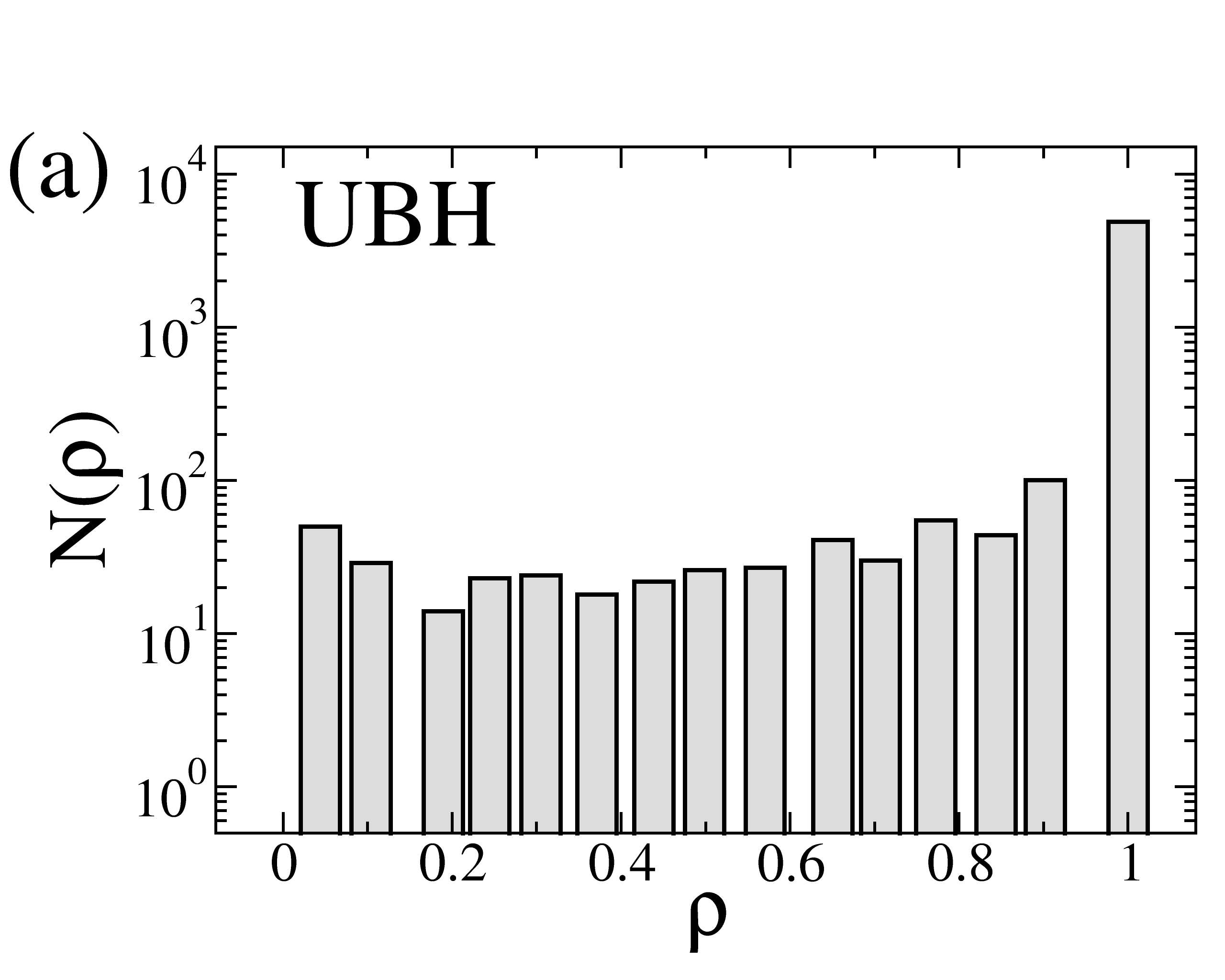}
\qquad
\includegraphics[width=7cm]{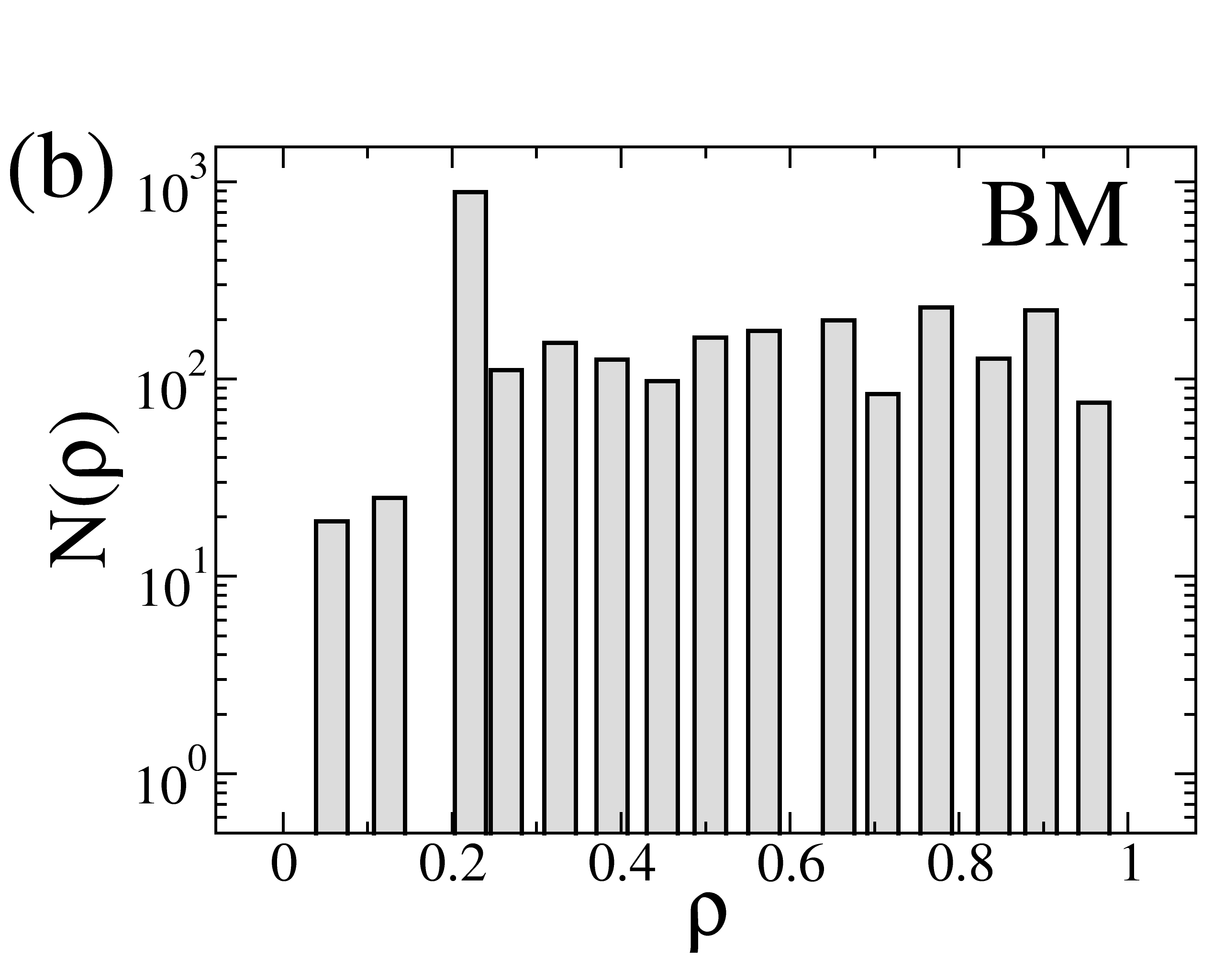}
\vskip .1cm
\caption{Number of agents $N\left(\rho \right)$
performing ``normal'' search strategies at rate $\rho$.}
\label{fig:ratio}
\end{center}
\end{figure}
\noindent As additional information, in Figure~\ref{fig:ratio}, 
we measure the number of agents $N\left(\rho\right)$ 
performing a ratio $\rho$ of bids made using
a normal strategy (i.e., after removing systematic coverages)
and the total number of bids placed. 
\\
In the following analysis, we consider 
only cleaned time series, where systematic coverages have
been deleted.

\clearpage

\subsection{Statistics of the length of the jumps}

\begin{figure}[!ht]
\begin{center}
\vskip .7cm
\includegraphics[width=7cm]{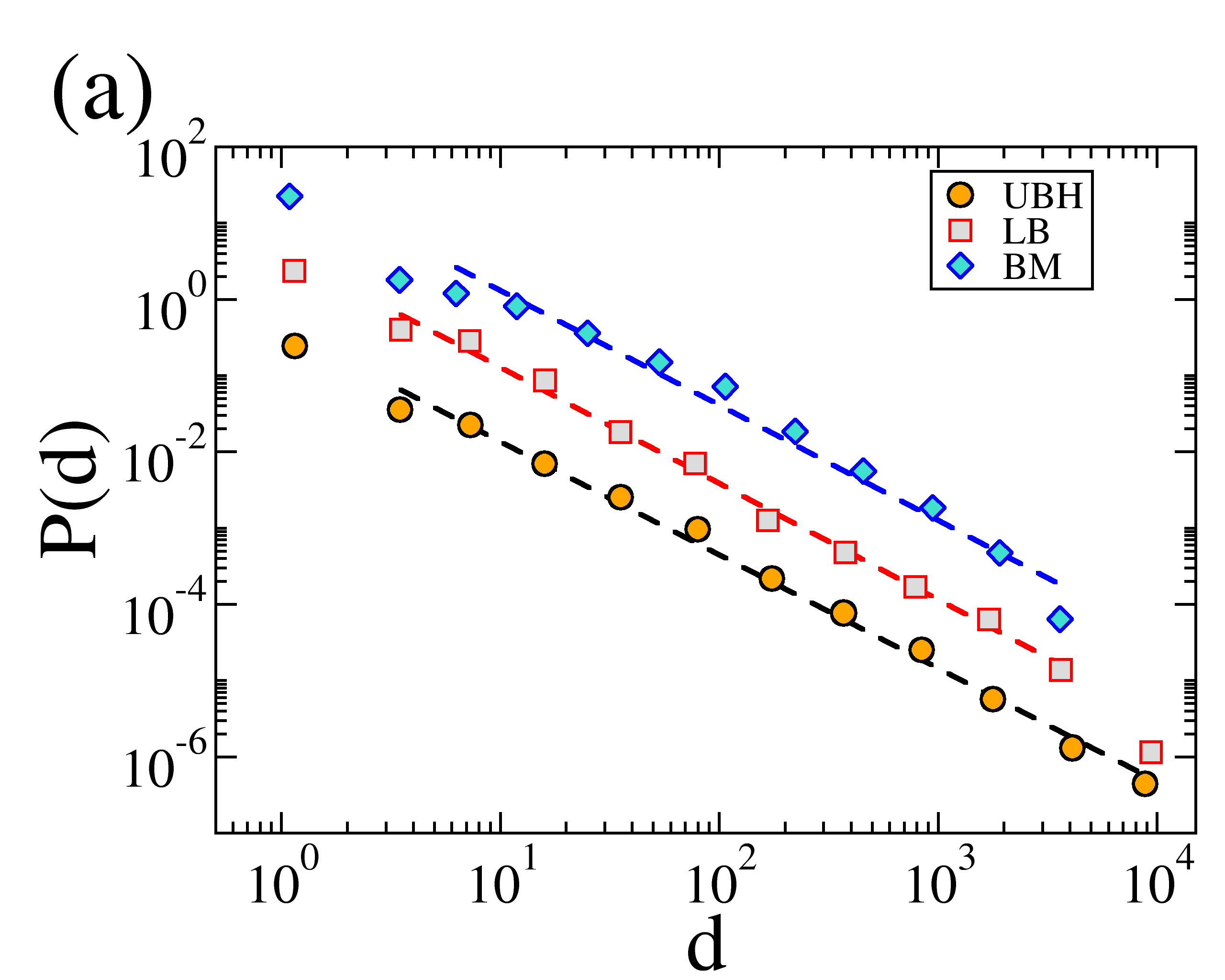}
\qquad
\includegraphics[width=7cm]{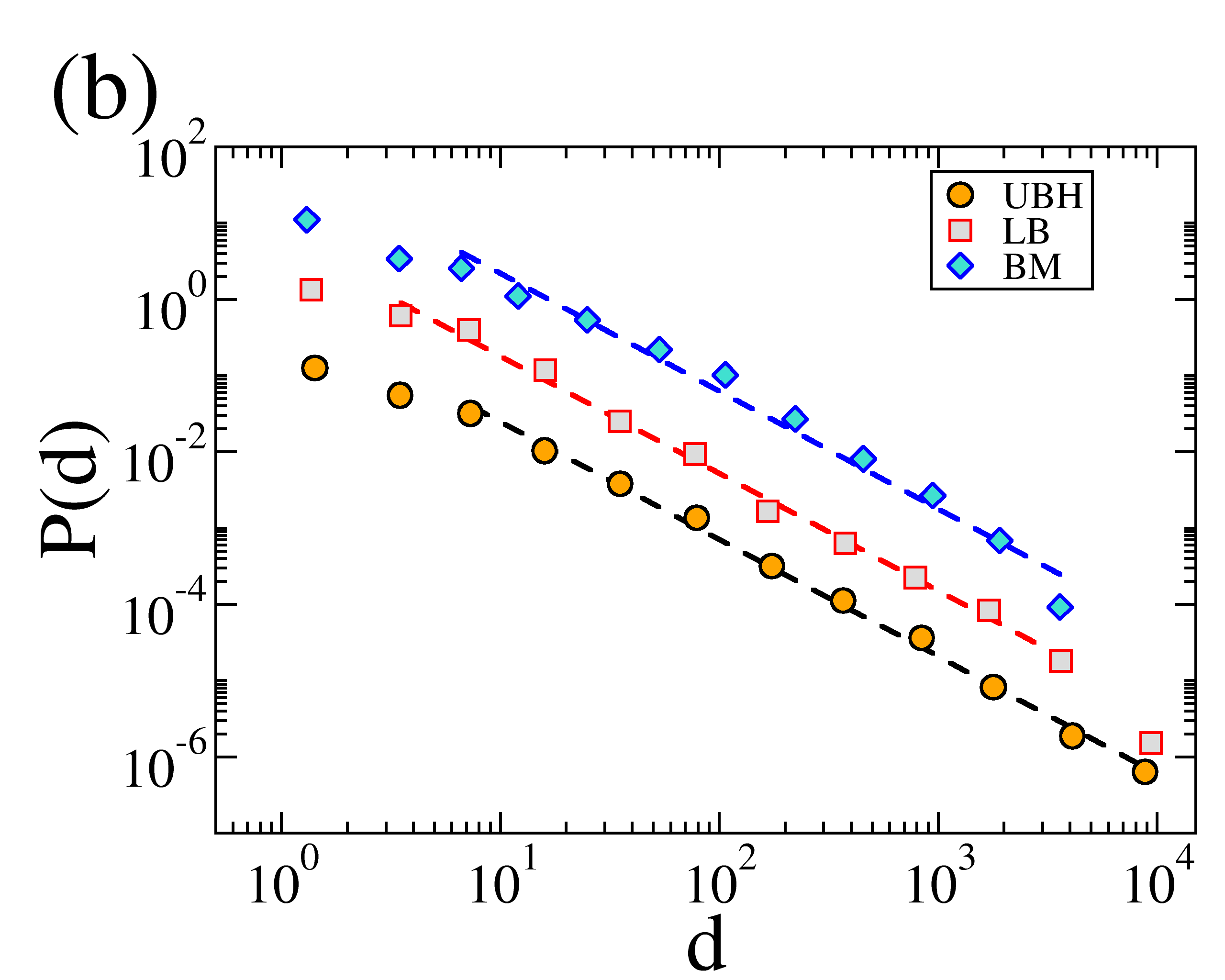}
\vskip .1cm
\caption{(a) Probability distribution function
$P\left(d\right)$ calculated over all agents and auctions
in the data sets UBH (orange circles), LB
(gray squares) and BM (turquoise diamonds). 
Dashed lines stand for best power-law fits (least square). We find
$\alpha=1.54(2)$ [black], $\alpha=1.54(3)$ [red] and $\alpha=1.54(5)$ [blue].
(b) Same as in panel a, but for cleaned
time series. Dashed lines have slopes $\alpha=1.49(2)$ [black],
$\alpha=1.52(2)$ [red] and $\alpha=1.51(5)$ [blue]. 
The data of this figure also reported in Figure~3A of the main text. Curves calculated 
for LB and BM data sets
have been vertically shifted for clarity.}
\label{fig:global}
\end{center}
\end{figure}

\begin{figure}[!ht]
\begin{center}
\vskip .7cm
\includegraphics[width=7cm]{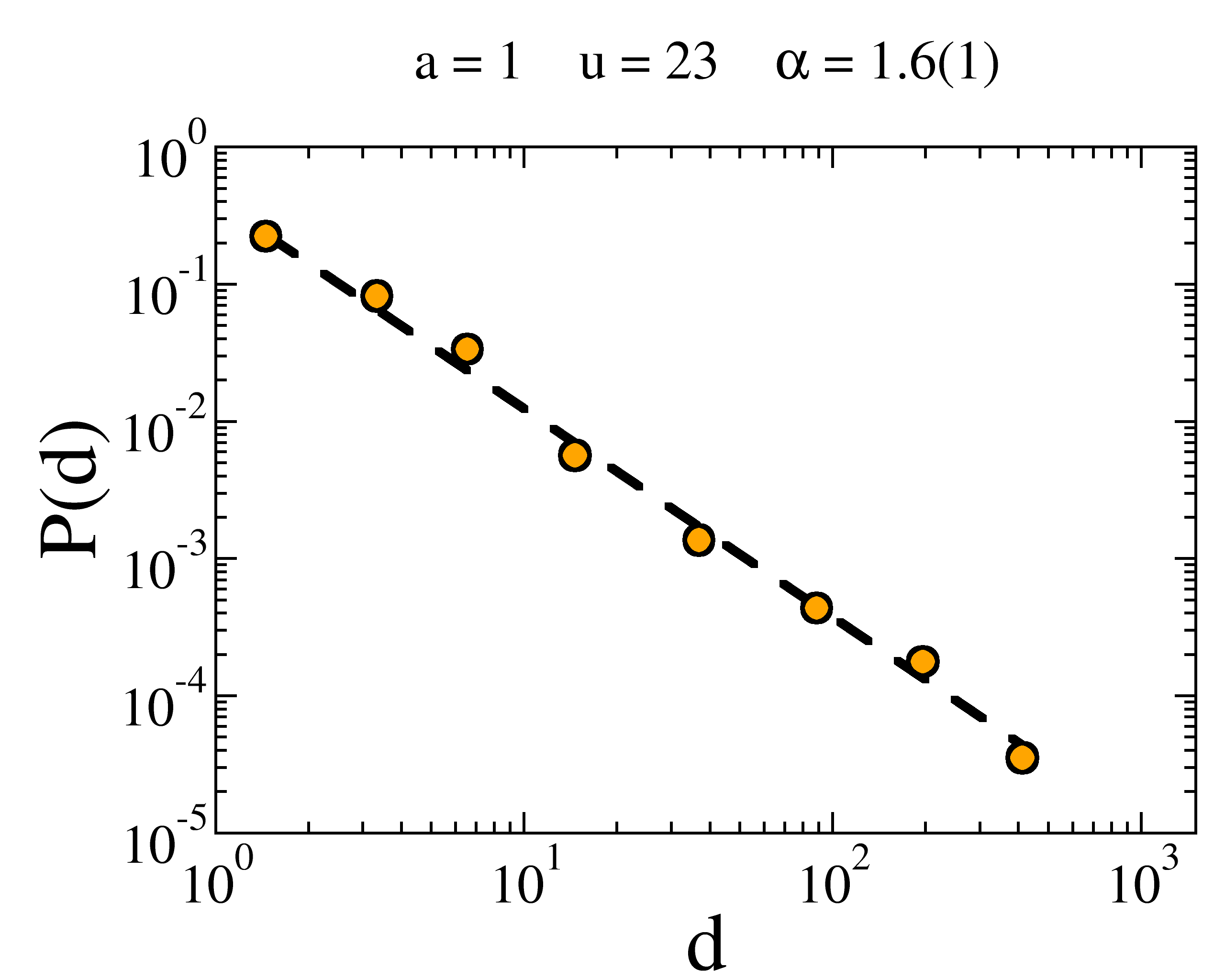}
\qquad
\includegraphics[width=7cm]{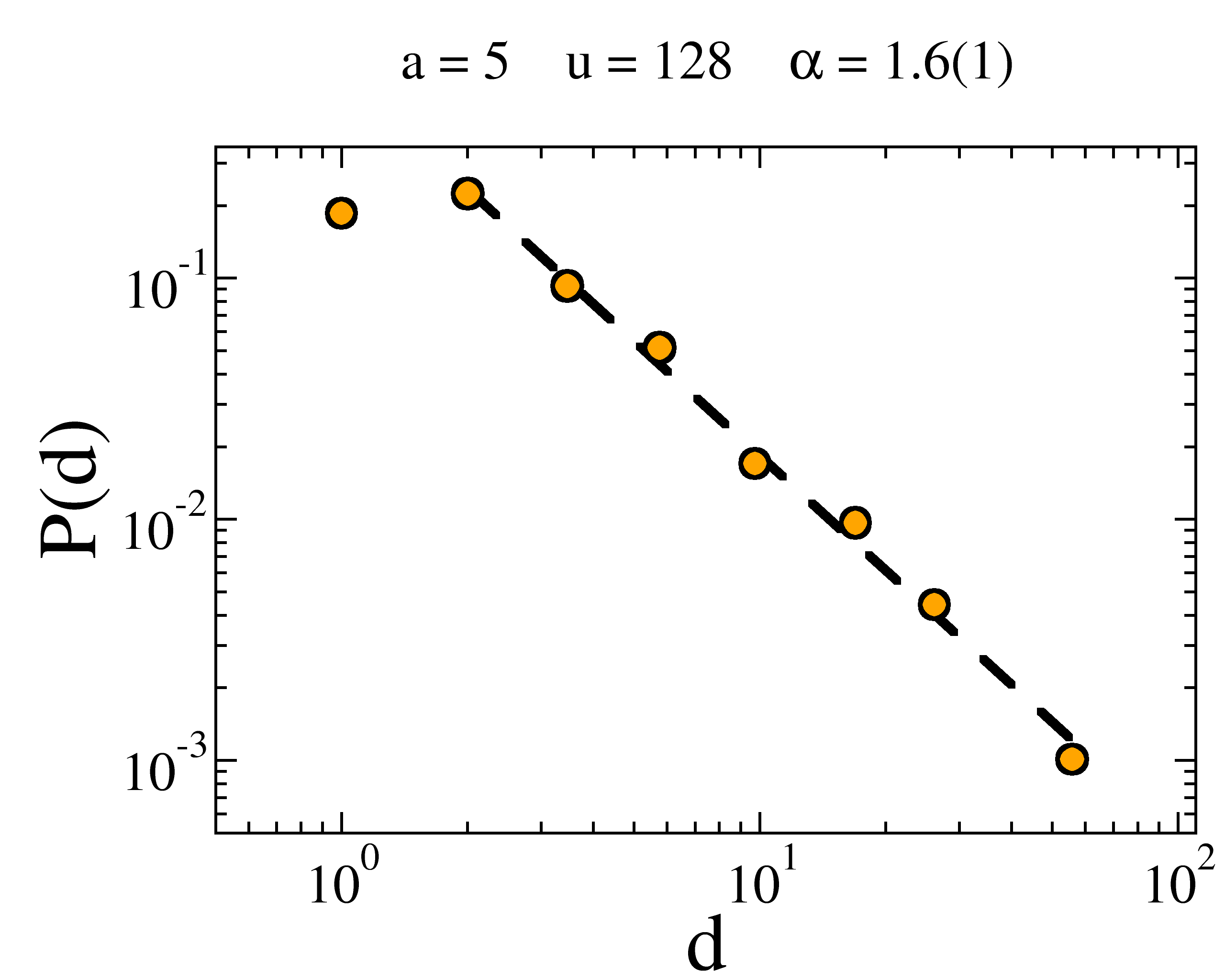}
\vskip .2cm
\includegraphics[width=7cm]{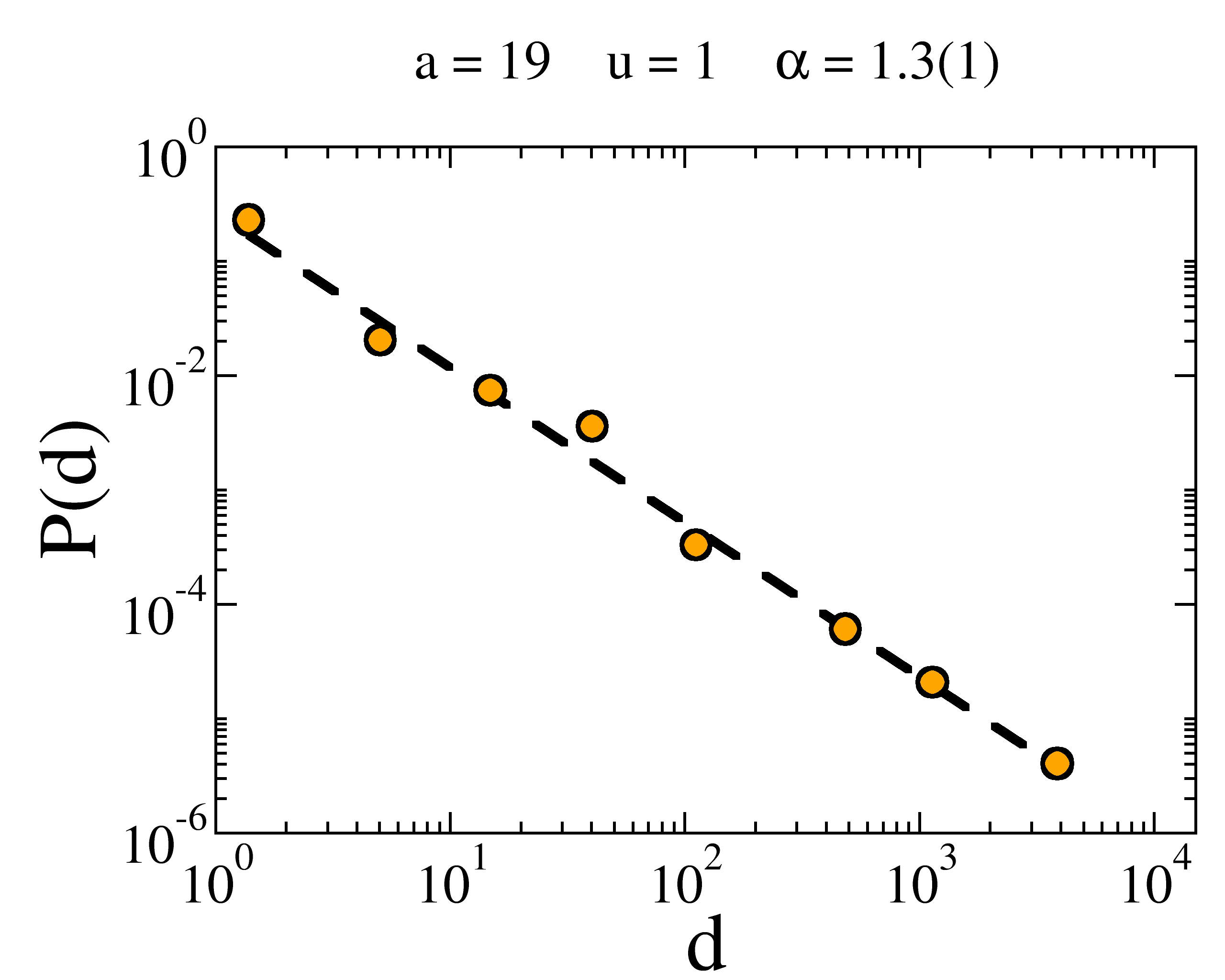}
\qquad
\includegraphics[width=7cm]{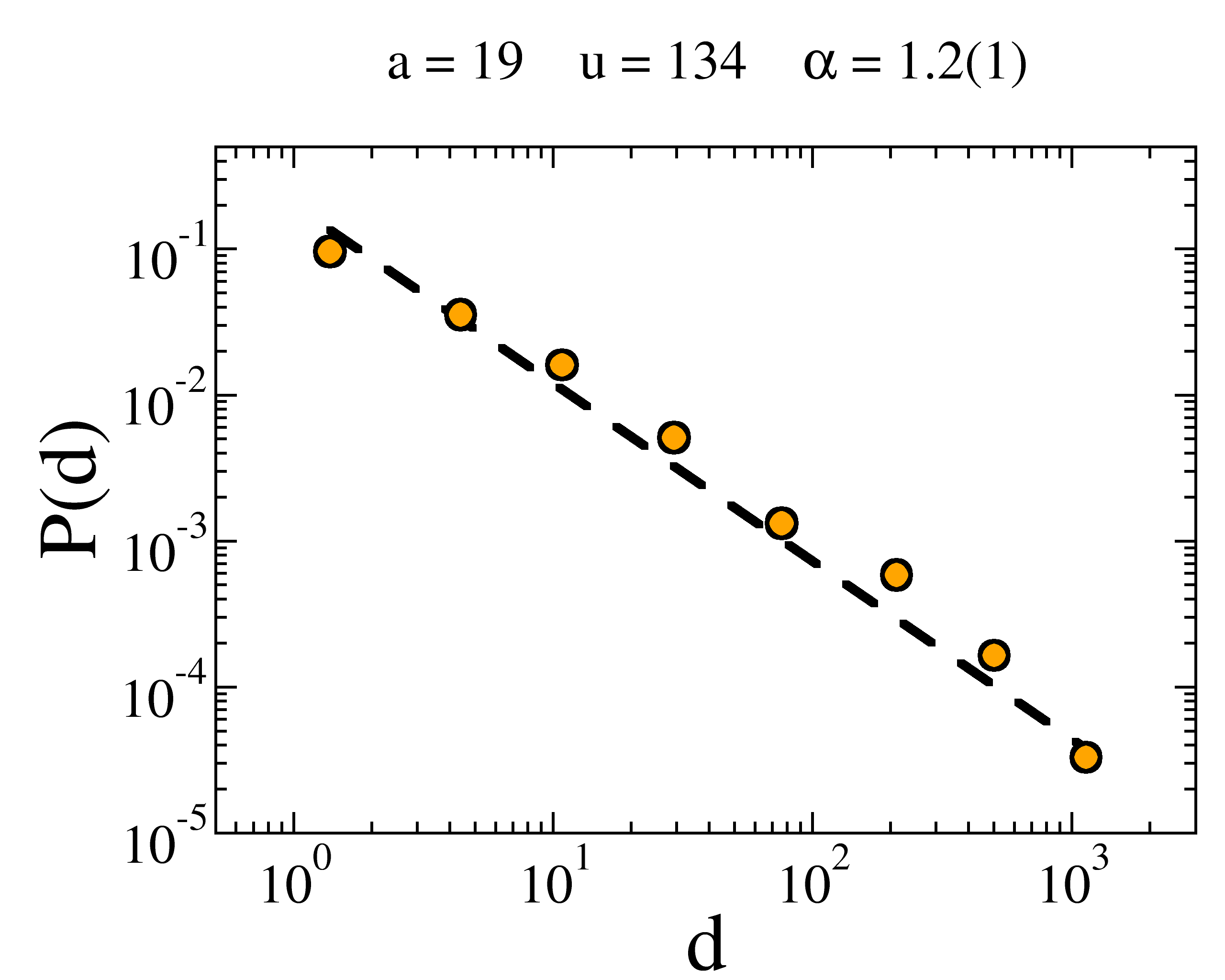}
\vskip .2cm
\includegraphics[width=7cm]{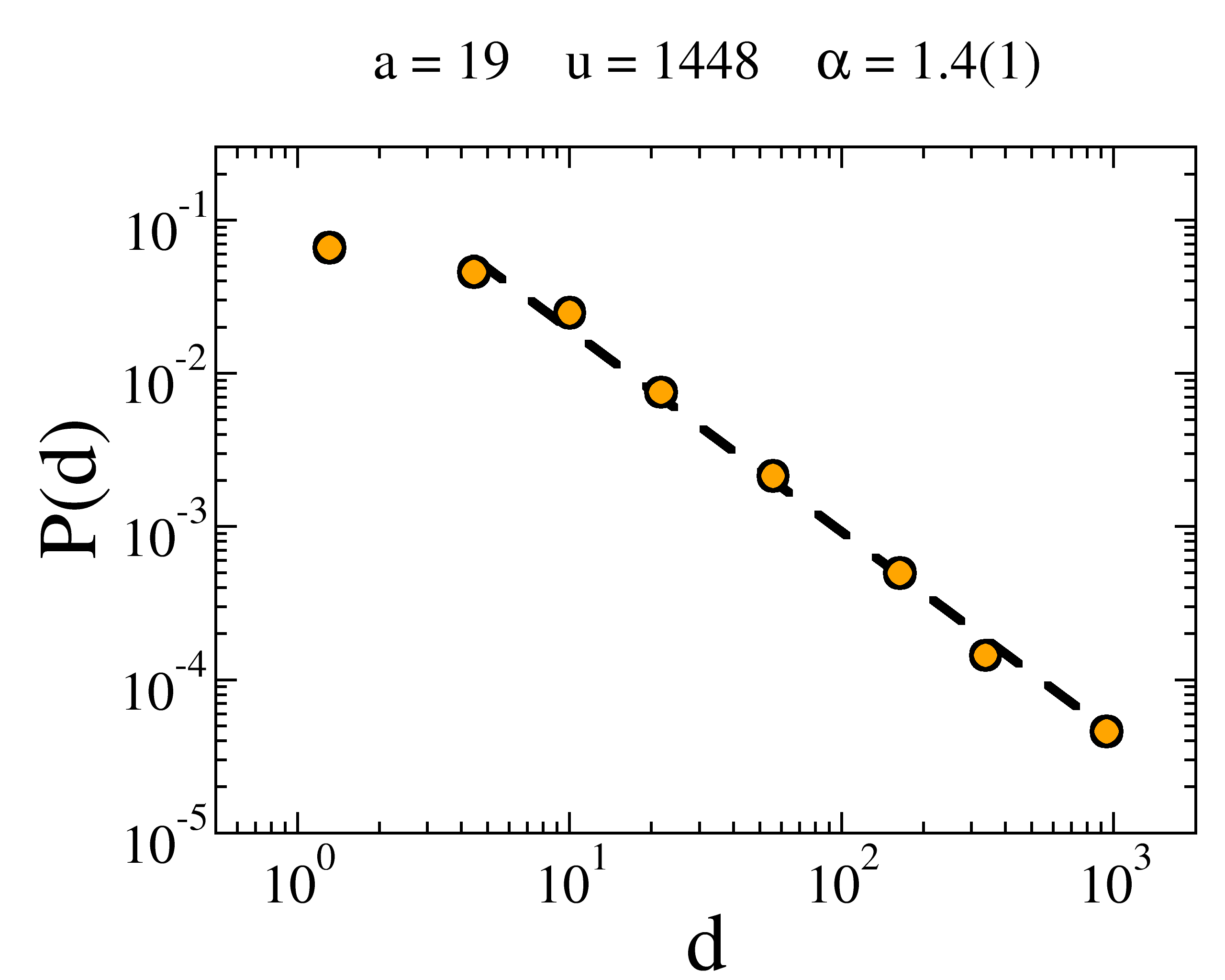}
\qquad
\includegraphics[width=7cm]{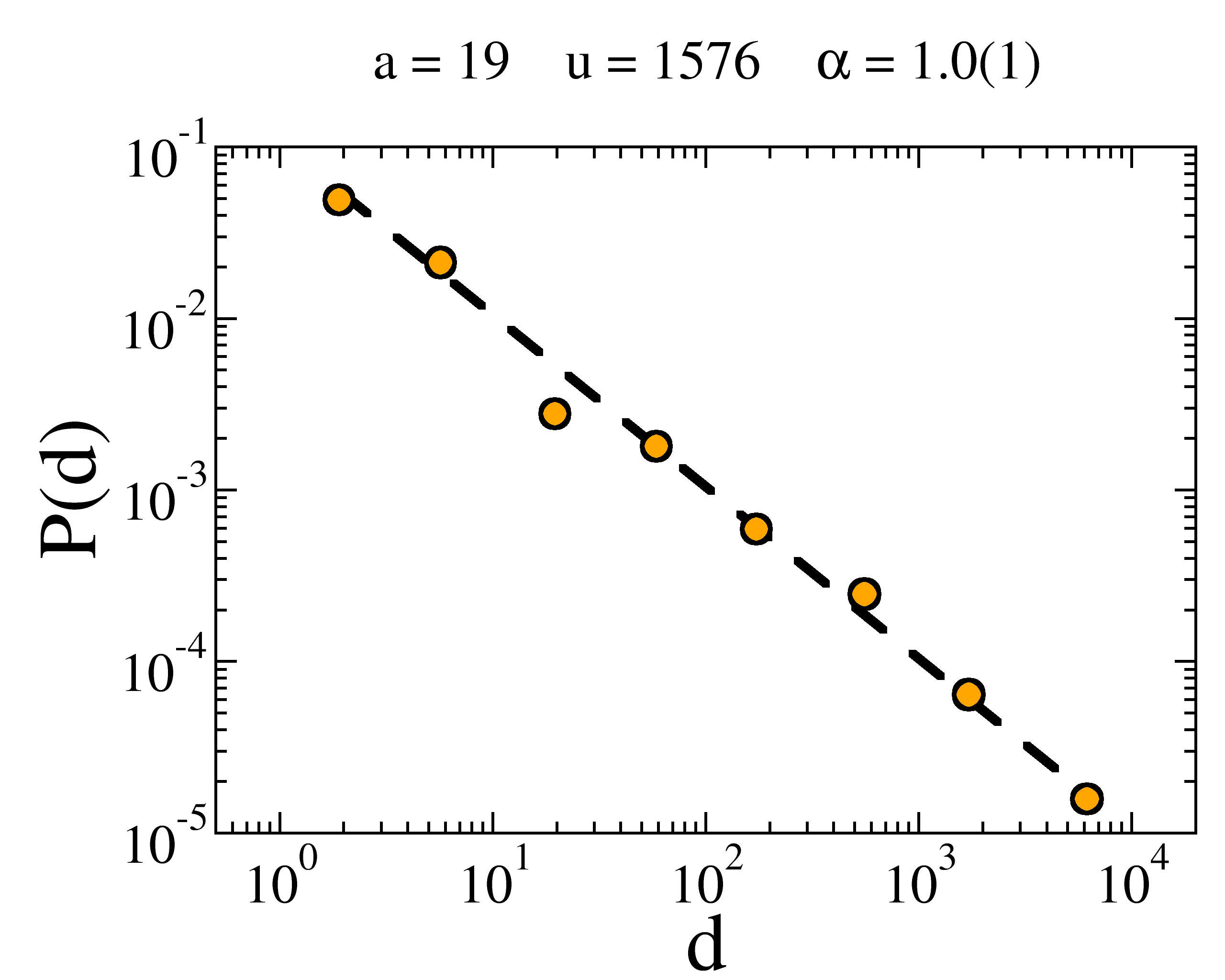}

\vskip .1cm
\caption{UBH data set. Probability distribution function $P\left(d\right)$
measured for agent $u$ in auction $a$. We show
several $P\left(d\right)$s for different
pairs $u$ and $a$. Dashed lines denote the best power-law fit 
(least square) $P\left(d\right) \sim d^{-\alpha} $ obtained
for the data.}
\label{fig:exponents_single_users_1}
\end{center}
\end{figure}

\begin{figure}[!ht]
\begin{center}
\vskip .7cm
\includegraphics[width=7cm]{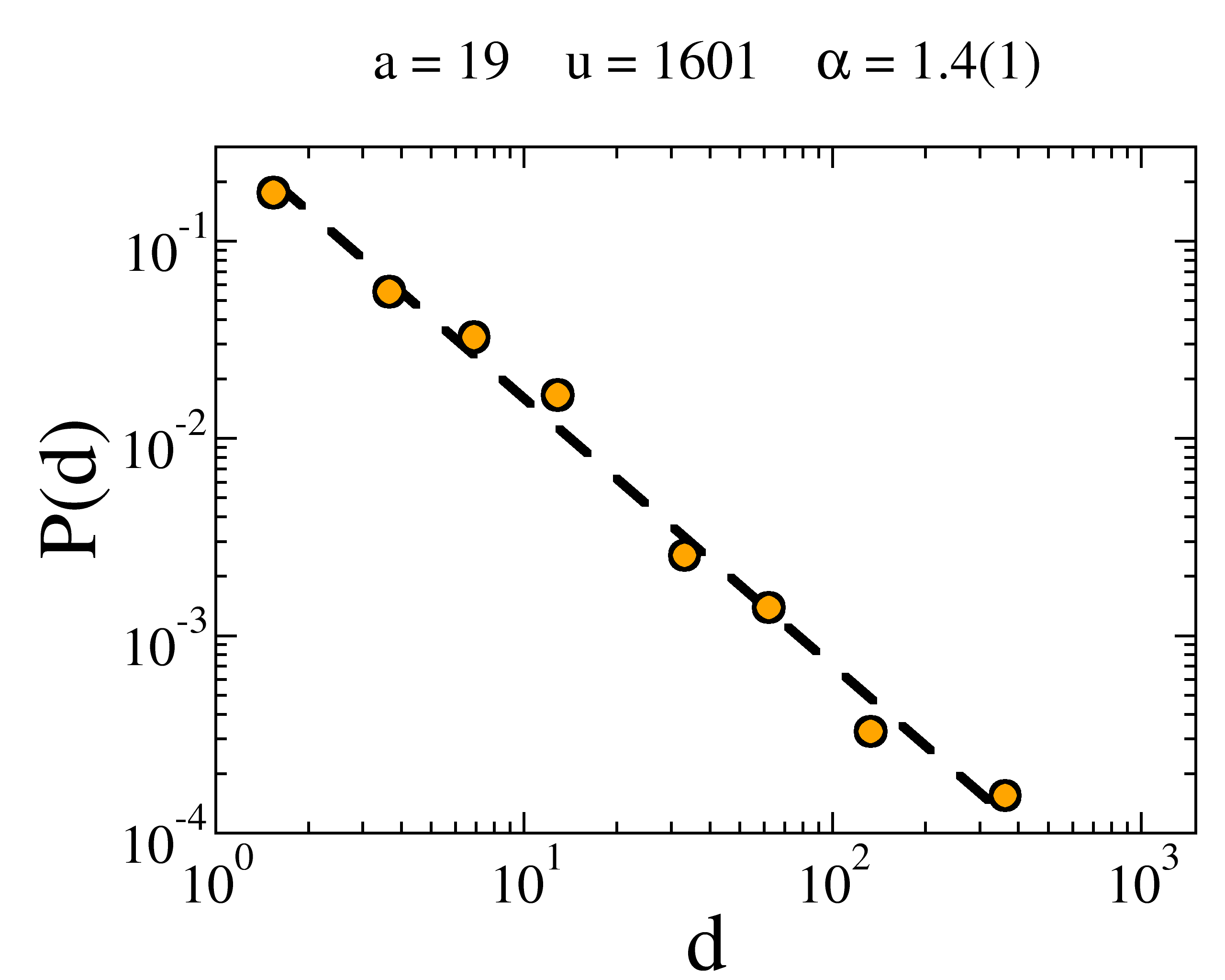}
\qquad
\includegraphics[width=7cm]{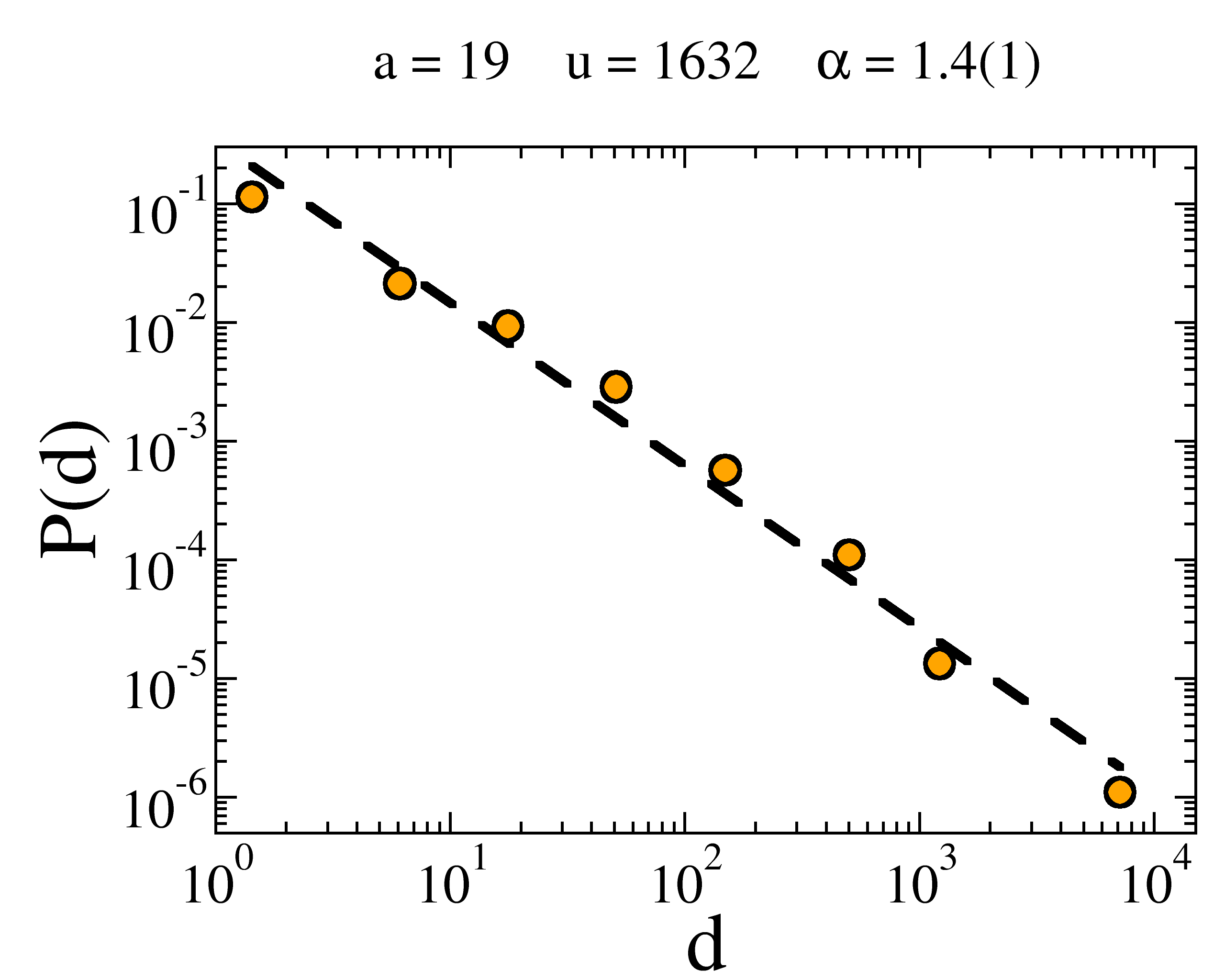}
\vskip .2cm
\includegraphics[width=7cm]{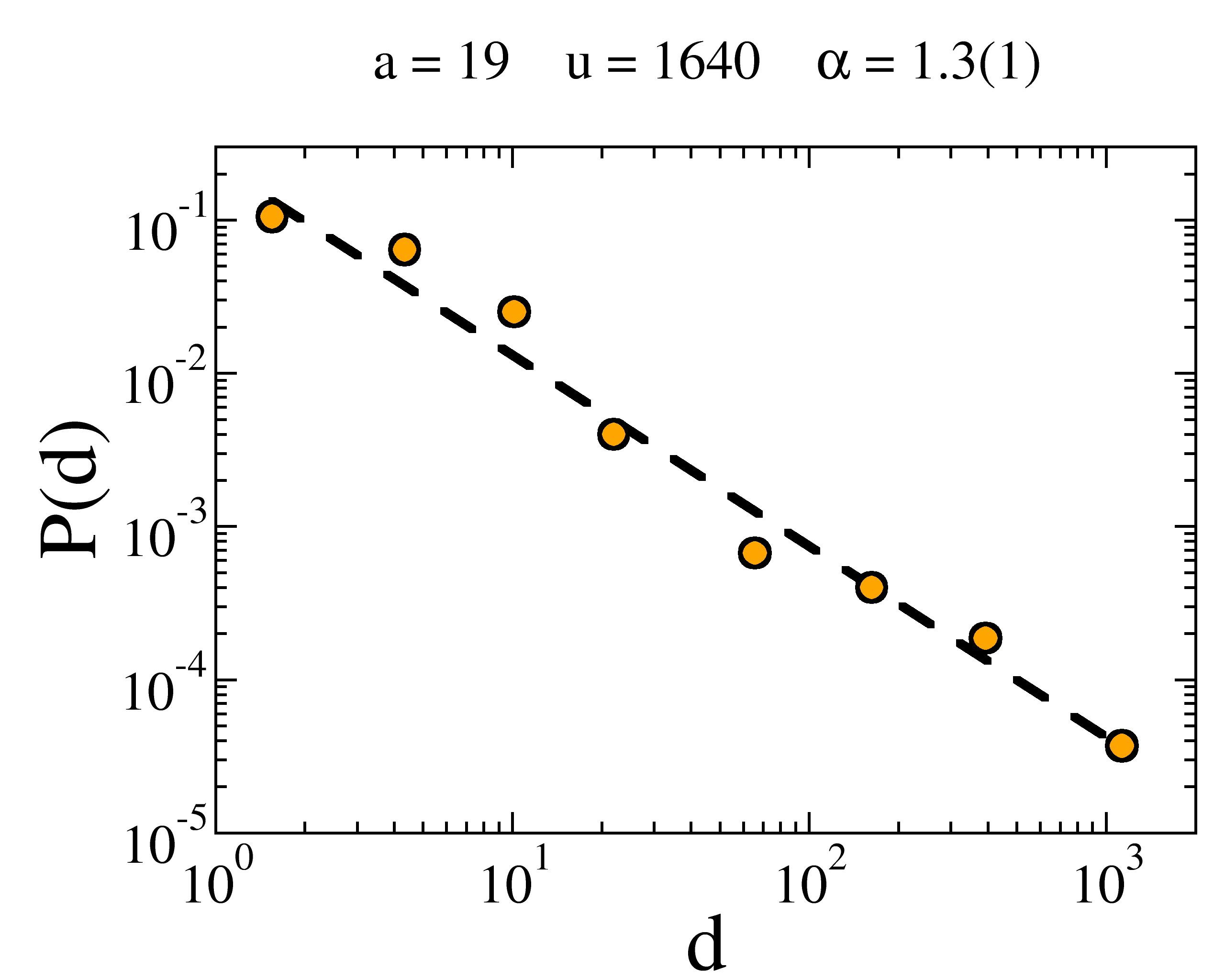}
\qquad
\includegraphics[width=7cm]{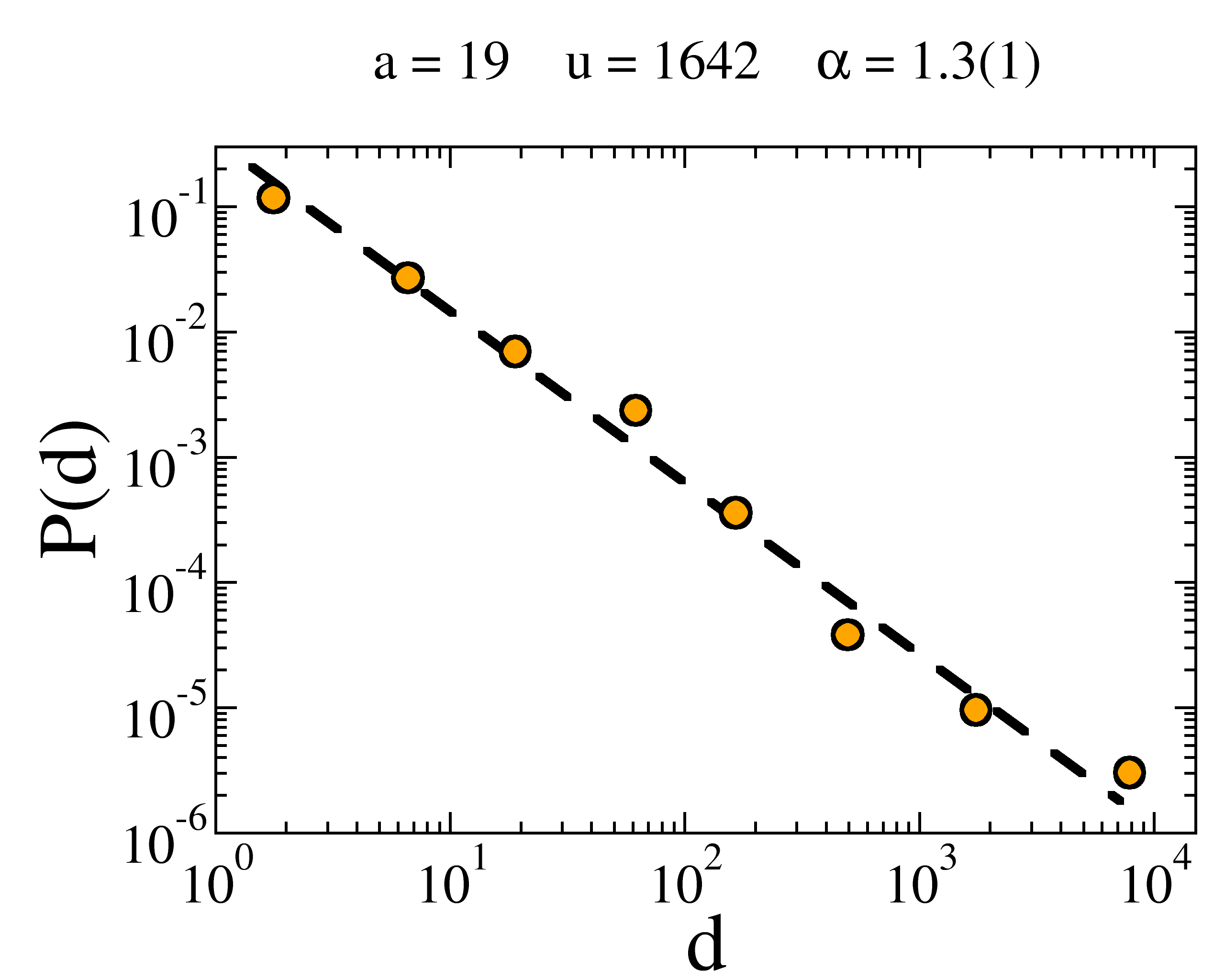}
\vskip .2cm
\includegraphics[width=7cm]{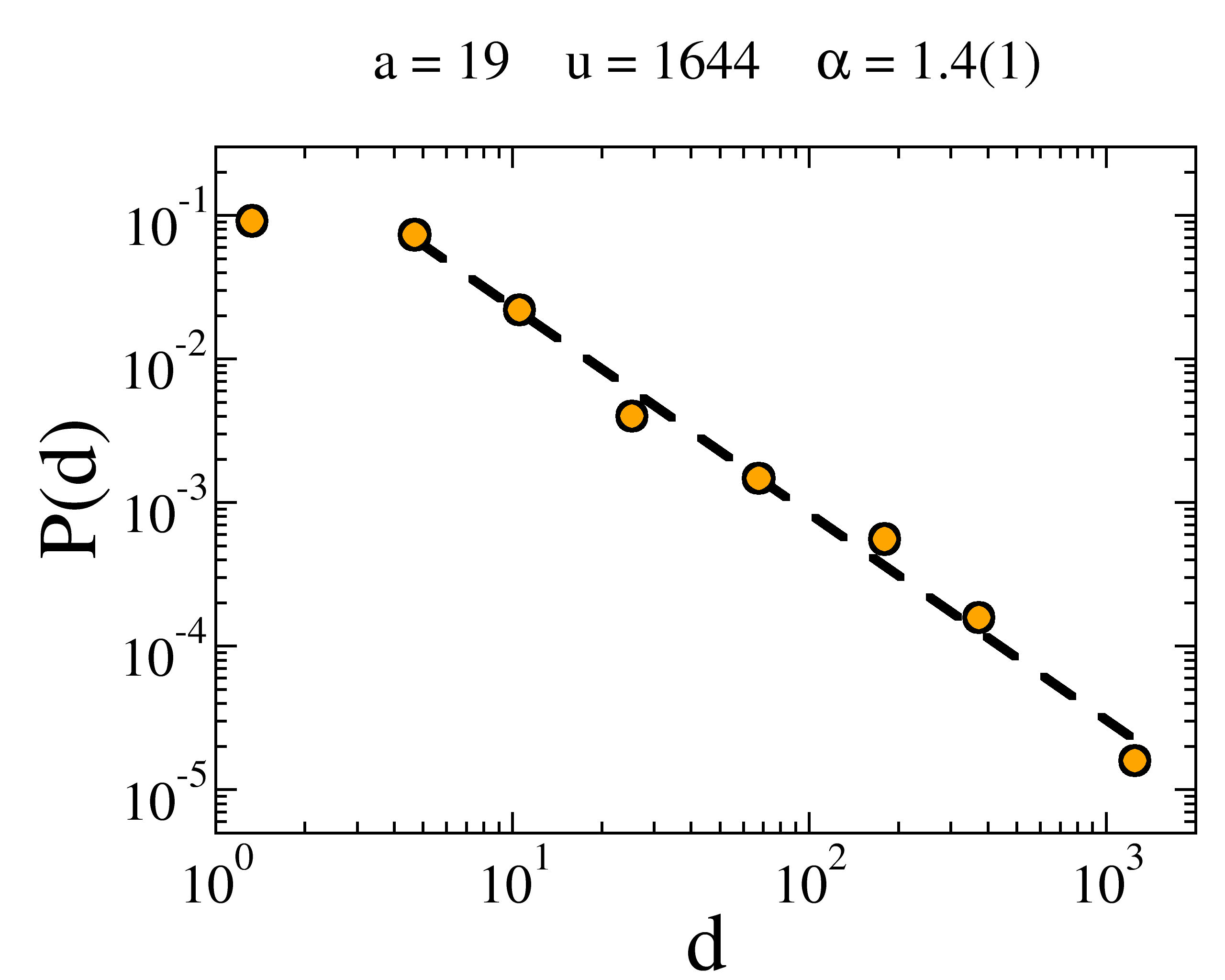}
\qquad
\includegraphics[width=7cm]{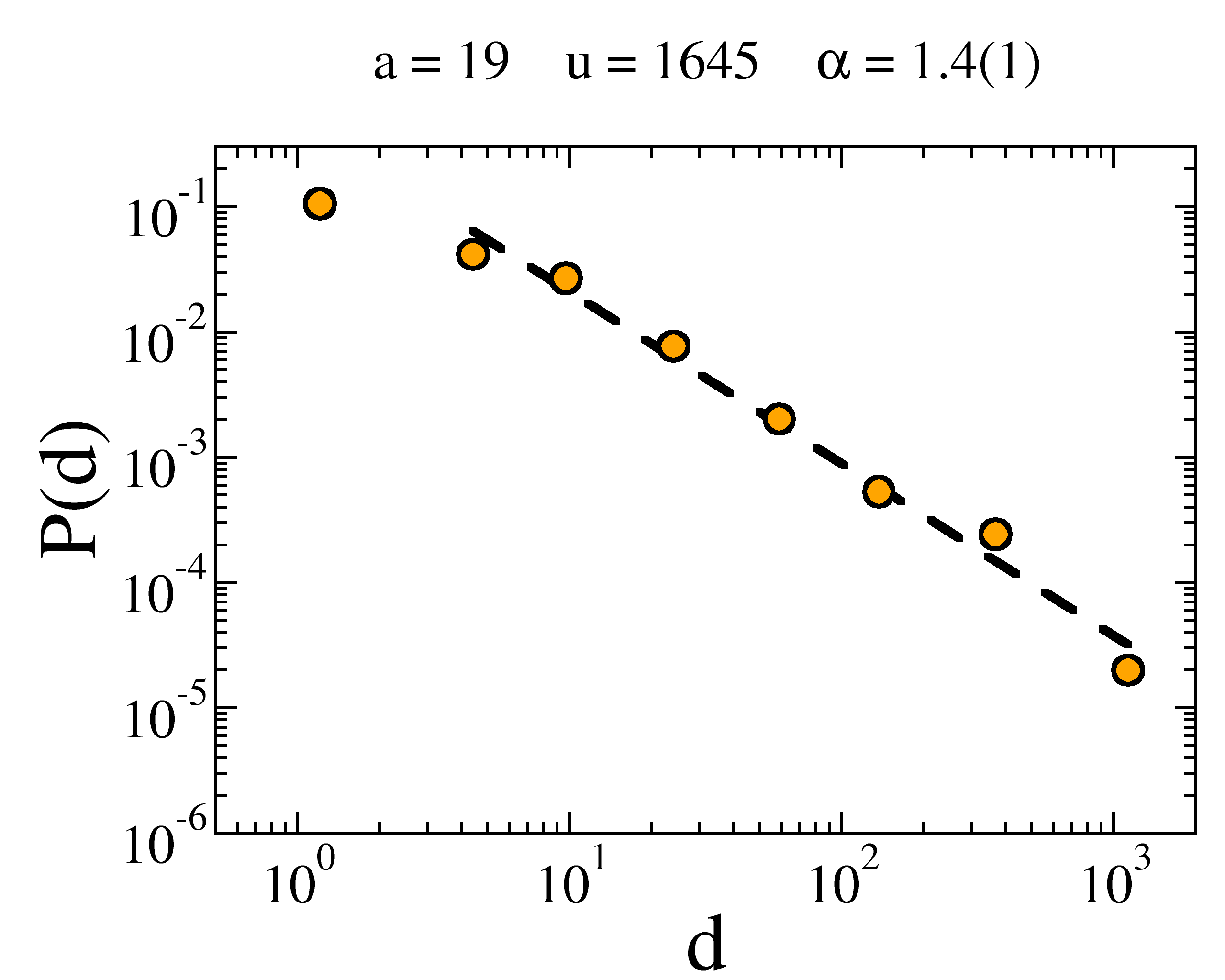}

\vskip .1cm
\caption{UBH data set. Same as Figure~\ref{fig:exponents_single_users_1}.}
\label{fig:exponents_single_users_2}
\end{center}
\end{figure}

\begin{figure}[!ht]
\begin{center}
\vskip .7cm
\includegraphics[width=7cm]{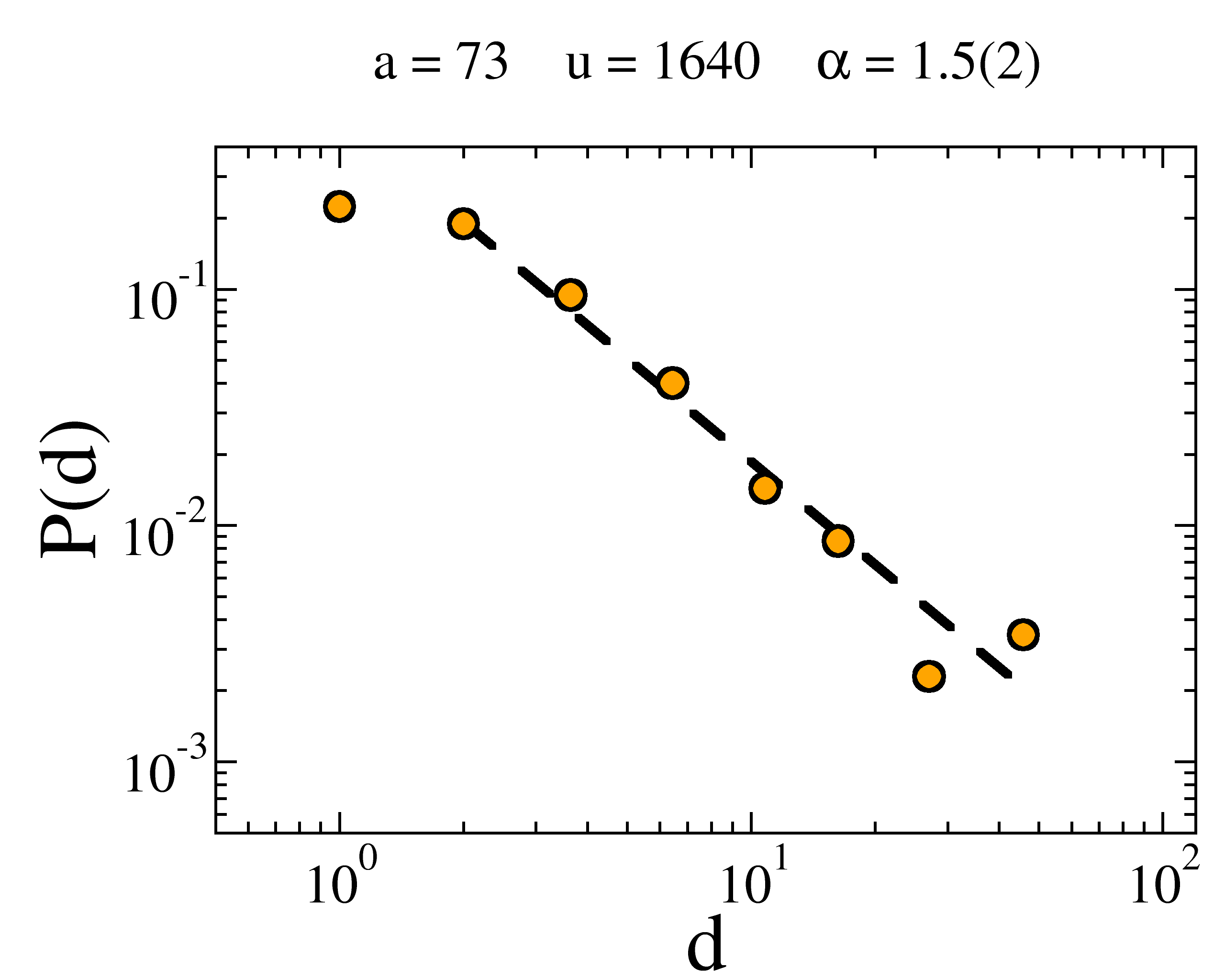}
\qquad
\includegraphics[width=7cm]{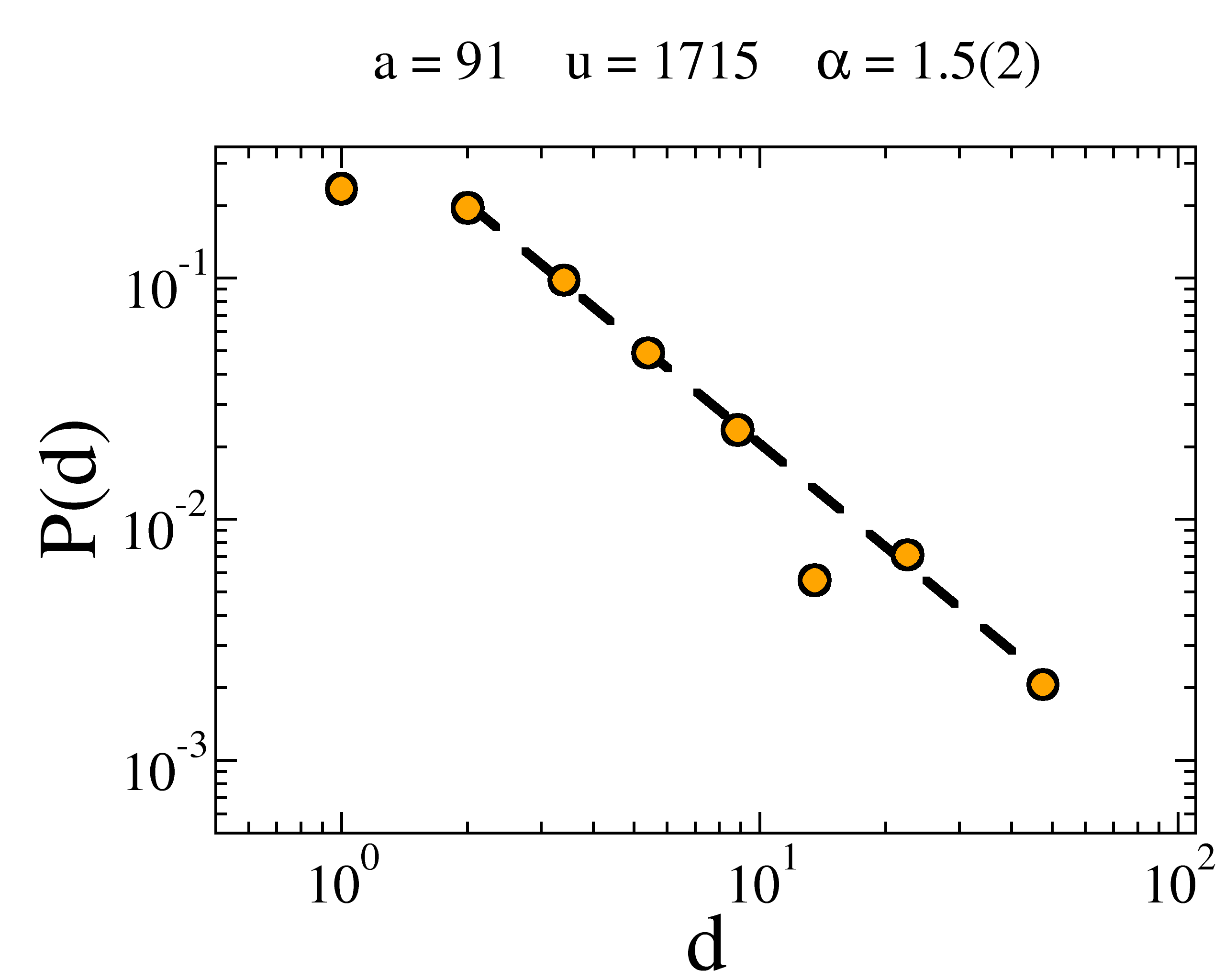}
\vskip .1cm
\caption{UBH data set. Same as Figure~\ref{fig:exponents_single_users_1} and 
~\ref{fig:exponents_single_users_2}.}
\label{fig:exponents_single_users_3}
\end{center}
\end{figure}

\begin{figure}[!ht]
\begin{center}
\vskip .7cm
\includegraphics[width=7cm]{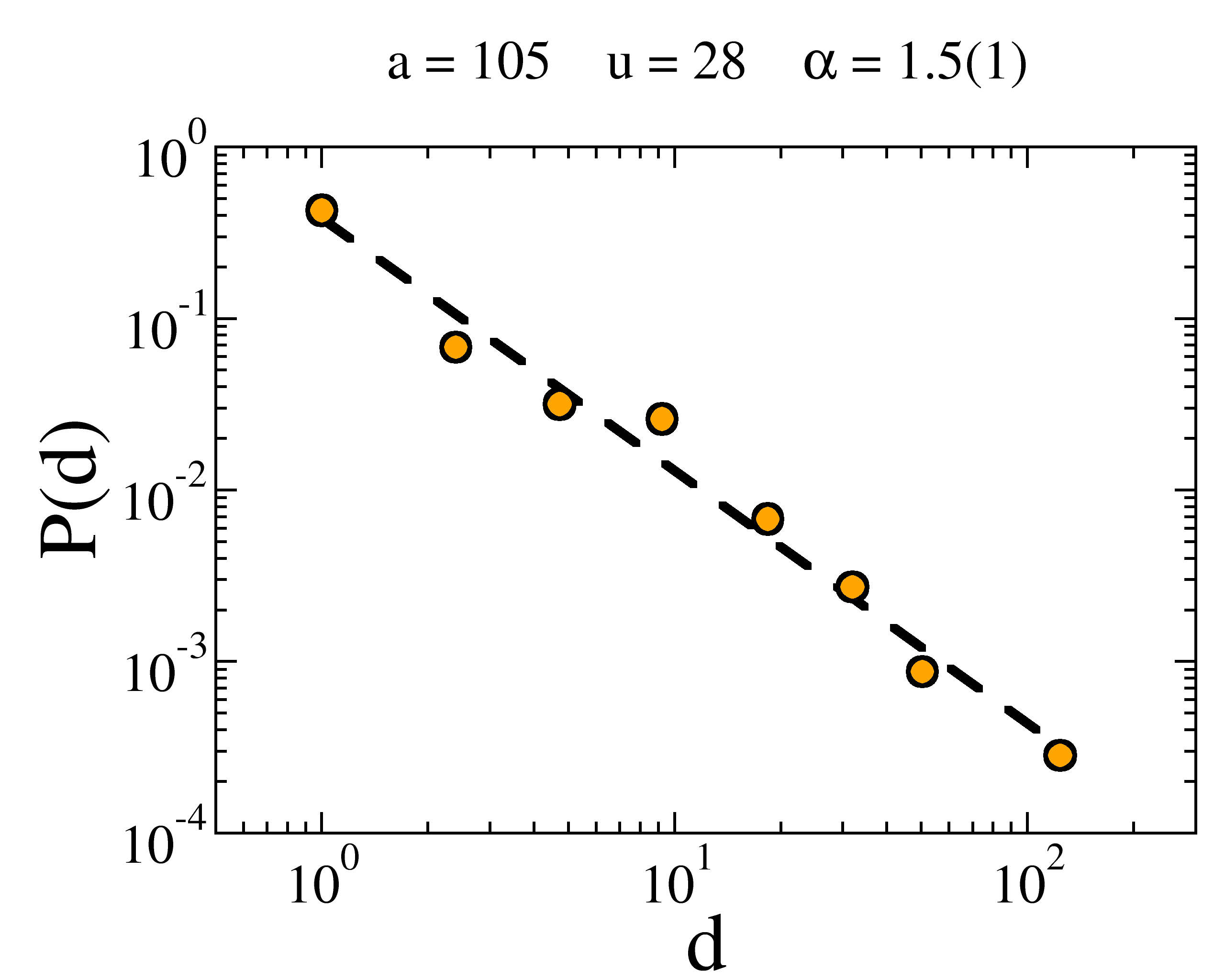}
\qquad
\includegraphics[width=7cm]{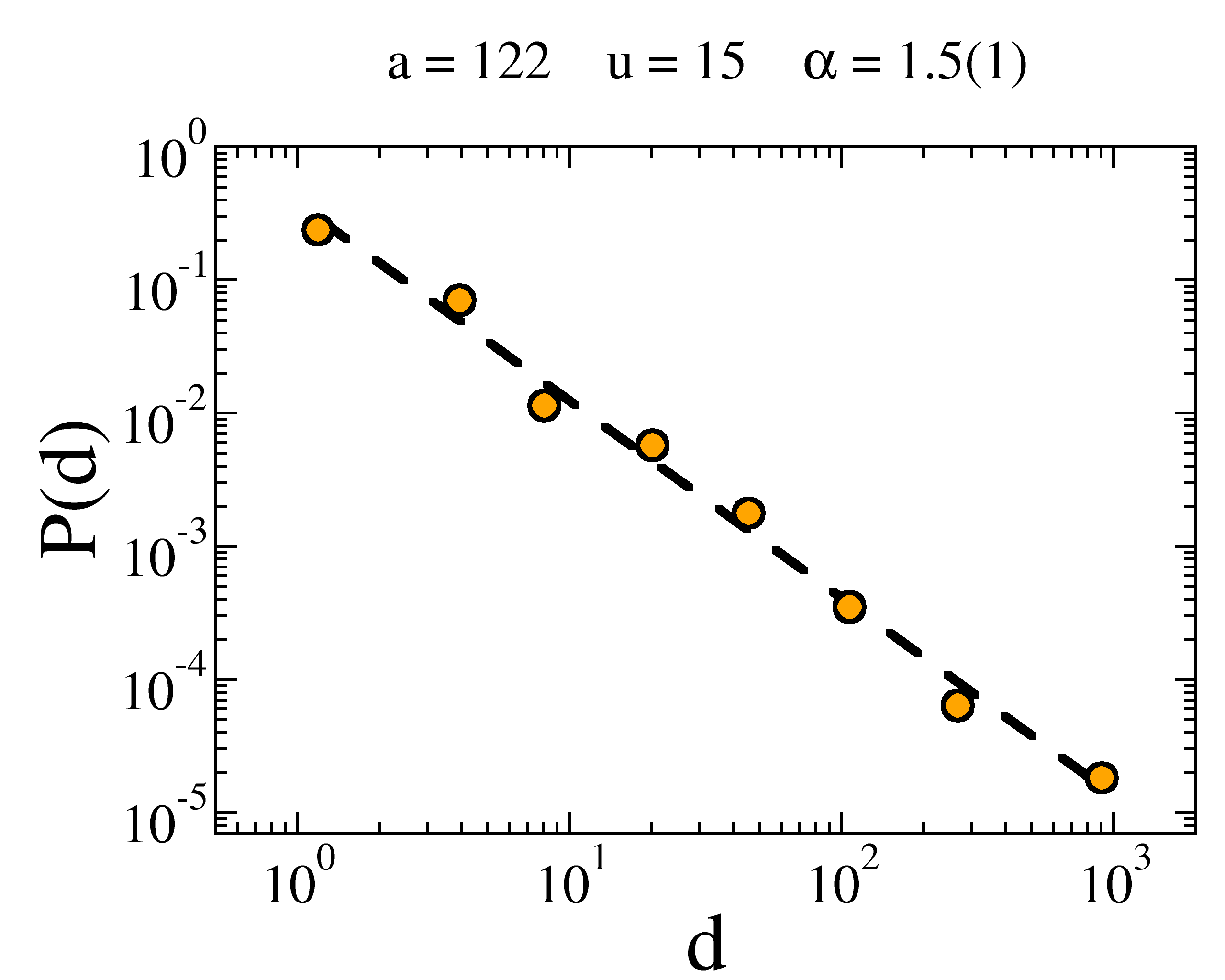}
\vskip .2cm
\includegraphics[width=7cm]{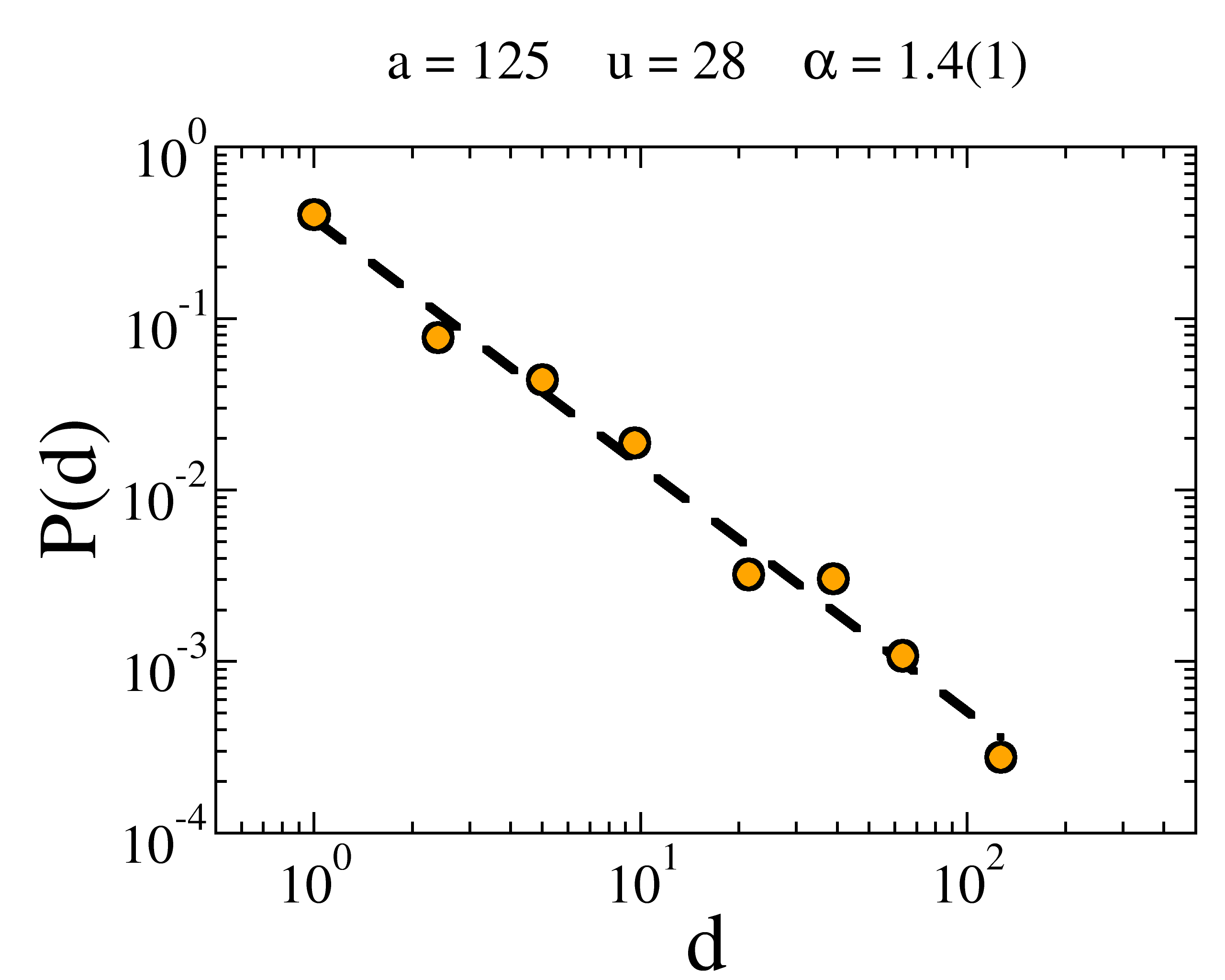}
\qquad
\includegraphics[width=7cm]{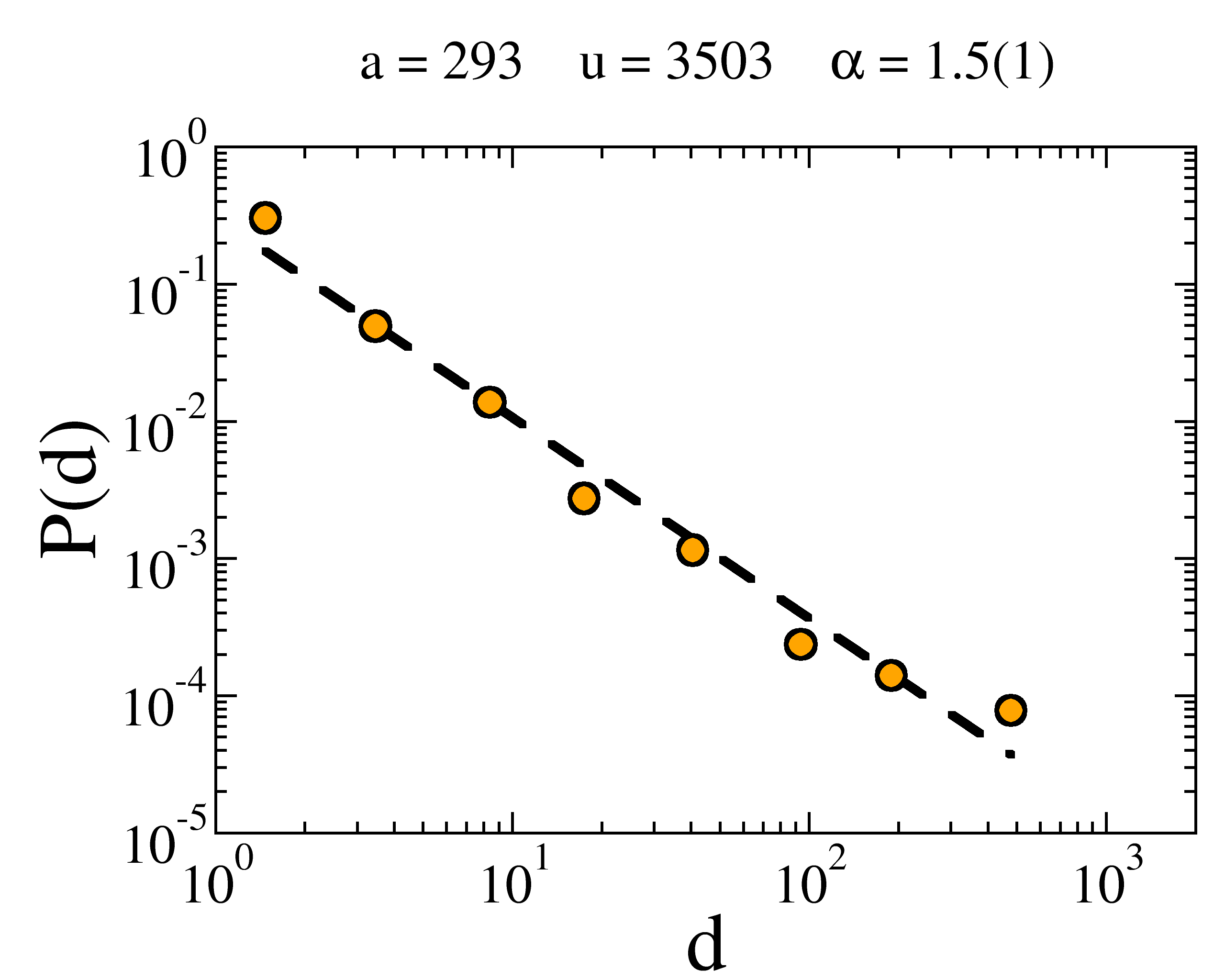}
\vskip .2cm
\includegraphics[width=7cm]{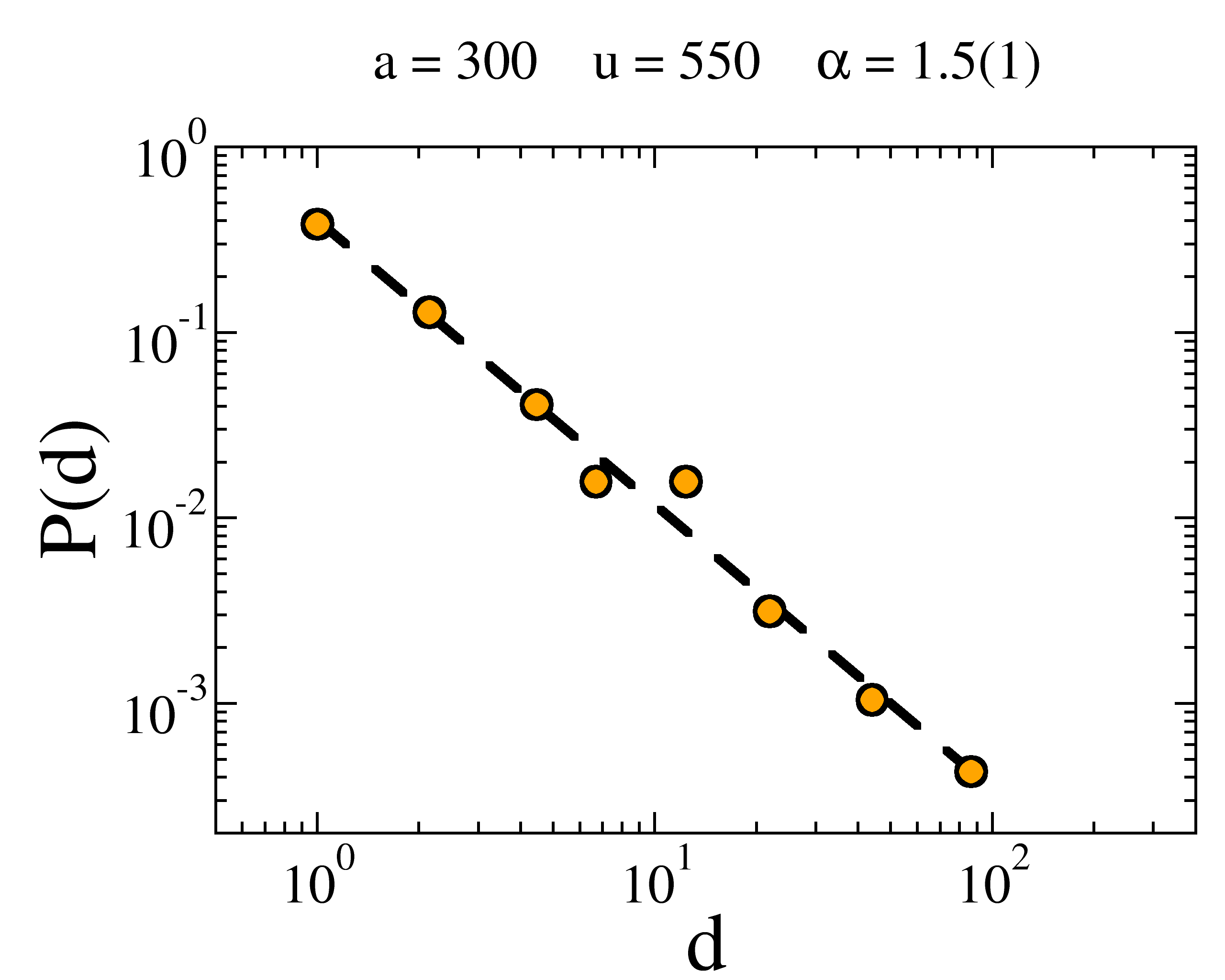}
\qquad
\includegraphics[width=7cm]{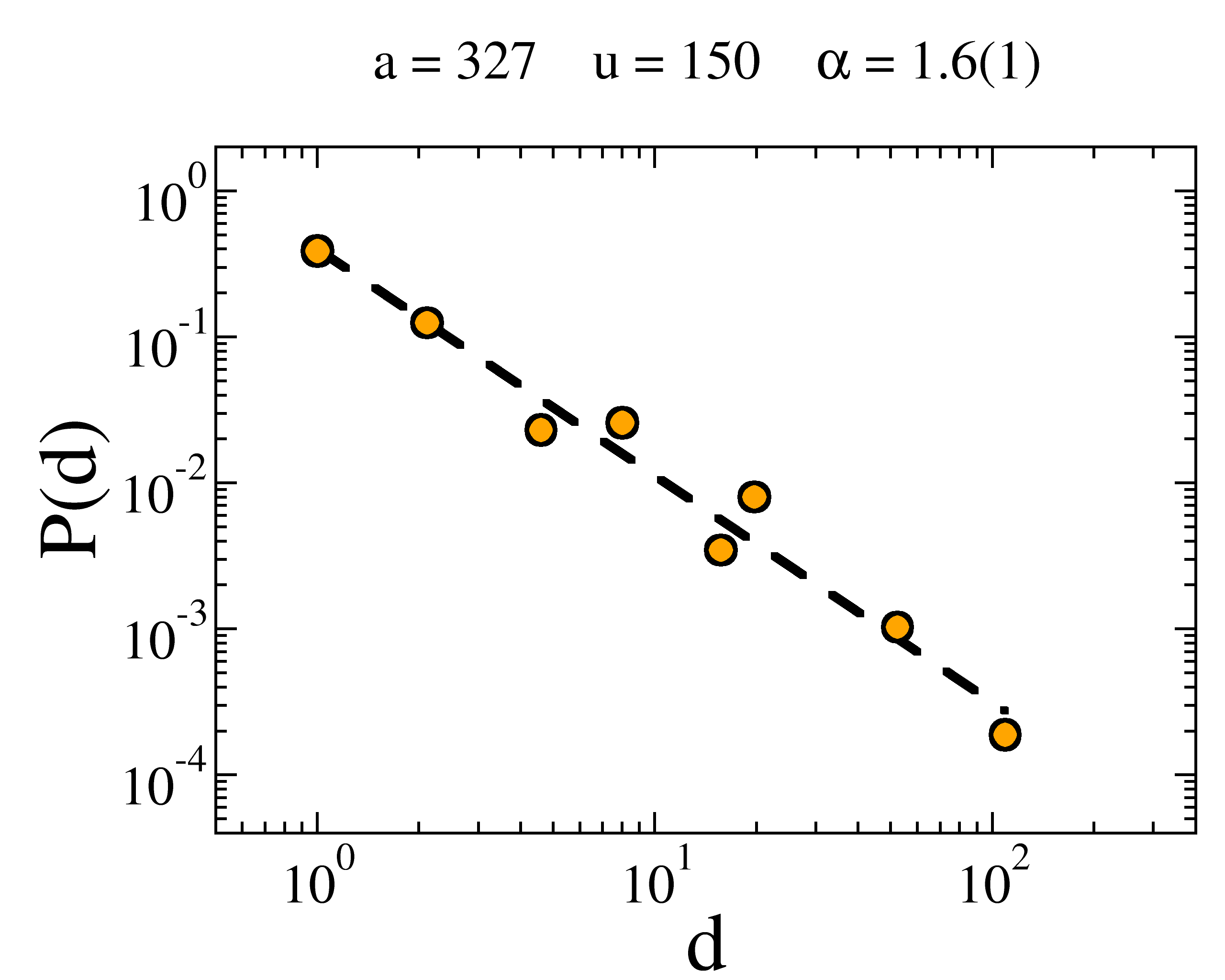}

\vskip .1cm
\caption{BM data set. Probability distribution function $P\left(d\right)$
measured for agent $u$ in auction $a$. We show
several $P\left(d\right)$s for different
pairs $u$ and $a$.}
\label{fig:bm_exponents_single_users_1}
\end{center}
\end{figure}

\begin{figure}[!ht]
\begin{center}
\vskip .7cm
\includegraphics[width=7cm]{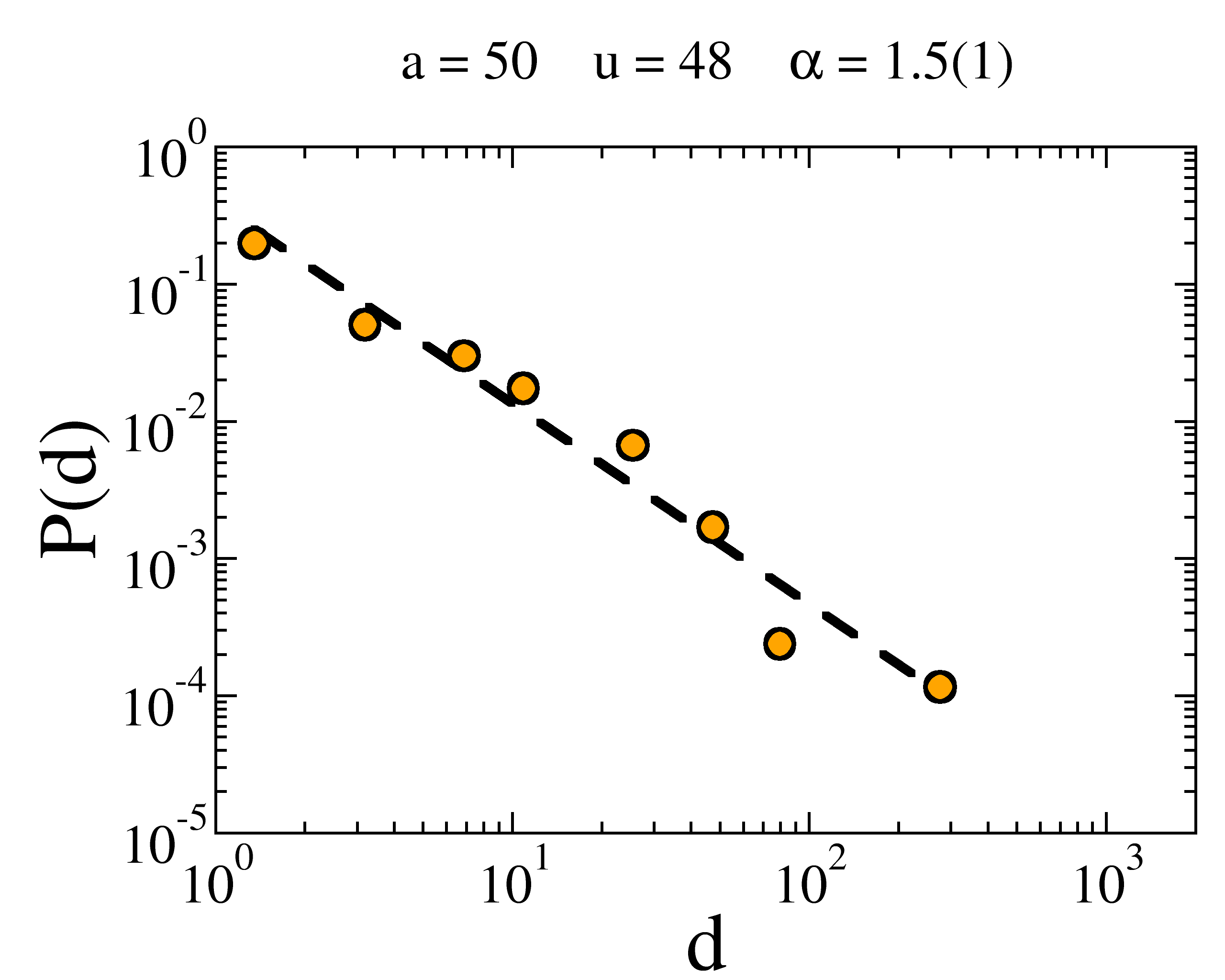}
\qquad
\includegraphics[width=7cm]{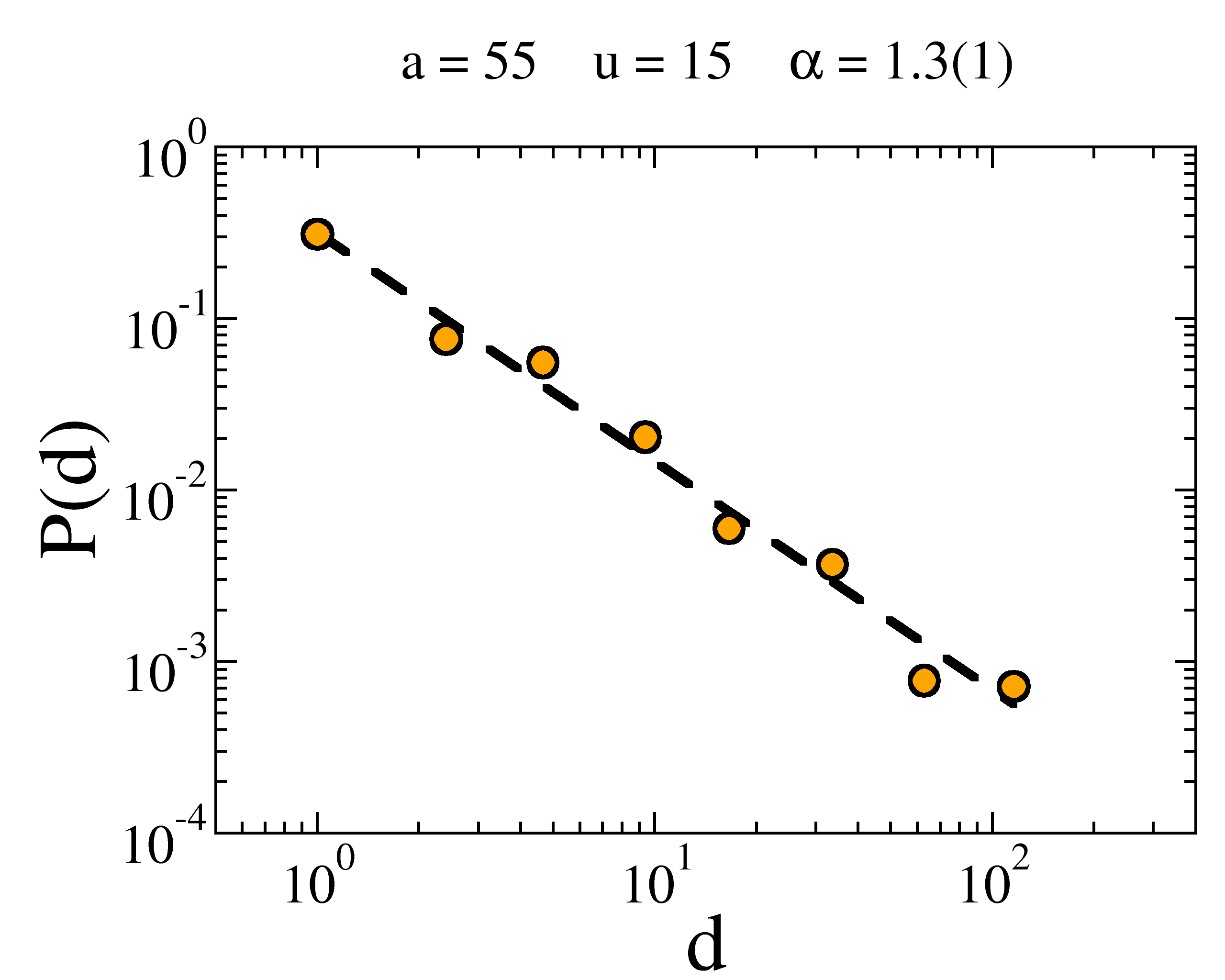}
\vskip .2cm
\includegraphics[width=7cm]{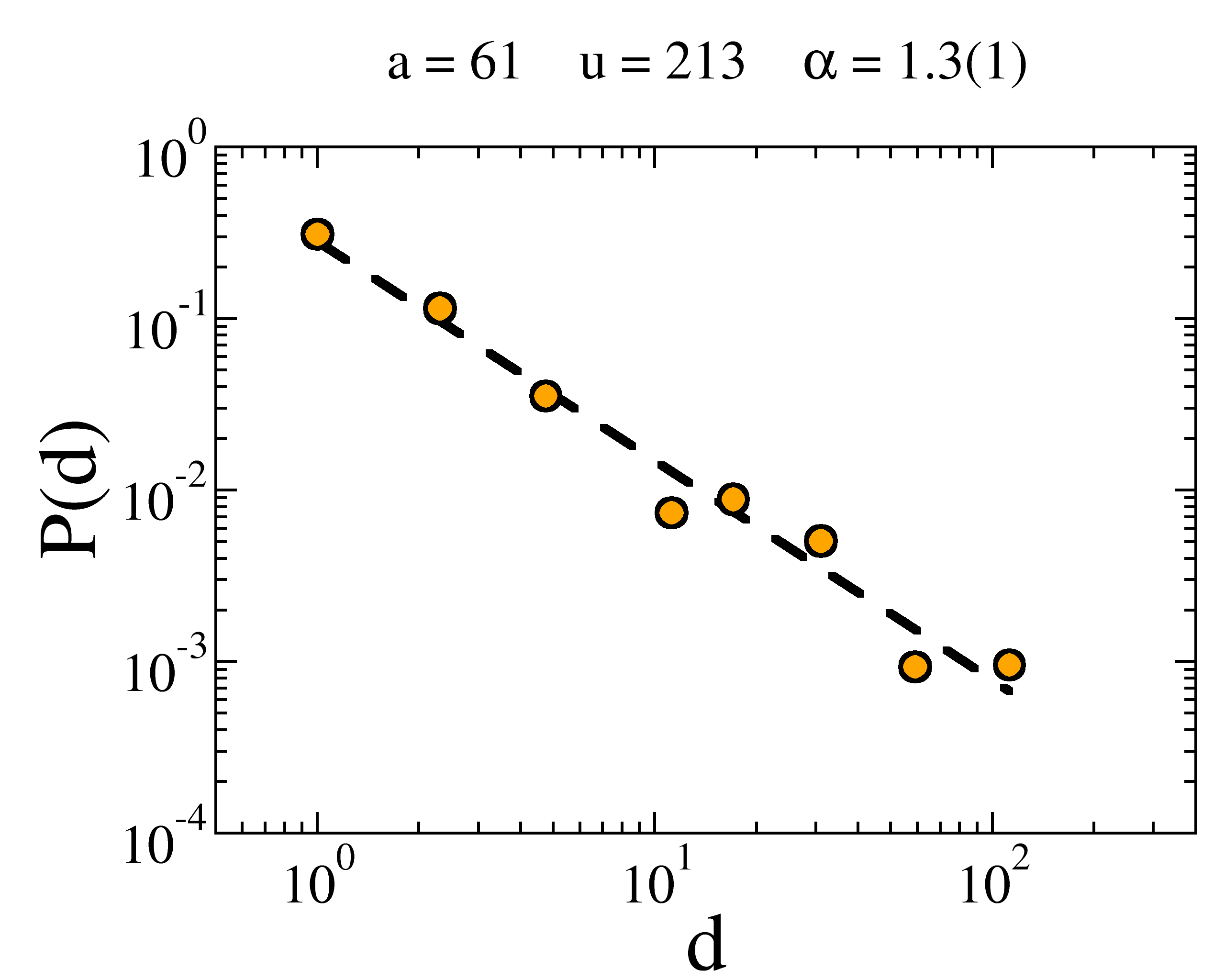}
\qquad
\includegraphics[width=7cm]{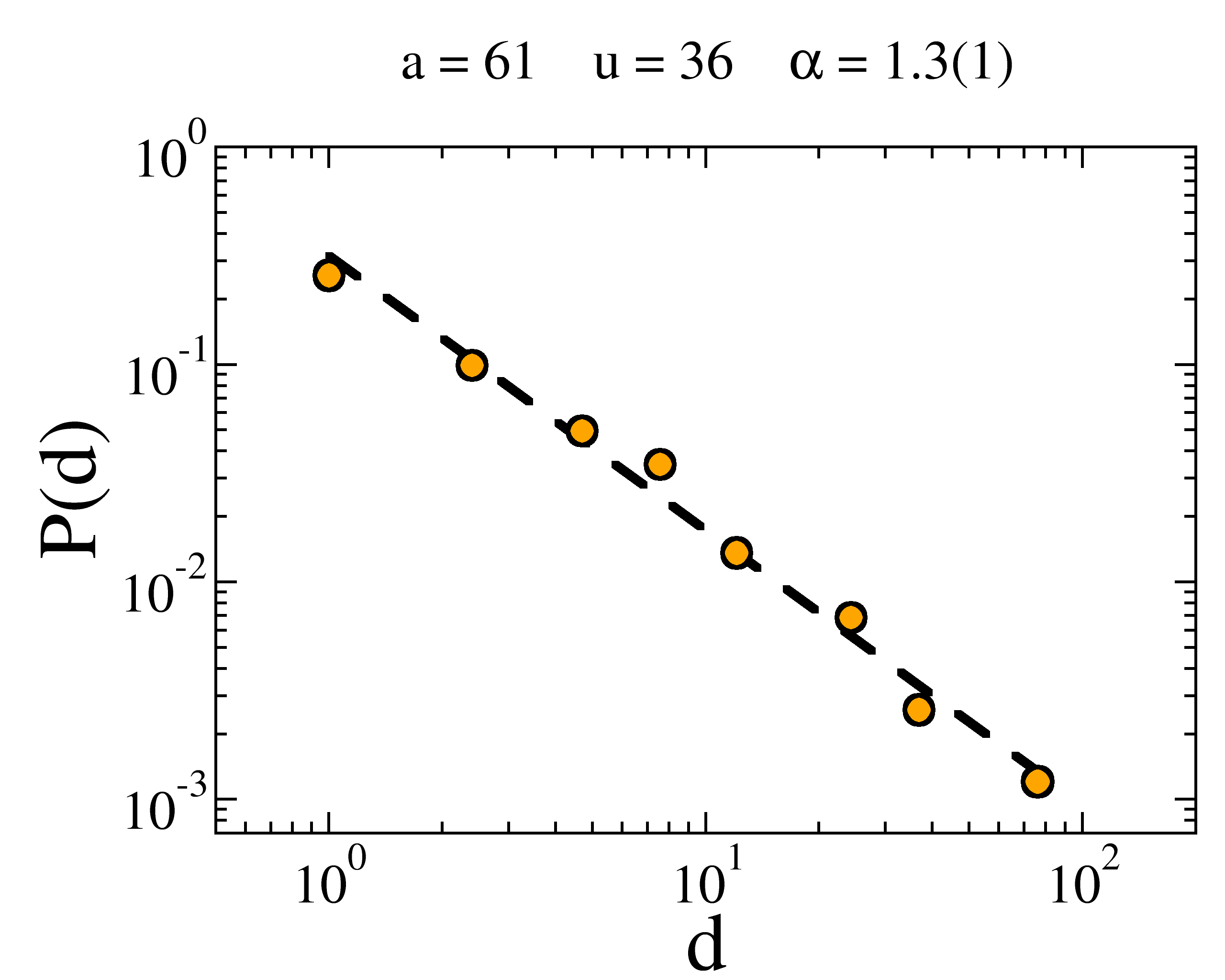}
\vskip .2cm
\includegraphics[width=7cm]{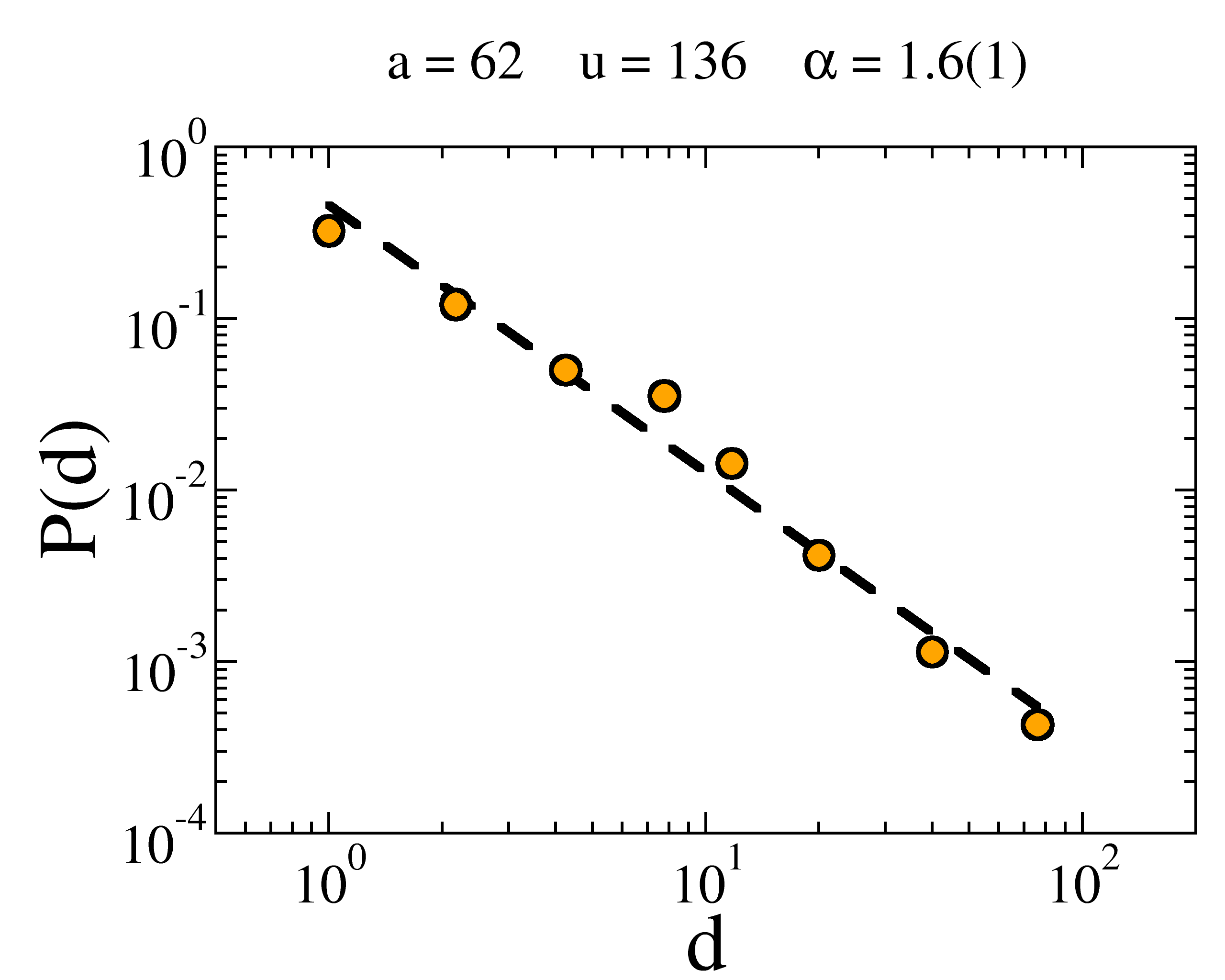}
\qquad
\includegraphics[width=7cm]{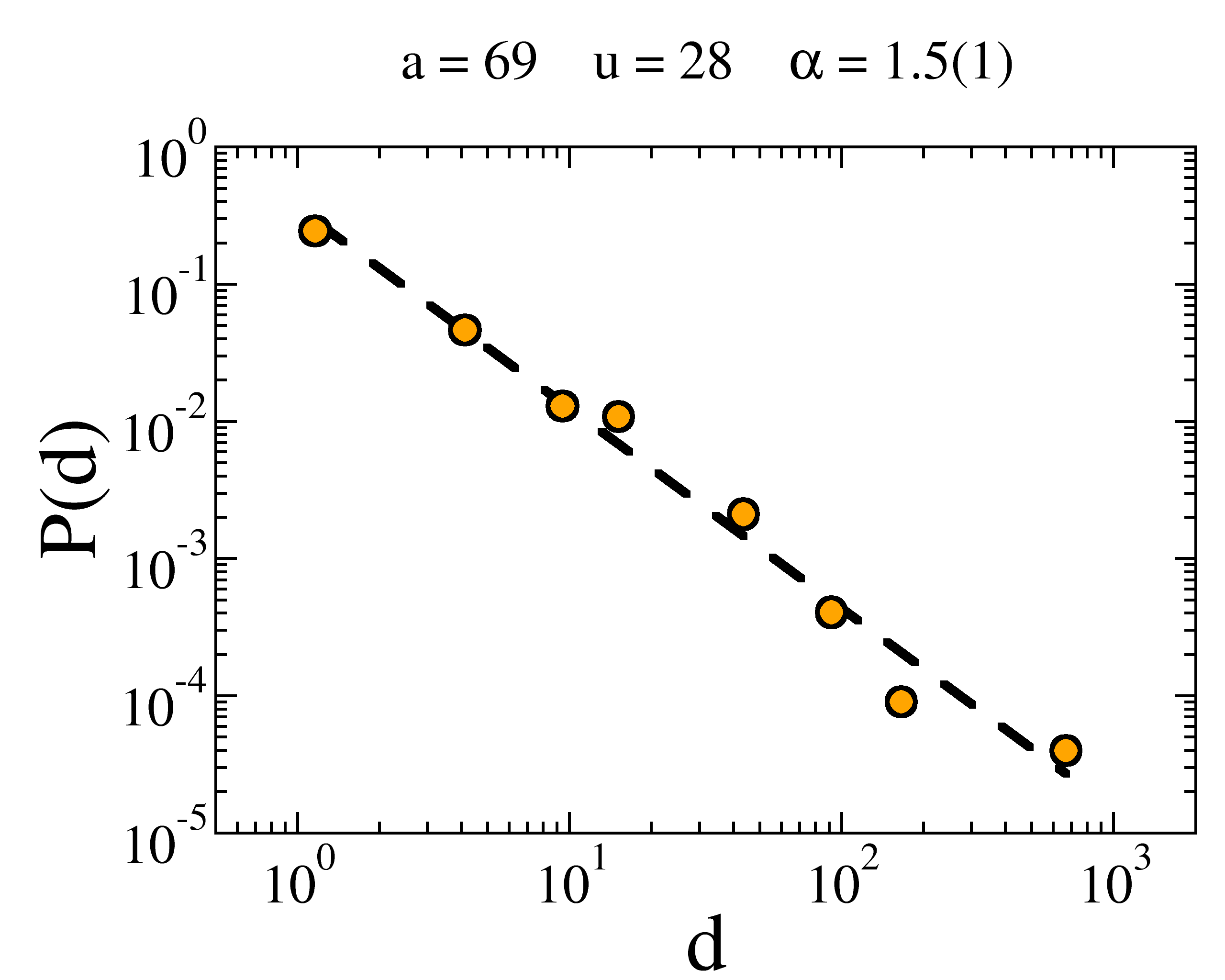}

\vskip .1cm
\caption{BM data set. Same as Figure~\ref{fig:bm_exponents_single_users_1}.}
\label{fig:bm_exponents_single_users_2}
\end{center}
\end{figure}

\begin{figure}[!ht]
\begin{center}
\vskip .7cm
\includegraphics[width=7cm]{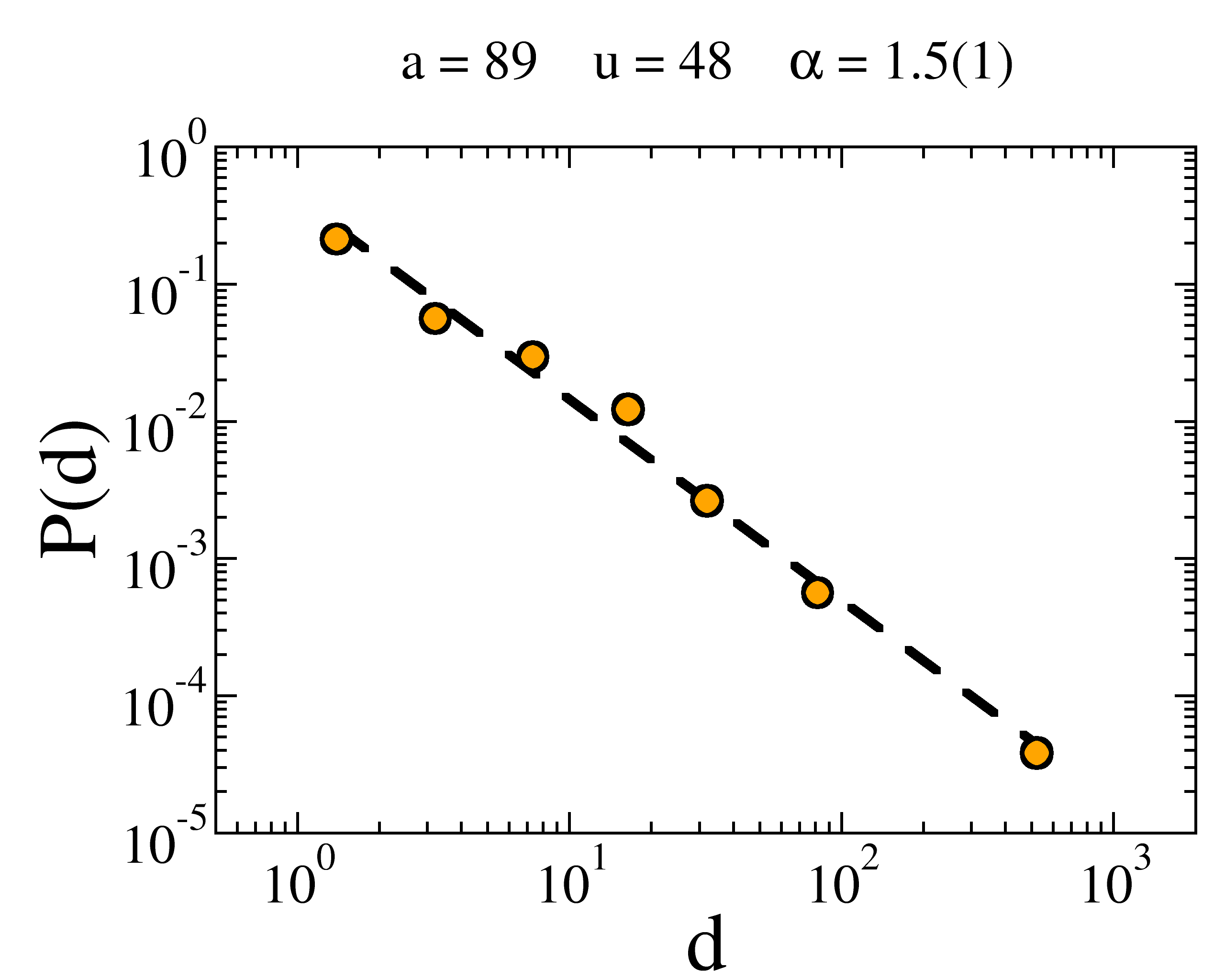}
\qquad
\includegraphics[width=7cm]{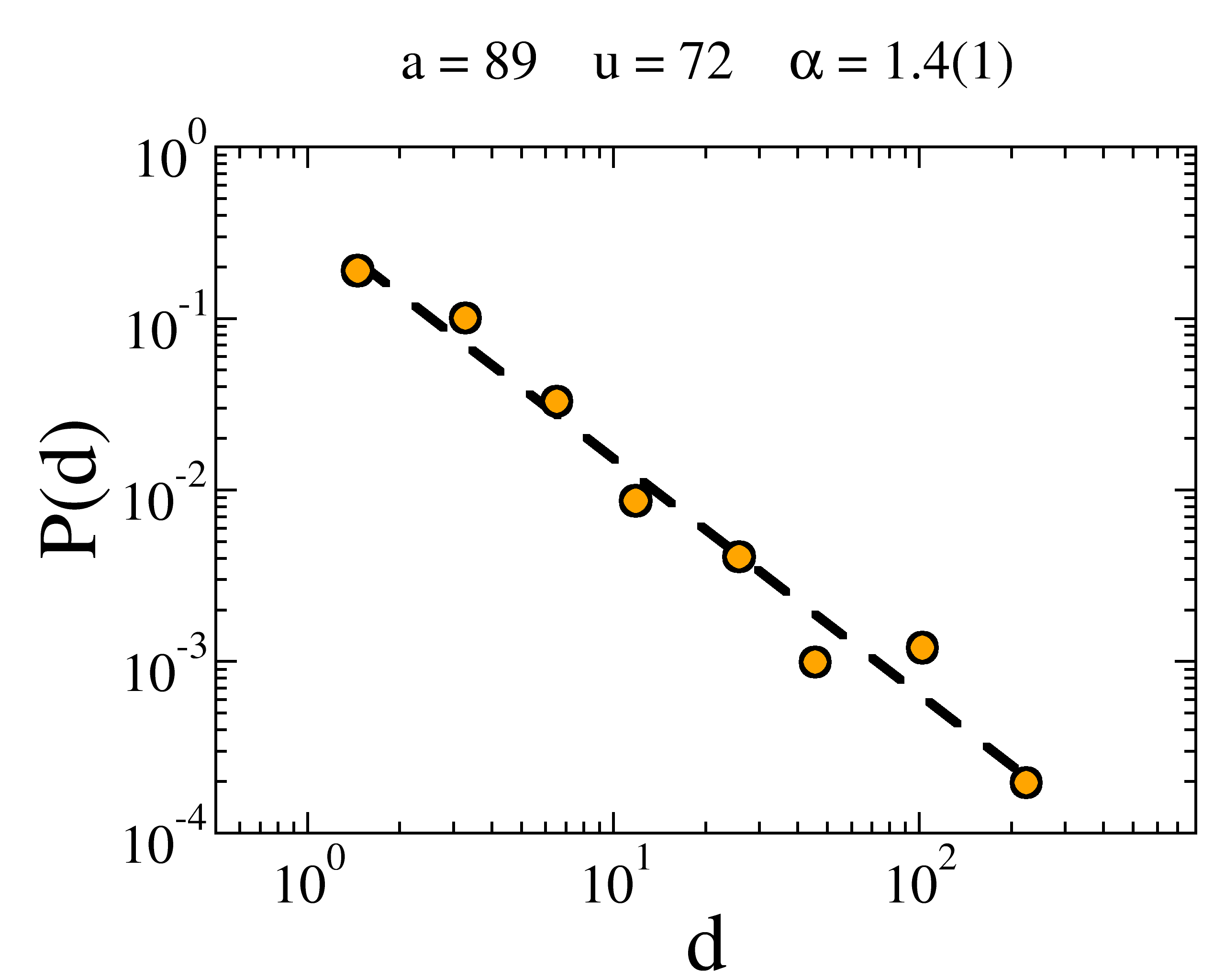}
\vskip .2cm
\includegraphics[width=7cm]{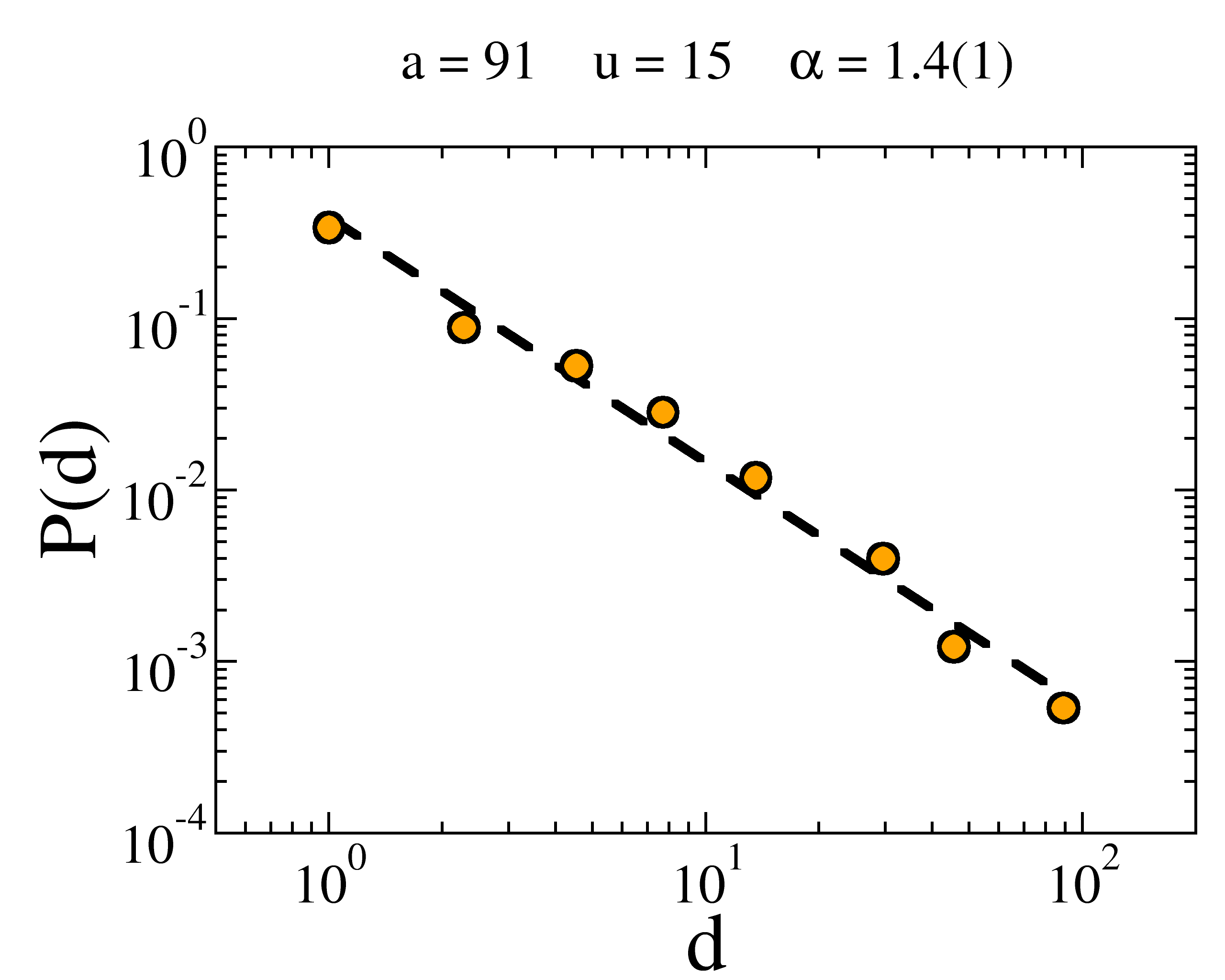}
\qquad
\includegraphics[width=7cm]{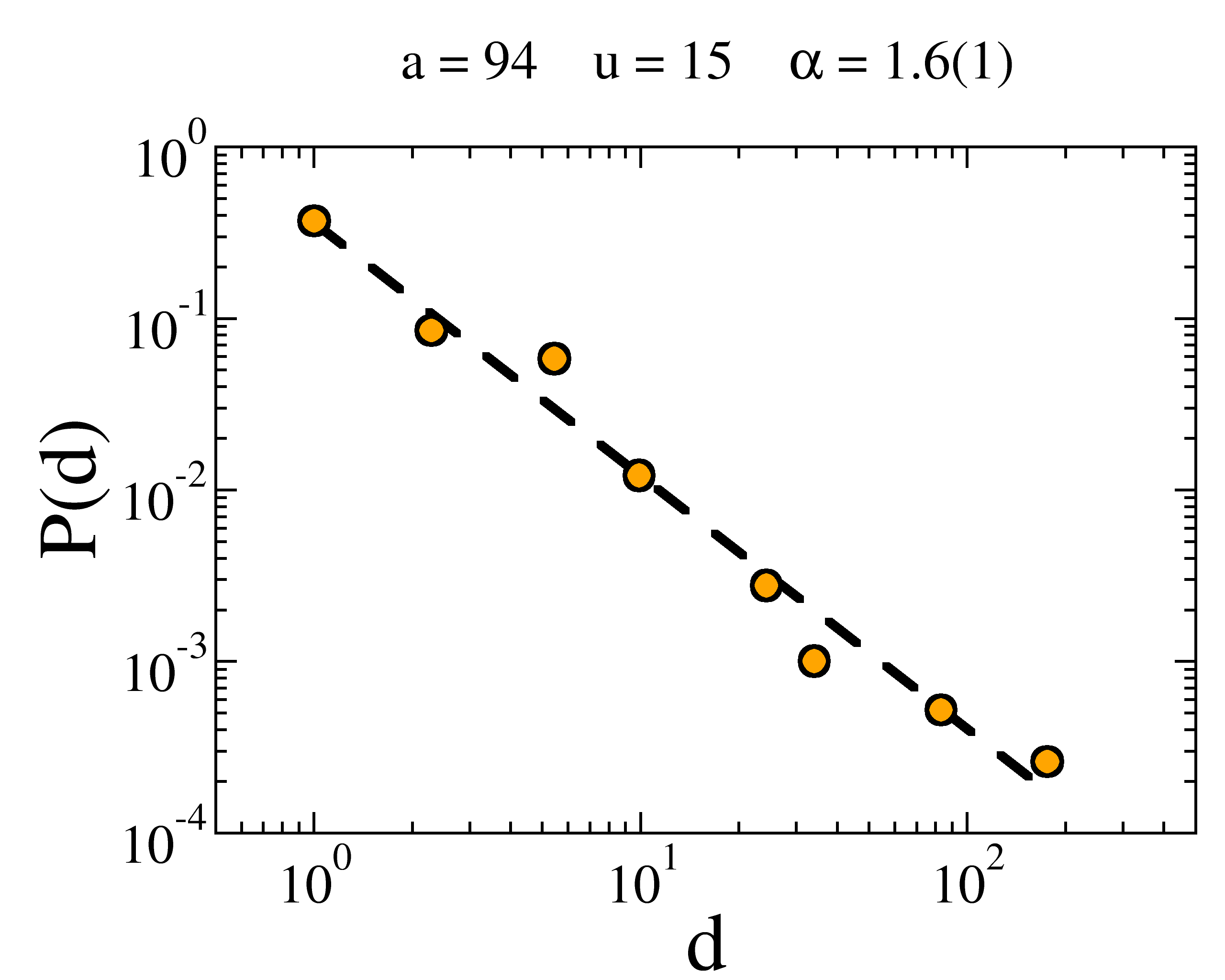}
\vskip .1cm
\caption{BM data set. Same as Figure~\ref{fig:bm_exponents_single_users_1} and 
~\ref{fig:bm_exponents_single_users_2}.}
\label{fig:bm_exponents_single_users_3}
\end{center}
\end{figure}

\noindent We measure the probability distribution
function (pdf) $P\left(d\right)$
of the difference between subsequent bids made by 
single agents in single auctions. 
Global (i.e., aggregated over all agents and auctions)
pdfs of both data sets
are plotted in Figure~\ref{fig:global}.
For agents with a sufficient number of bids
in the same auction, we also compute
individual pdfs. Some examples
are reported in the various panels of Figures~\ref{fig:exponents_single_users_1},~\ref{fig:exponents_single_users_2} and ~\ref{fig:exponents_single_users_3} for
UBH data set and in Figures~\ref{fig:bm_exponents_single_users_1},~\ref{fig:bm_exponents_single_users_2} and ~\ref{fig:bm_exponents_single_users_3} for
BM data set.
In each panel of these figures, we explicitly indicate the 
id of the auction $a$ and the id of the agent $u$
as they appear in our anonymised version of the data sets.
In all cases, we find a behavior compatible with
\begin{equation}
P\left(d\right) \sim d^{-\alpha} \;\; .
\end{equation}
The search strategy adopted by agents is therefore
given by L\'evy flights with 
characteristic exponent $\alpha$.
The best fits with power-laws are plotted in 
Figures~\ref{fig:exponents_single_users_1},~\ref{fig:exponents_single_users_2},~\ref{fig:exponents_single_users_3},~\ref{fig:bm_exponents_single_users_1},~\ref{fig:bm_exponents_single_users_2} and ~\ref{fig:bm_exponents_single_users_3} with black dashed lines
and the value of $\alpha$, plus the associated
error, corresponding to the best fit
is reported at the top of each panel.
\\
In order to calculate a pdf, we divide
the range of possible values of $d$ in bins
equally spaced on the logarithmic scale.
We then drawn the pdf by
associating to each bin the number of
$d$s falling in that bin divided by the number
of integers that could enter in the bin (this
because $d$ can assume only integer values).
Everything is then normalized by simply
dividing by the total number of points.
We compute the best power-law exponent by
performing a linear least square fit
in double logarithmic scale.
Despite the binning procedure
may introduce a certain amount of arbitrariness
in the evaluation of the pdfs, we checked the consistency
of our results by varying the number of bins.
Moreover, we additionally make use of
a different fit method (maximum likelihood)
about which we will discuss later.

\clearpage

\subsection{Independence of the direction of the jumps}

\begin{figure}[!ht]
\begin{center}
\vskip .7cm
\includegraphics[width=7cm]{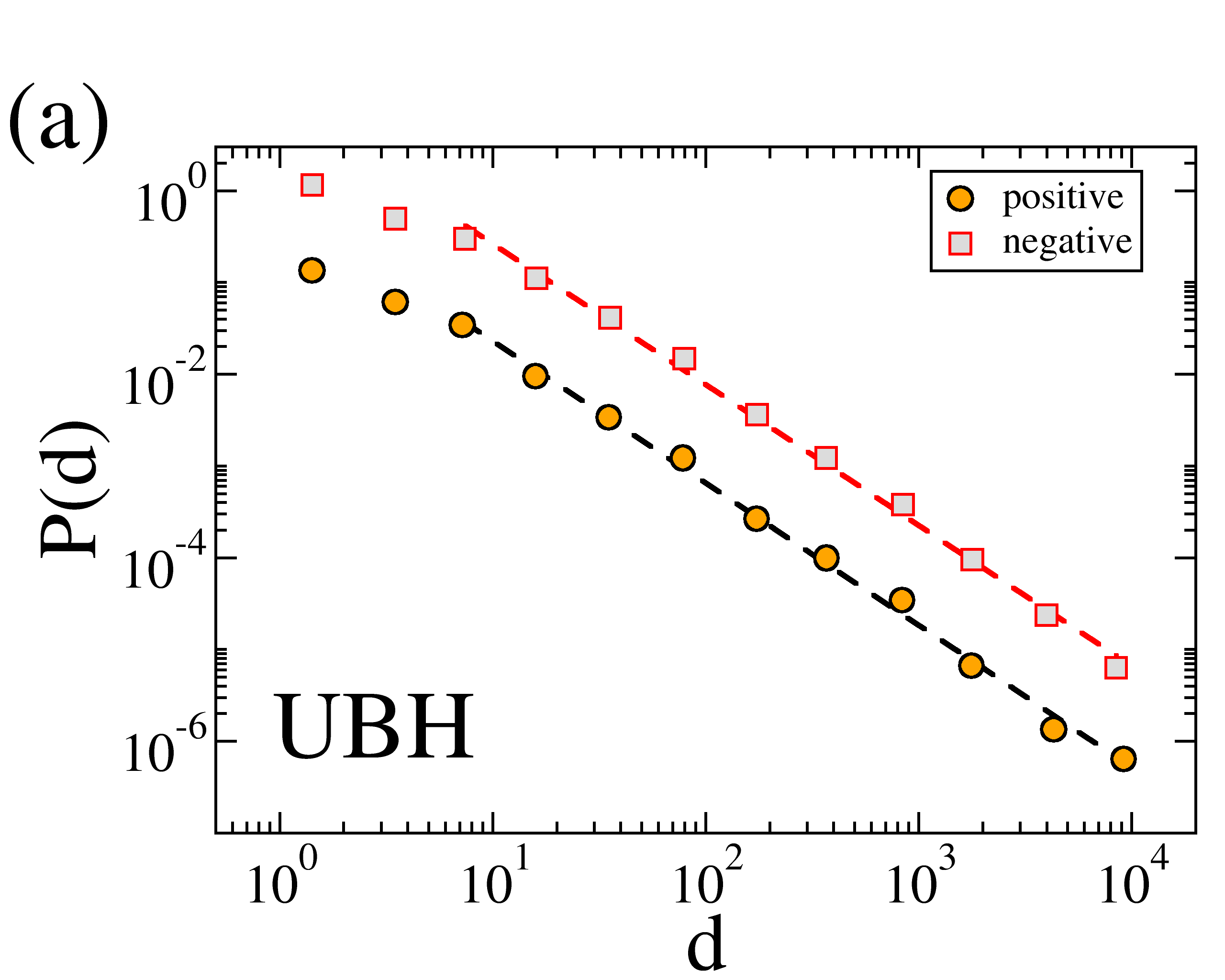}
\qquad
\includegraphics[width=7cm]{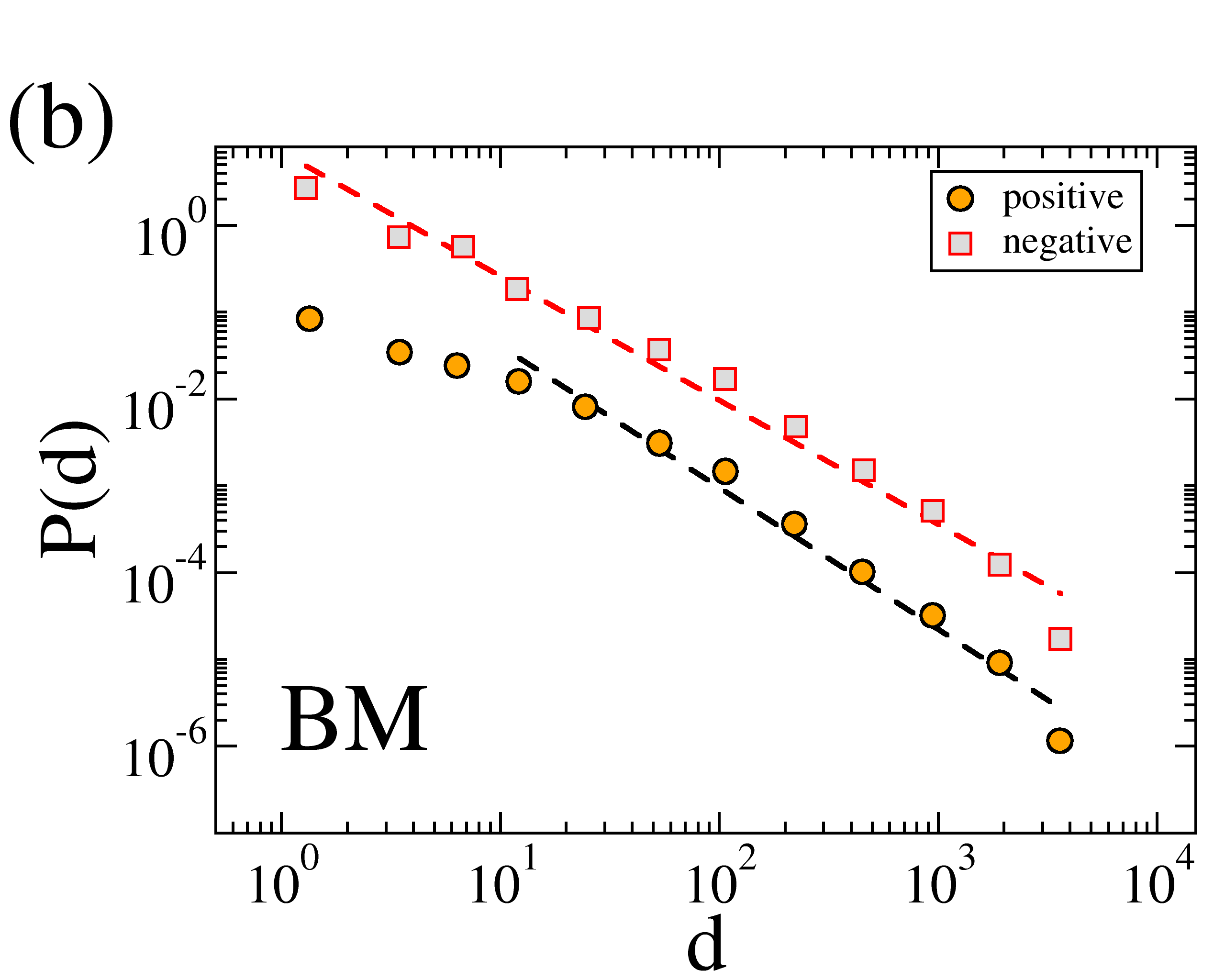}
\vskip .1cm
\caption{(a) UBH data set. Probability distribution function $P\left(d\right)$
calculated over all agents and auctions. We have separated
positive from  negative variations. The measured
decay exponents of the best fits with power-laws (dashed lines)
are $\alpha=1.53(2)$ [black] and $\alpha=1.54(2)$ [red], respectively.
The curve corresponding to negative variation
has been vertically shifted for clarity. 
This figure appears also in Fig.~3B of the main text. (b) BM data set.
here the best power-law fit are $\alpha=1.61(8)$ [black] and  $\alpha=1.45(4)$ [red].}
\label{fig:posneg}
\end{center}
\end{figure}

\noindent Since the rules of the auction  naturally bring agents
to move towards low bid values, it is important to 
stress any eventual difference
between the statistics associated with the length of gaps between
consecutive bid values. We aggregated data from all agents and auctions
and separate positive (i.e., $b_{t+1} > b_{t}$) from negative 
(i.e., $b_{t+1} < b_{t}$) variations. In Figure~\ref{fig:posneg}
the pdfs of positive and negative variations are plotted 
together. As one may notice,
there is not a significant difference between them and both
show a clear power-law decay with compatible exponents.

\clearpage

\subsection{Independence of the agents' activity}

\begin{figure}[!ht]
\begin{center}
\vskip .7cm
\includegraphics[width=7cm]{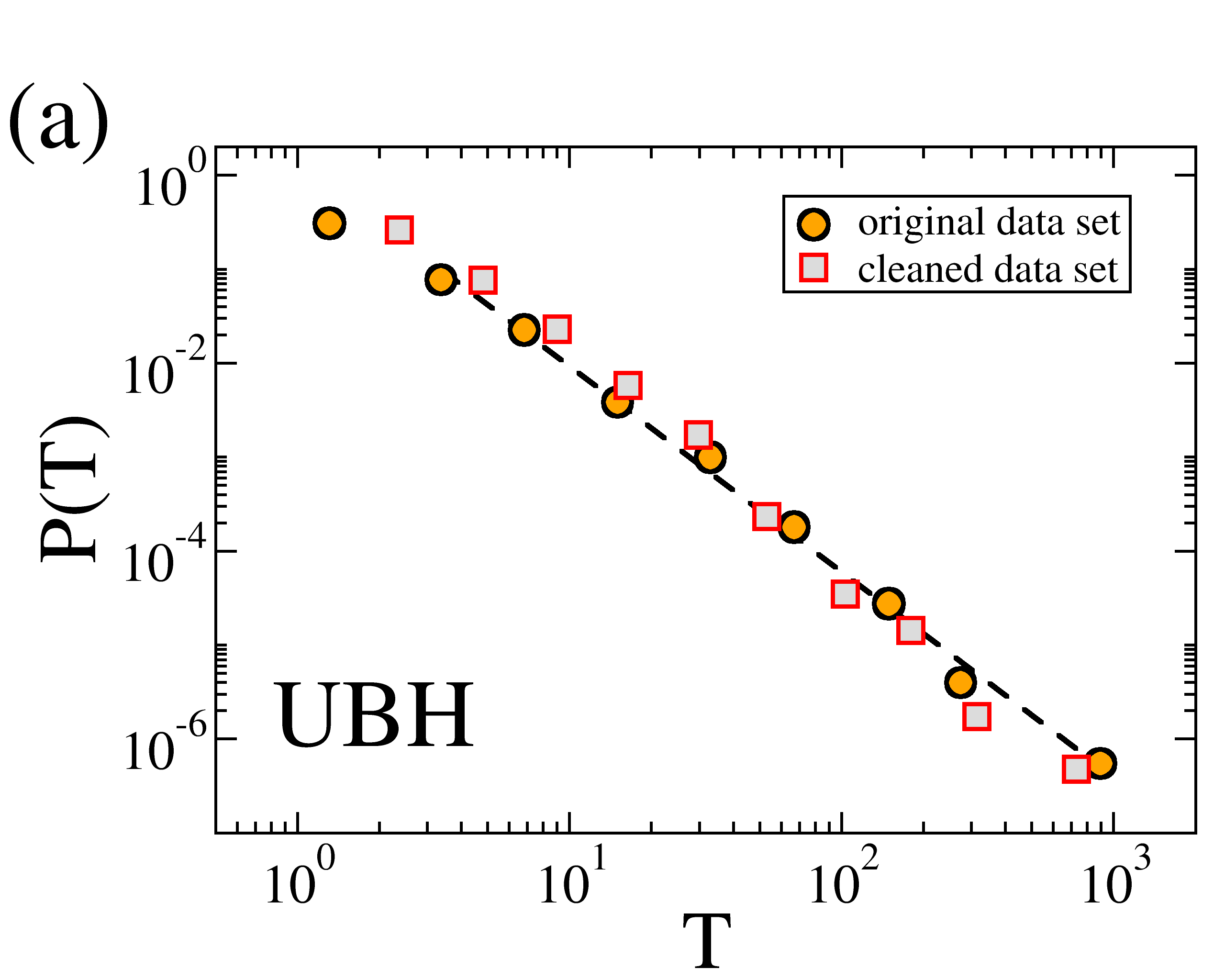}
\qquad
\includegraphics[width=7cm]{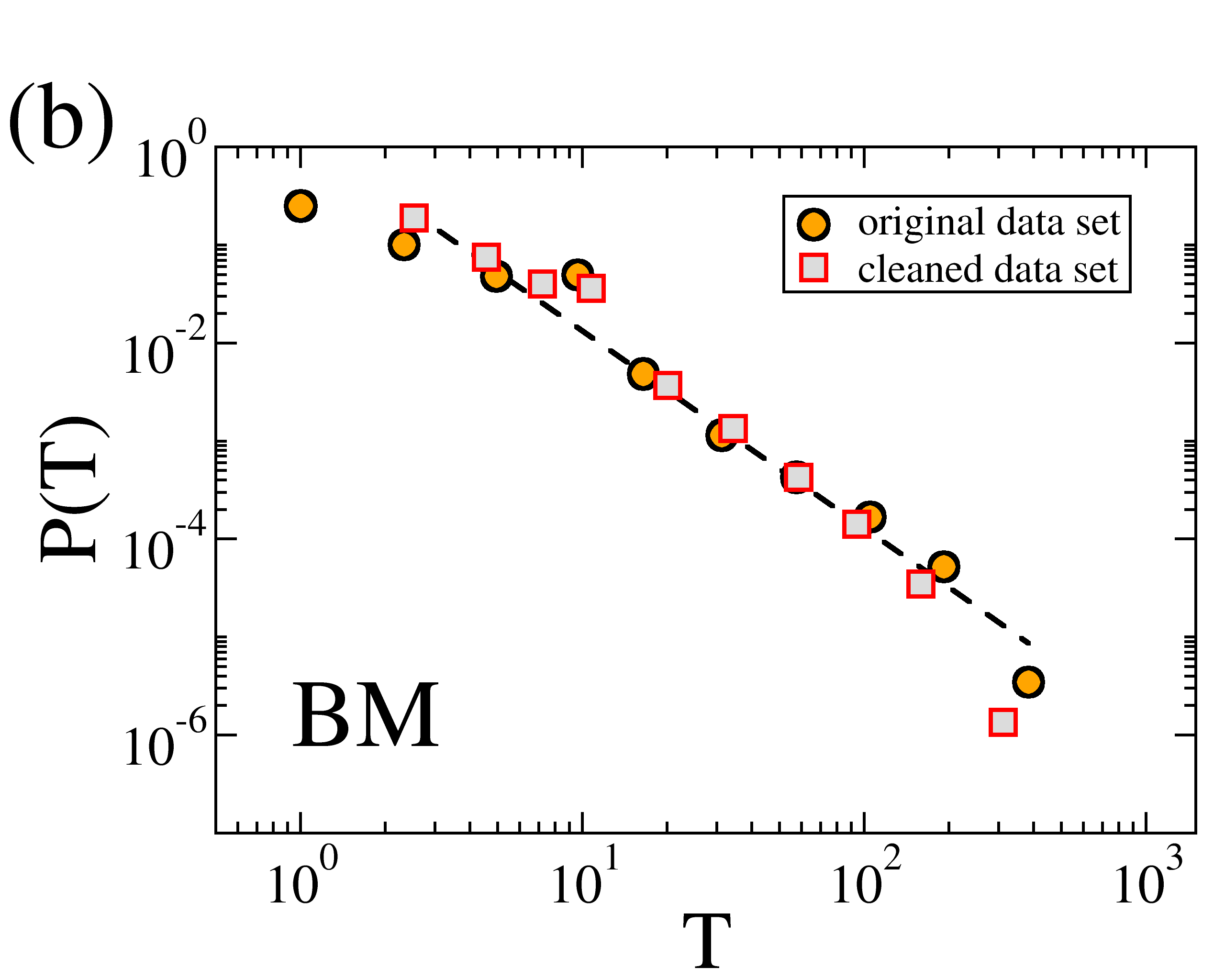}
\vskip .1cm
\caption{(a) UBH data set. Probability distribution 
function $P\left(T\right)$ of the number of bids $T$
performed by single agents in single auctions.
We calculate $P\left(T\right)$ on the original data set
(orange circles) and on the cleaned version of the same
data set (gray squares). In both cases,
the distribution scales power-like with a 
decay exponent equal to $2.2(2)$. (b) BM data set.
The exponent of the best fit is $2.0(4)$ [dashed line].}
\label{fig:activity}
\end{center}
\end{figure}

\noindent The evidence of L\'evy flights for single agents
in single auctions can be directly verified
only for agents with a sufficient number of bids $T$
in the same auction.
For values of $T$ smaller than $50$ is 
practically impossible to construct the histogram $P\left(d\right)$
and therefore no exponent can be measured.
Unfortunately, this situation is very frequent in our data sets.
We measure the number of bids made each agent in each
auction and plot the pdf of the number of bids
in a single auction in Figure~\ref{fig:activity}. 
The level of activity
is quite heterogeneous and decays power-like
with an exponent close to $2.2$. For completeness,
we measure the same pdf for both original and cleaned data sets
without noticing appreciable differences.

\begin{figure}[!ht]
\begin{center}
\vskip .7cm
\includegraphics[width=7cm]{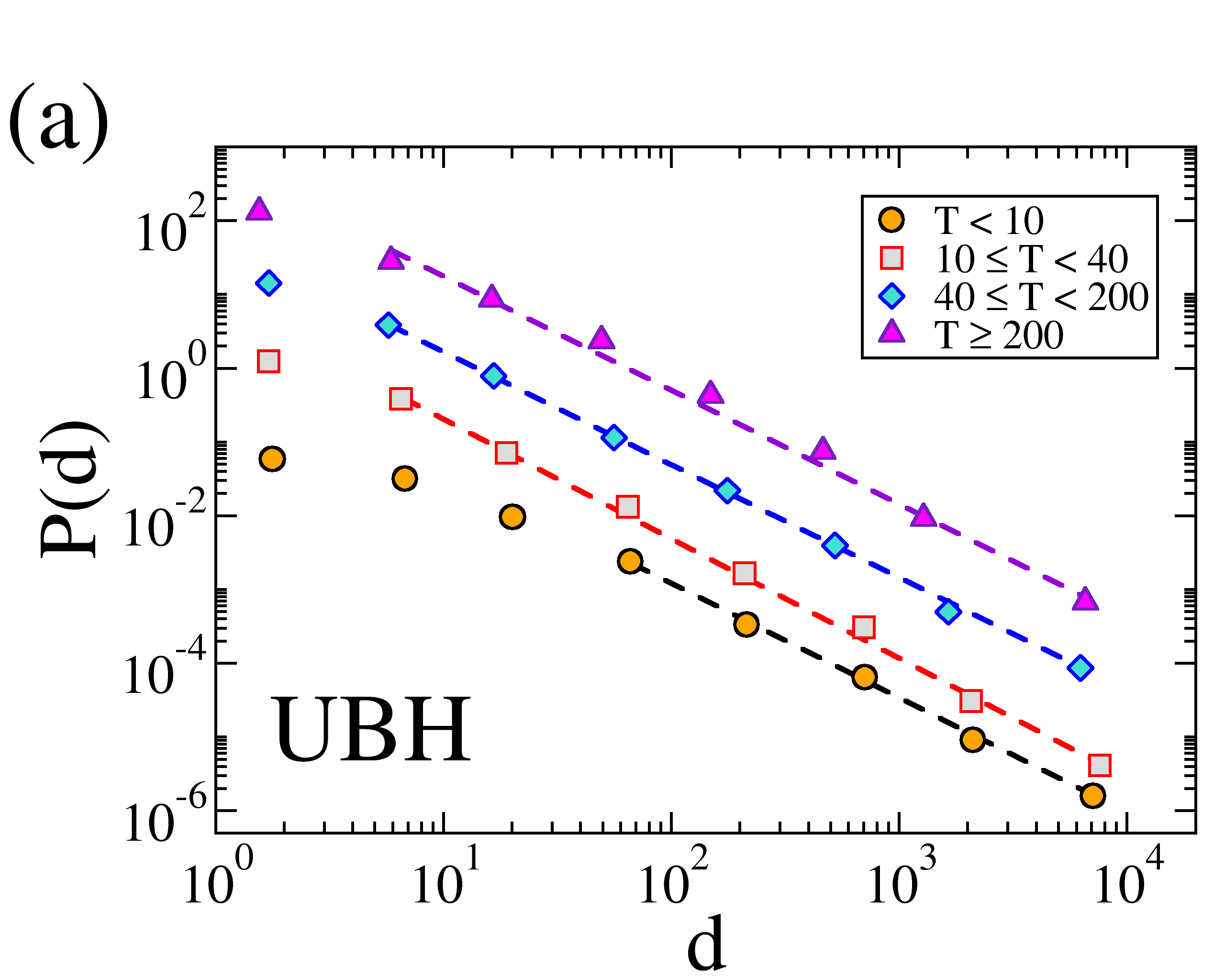}
\qquad
\includegraphics[width=7cm]{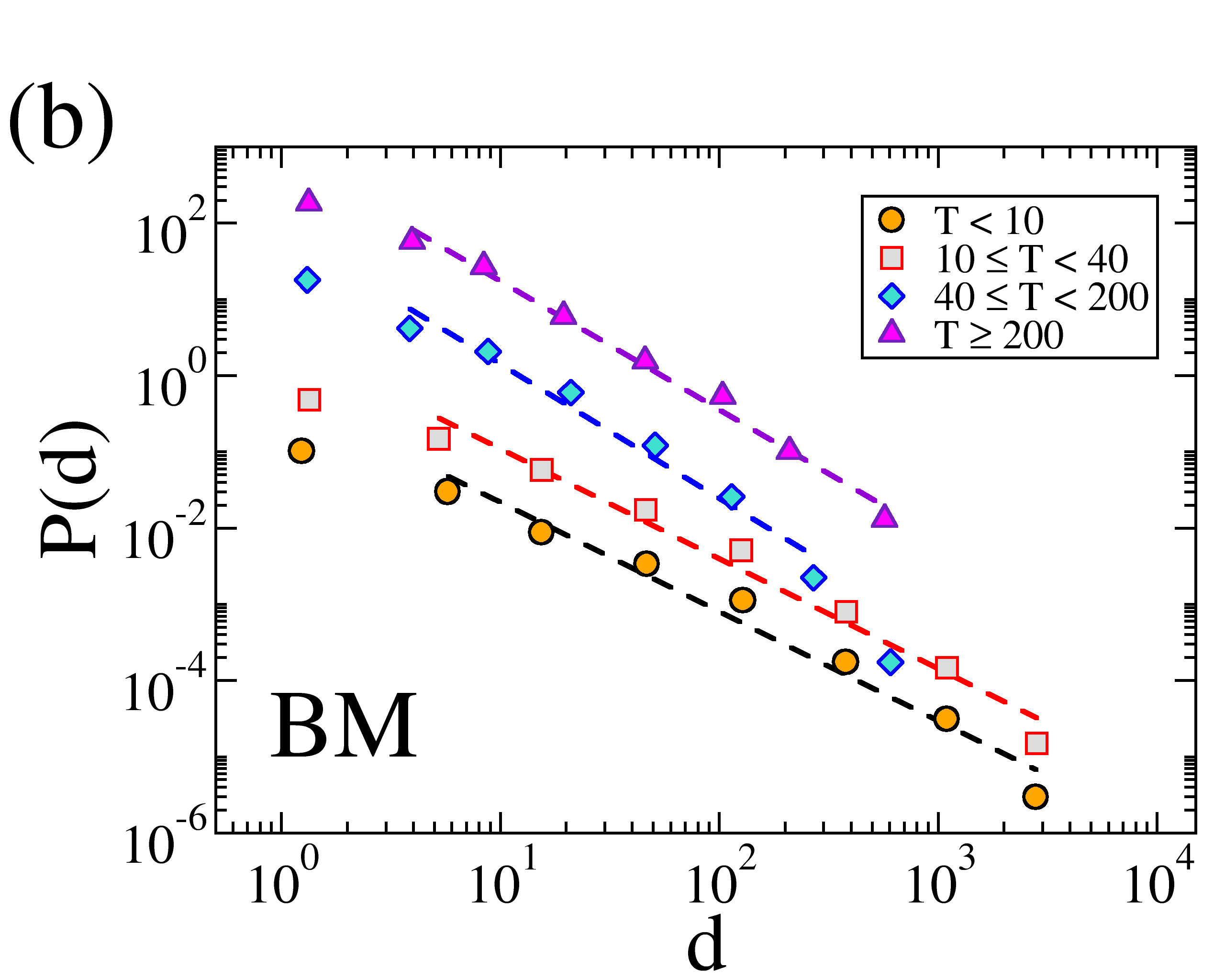}
\vskip .1cm
\caption{(a) UBH data set. Probability distribution function $P\left(d\right)$
of the length $d$ of the jumps performed by agents
in the bid space.
All auctions have been aggregated together. Different
curves correspond to agents with different
levels of activity. Their activity is measured as the number
of bids made in the same auction. We divide the population into four
subsets: $T < 10$ (orange circles), $10 \leq T < 40$ (gray squares),
 $40 \leq T < 200$ (blue diamonds) and $T \geq 200$ (violet triangles).
Dashed lines represent the best power-law fits. 
The value of the measured exponents are: $\alpha=1.55(2)$ (black), $\alpha=1.62(4)$ (red),
$\alpha=1.53(2)$ (blue) and $\alpha=1.54(3)$ (violet). Curves have been
vertically shifted for clarity. This figure
is also reported in Fig.~3C of the main text. (b) BM data set. Best 
power-law fits (dashed lines) have exponents: $\alpha=1.43(4)$ (black), 
$\alpha=1.44(5)$ (red),
$\alpha=1.7(1)$ (blue) and $\alpha=1.7(1)$ (violet).}
\label{fig:number_bid}
\end{center}
\end{figure}

\noindent We divide the population in different ranges
of activity. We aggregate the length of the jumps
performed by all agents in a given bin and measure
the resulting $P\left(d\right)$. The results of this analysis
are reported in Figure~\ref{fig:number_bid}. Independently
of the activity level, the aggregated pdfs 
decay power-like and have similar exponents $\alpha$.
This means that the presence of L\'evy flights
is typical for every agent independently of
how many bids the agent has made.

\clearpage

\subsection{Independence of the bidding time}

\begin{figure}[!ht]
\begin{center}
\vskip .7cm
\includegraphics[width=7cm]{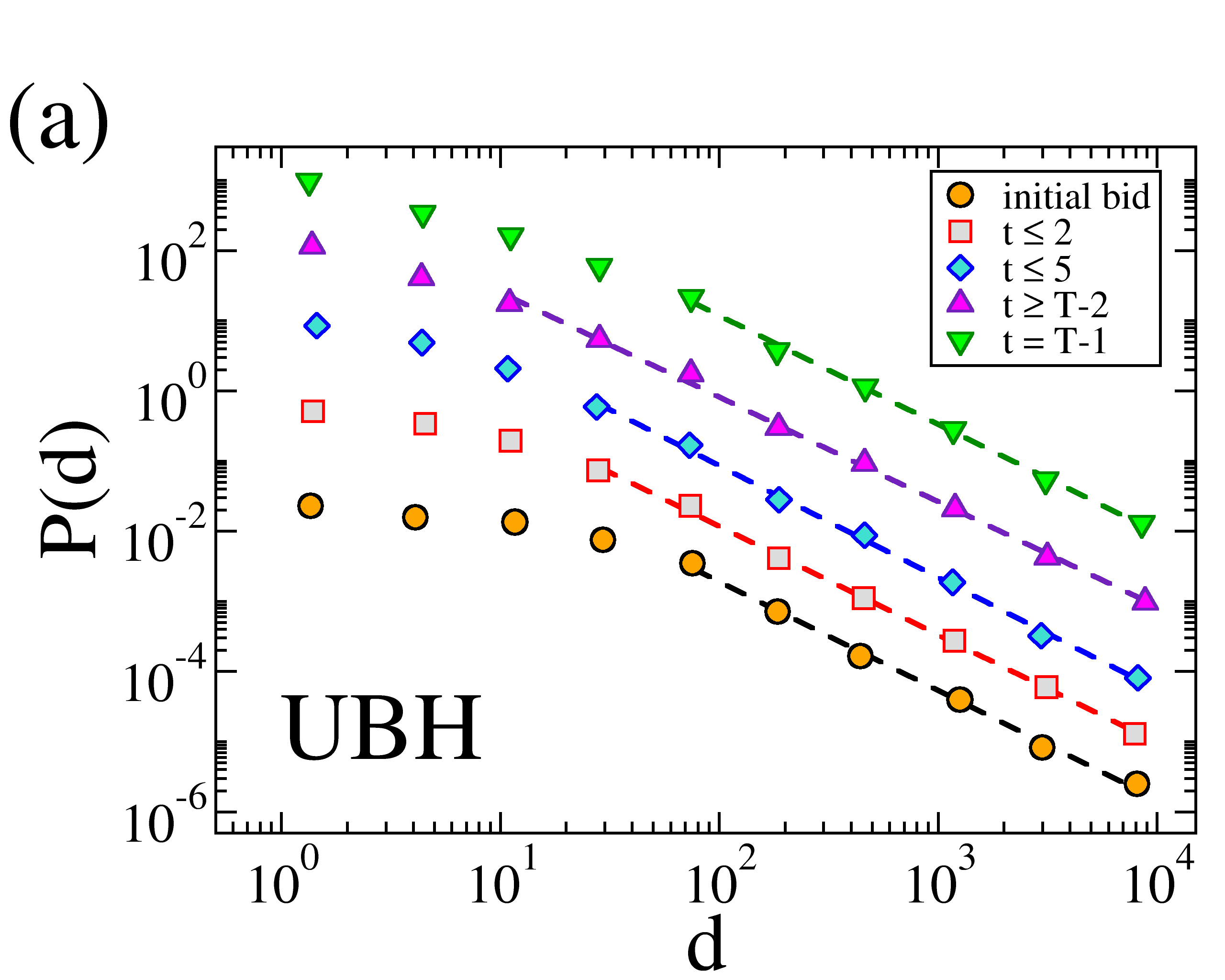}
\qquad
\includegraphics[width=7cm]{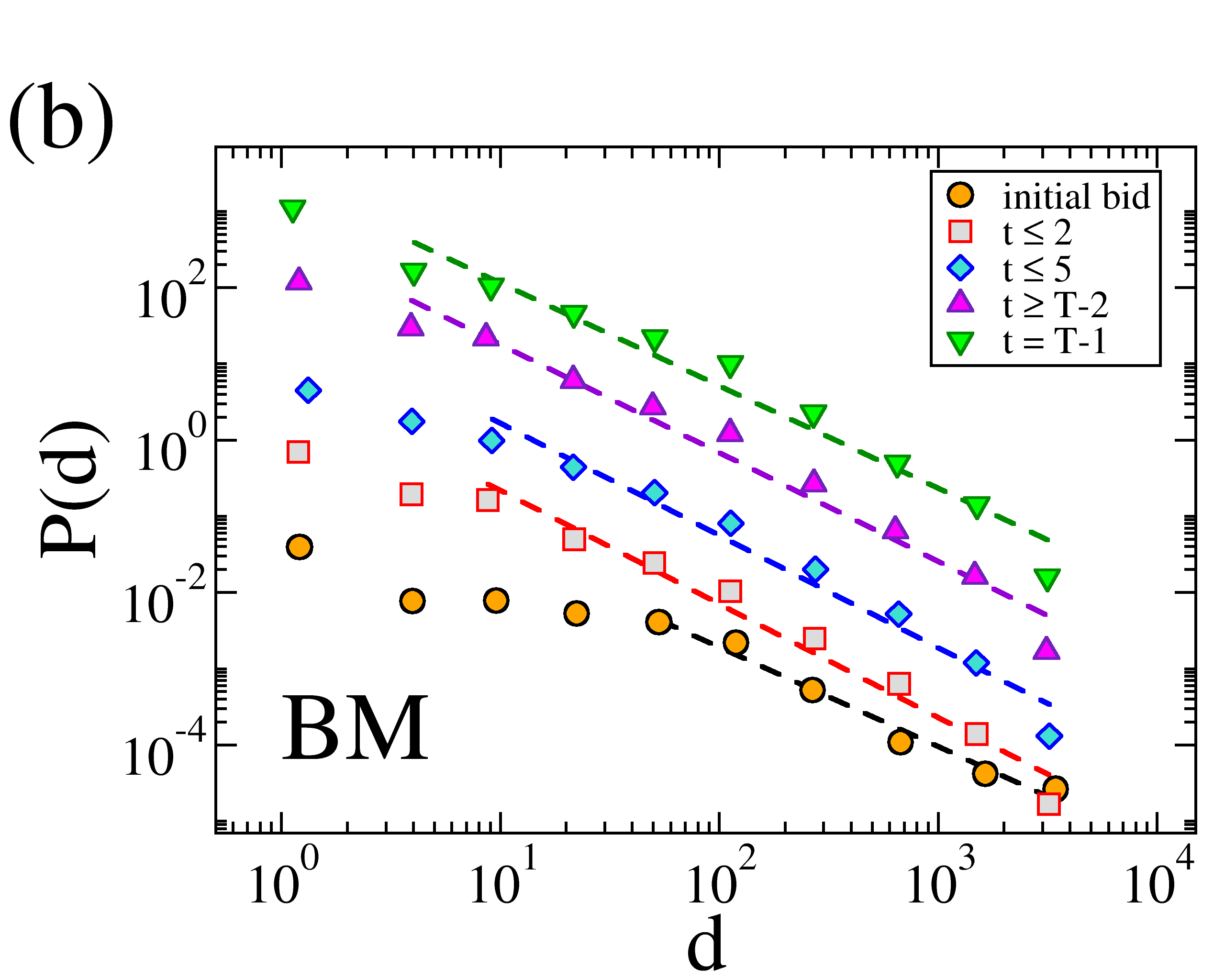}
\vskip .1cm
\caption{(a) UBH data set. Probability distribution function $P\left(d\right)$
of the length $d$ of the flights performed by agents
in the bid space.
All auctions have been aggregated together. Different
curves correspond to
different periods of activity for agents:
orange circles correspond to the initial guess
made by agents;
bids gaps with $t\leq 2$ (gray squares) and $t \leq 5$ (blue diamonds)
aggregate the data corresponding to the early
activity of agents; $t \geq T -2$ (violet
up triangles) and $t=T-1$ (green down triangles)
corresponds to the jumps made by agents at the end of their own activity.
Dashed lines have been obtained as best power-law fits with
data points. The value of the measured exponents are: $\alpha=1.55(4)$ (black), $\alpha=1.54(3)$ (red),
$\alpha=1.59(4)$ (blue), $\alpha=1.49(5)$ (violet) and
$\alpha=1.54(6)$ (green). Curves have been
vertically shifted for clarity. This figure is also reported in Fig.~2d of the main
text.
(b) BM data set. Best power-law fits (dashed lines)
have exponents: $\alpha=1.3(1)$ (black), $\alpha=1.50(5)$ (red),
$\alpha=1.47(4)$ (blue), $\alpha=1.43(6)$ (violet) and
$\alpha=1.35(6)$ (green). For the initial bid $b_1$, we compute the length
of the jump as $d=M-b_1+1$, with $M$ being the maximal bid value
in the HUB auctions.}
\label{fig:time_exponents}
\end{center}
\end{figure}

\noindent Another fundamental point is to understand
whether the L\'evy flight strategy is emergent
or {\it a priori} given. We test these hypotheses
by measuring the pdfs $P\left(d\right)$ corresponding
to a certain range during the activity of the agents.
The results are reported in Figure~\ref{fig:time_exponents}.
We consider ranges of activity periods corresponding 
to $t \leq 2$, $t \leq 5$, $t \geq T -2$ and $t = T-1$. 
Additionally, we consider the distribution of the 
initial bid values (i.e.,
the first bid made by all agents in all auctions). In every case,
we are able to fit the curves with power-laws and the resulting
exponents are compatible each other.
We can effectively conclude that the strategy to adopt a L\'evy flight
is not an emergent property induced by the evolution
of the auction. Instead the strategy 
to follow L\'evy flights is intrinsically present, in each agent,
during the whole duration of the auction.

\clearpage

\subsection{Independence between jumps}

\begin{figure}[!ht]
\begin{center}
\vskip .7cm
\includegraphics[width=7cm]{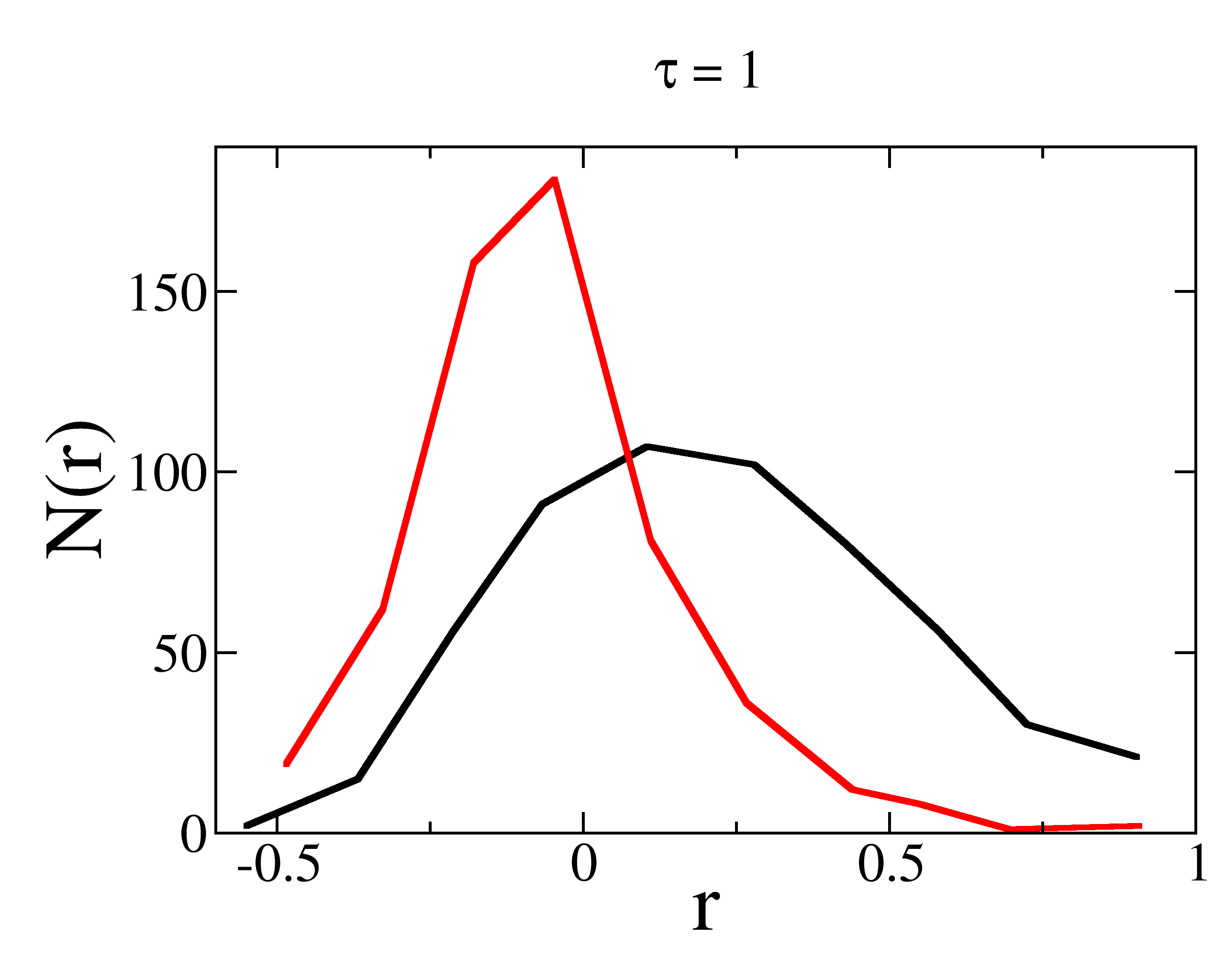}
\qquad
\includegraphics[width=7cm]{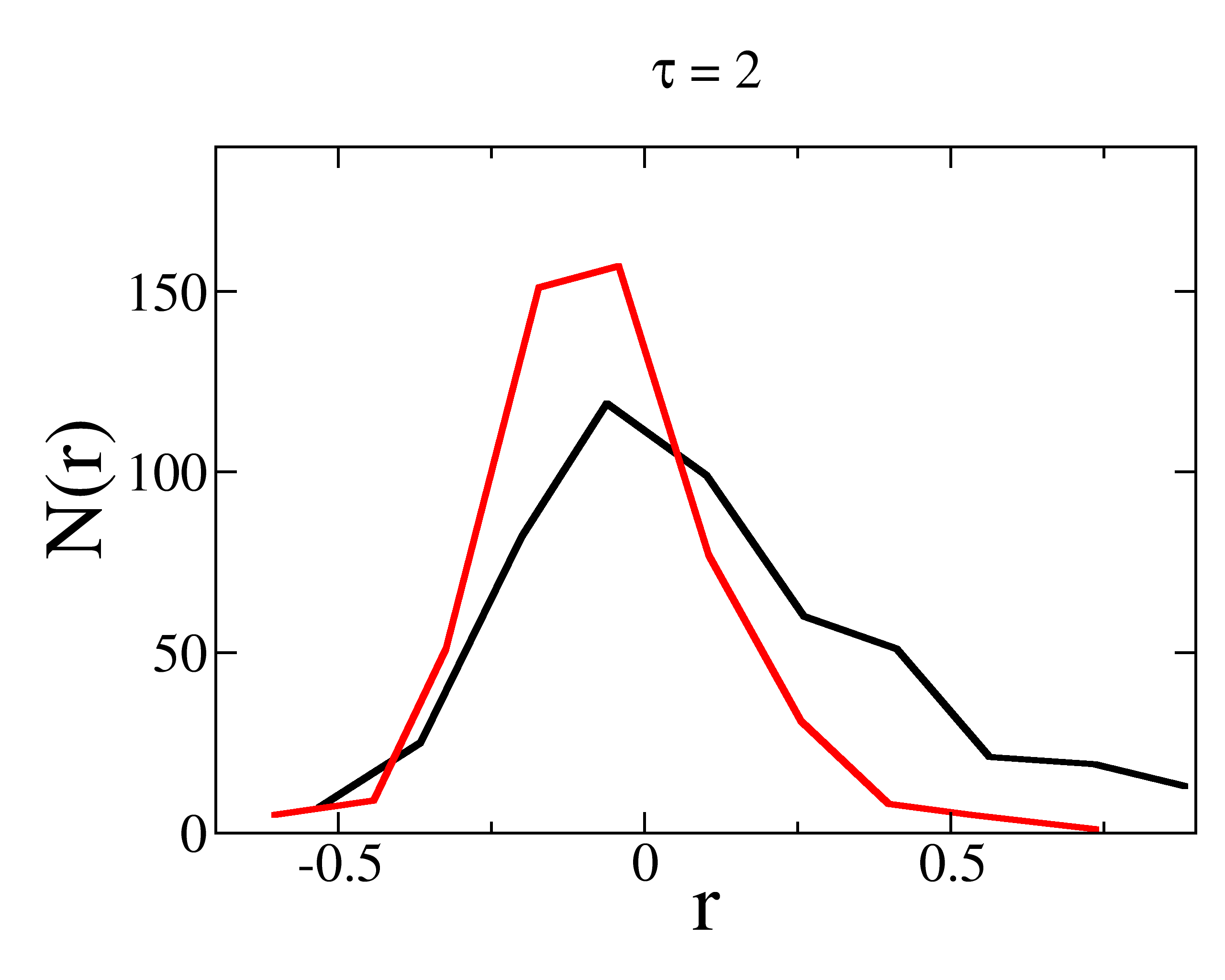}
\vskip .2cm
\includegraphics[width=7cm]{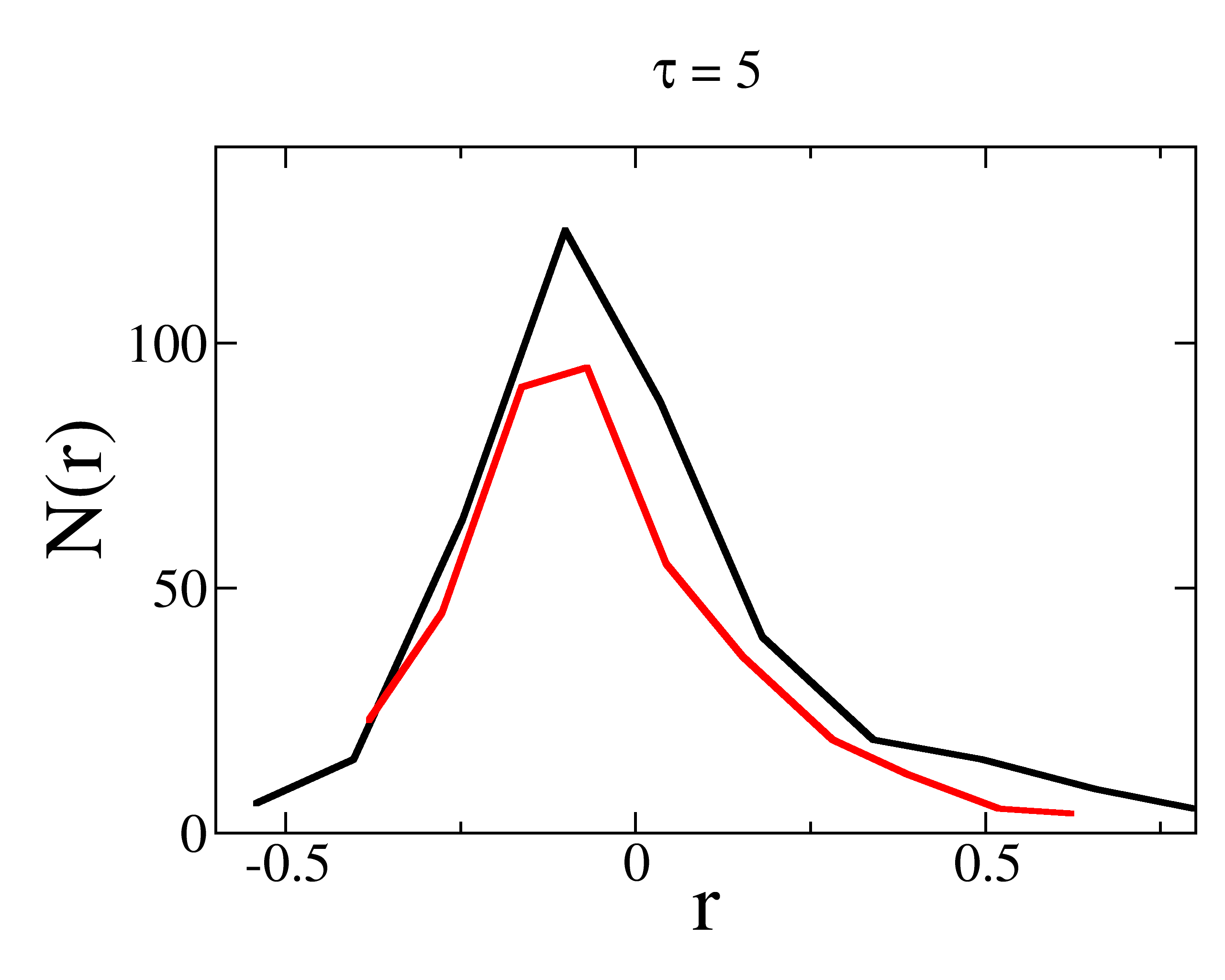}
\qquad
\includegraphics[width=7cm]{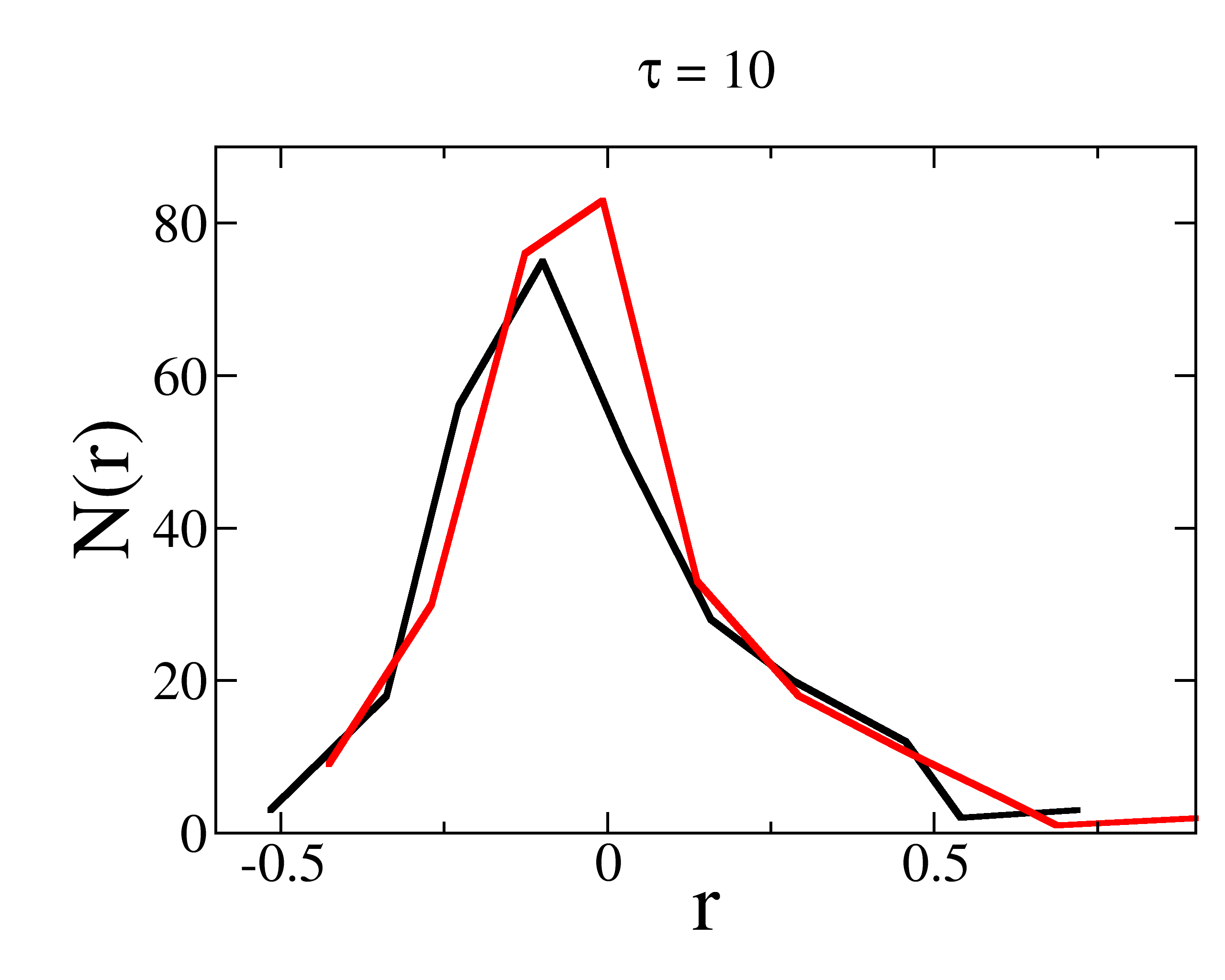}
\vskip .1cm
\caption{UBH data set. Number of
agents $N\left(r\right)$ whose bid gaps at position
$t$ and $t+\tau$ have Pearson's correlation
coefficient equal to $r$. Black curves are measured
on real data, while the red ones are calculated
over a reshuffled version of the same data.
The reshuffling is made by randomly exchange
pairs of entries in the time series of the time gaps
with the only prescription that the sum of them
is not lower than one and not larger than $M$. 
We consider different values of $\tau$. In each plot only
agents with at least $10+\tau$ bids in the same auction are
considered.}
\label{fig:correlation}
\end{center}
\end{figure}

\begin{figure}[!ht]
\begin{center}
\vskip .7cm
\includegraphics[width=7cm]{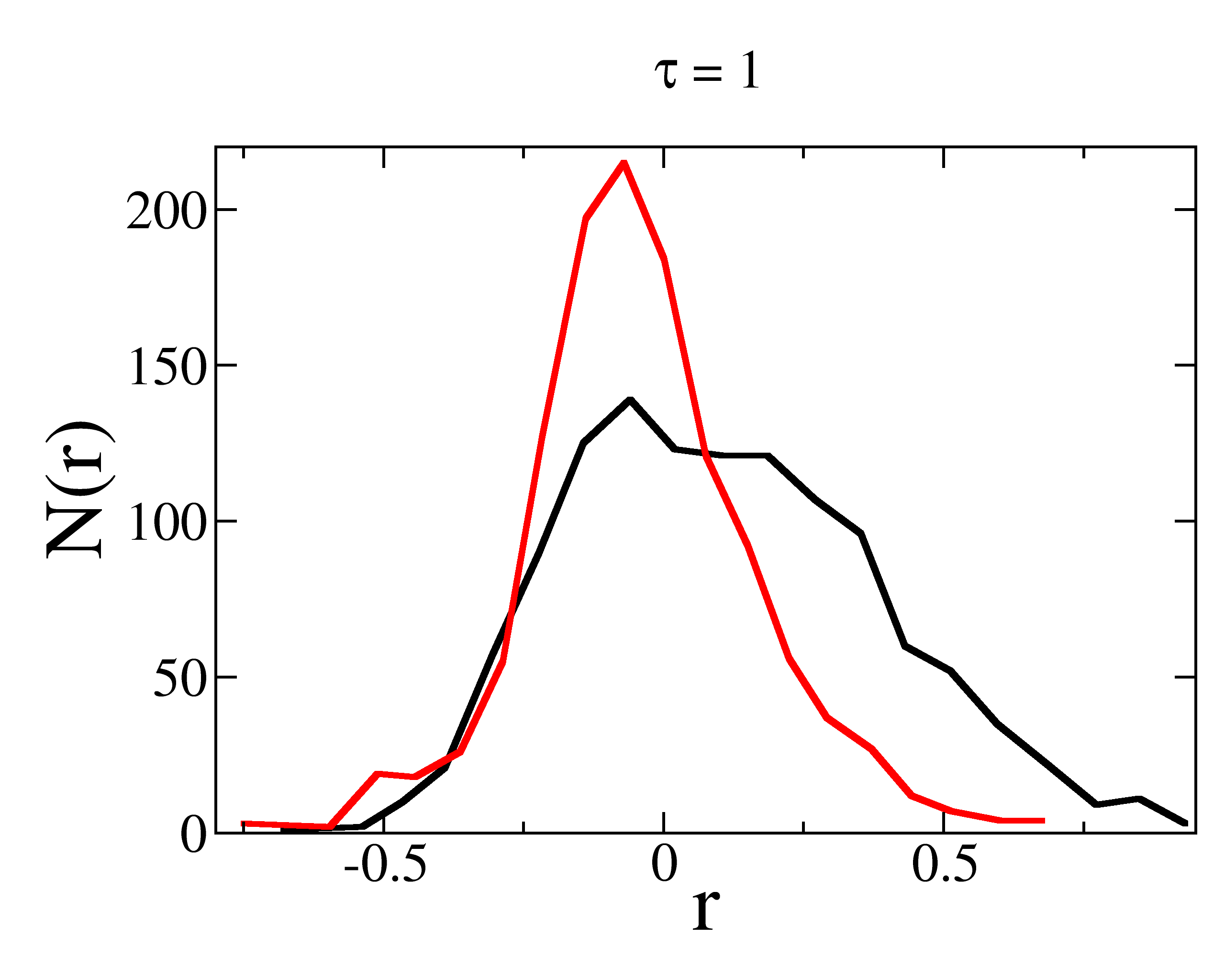}
\qquad
\includegraphics[width=7cm]{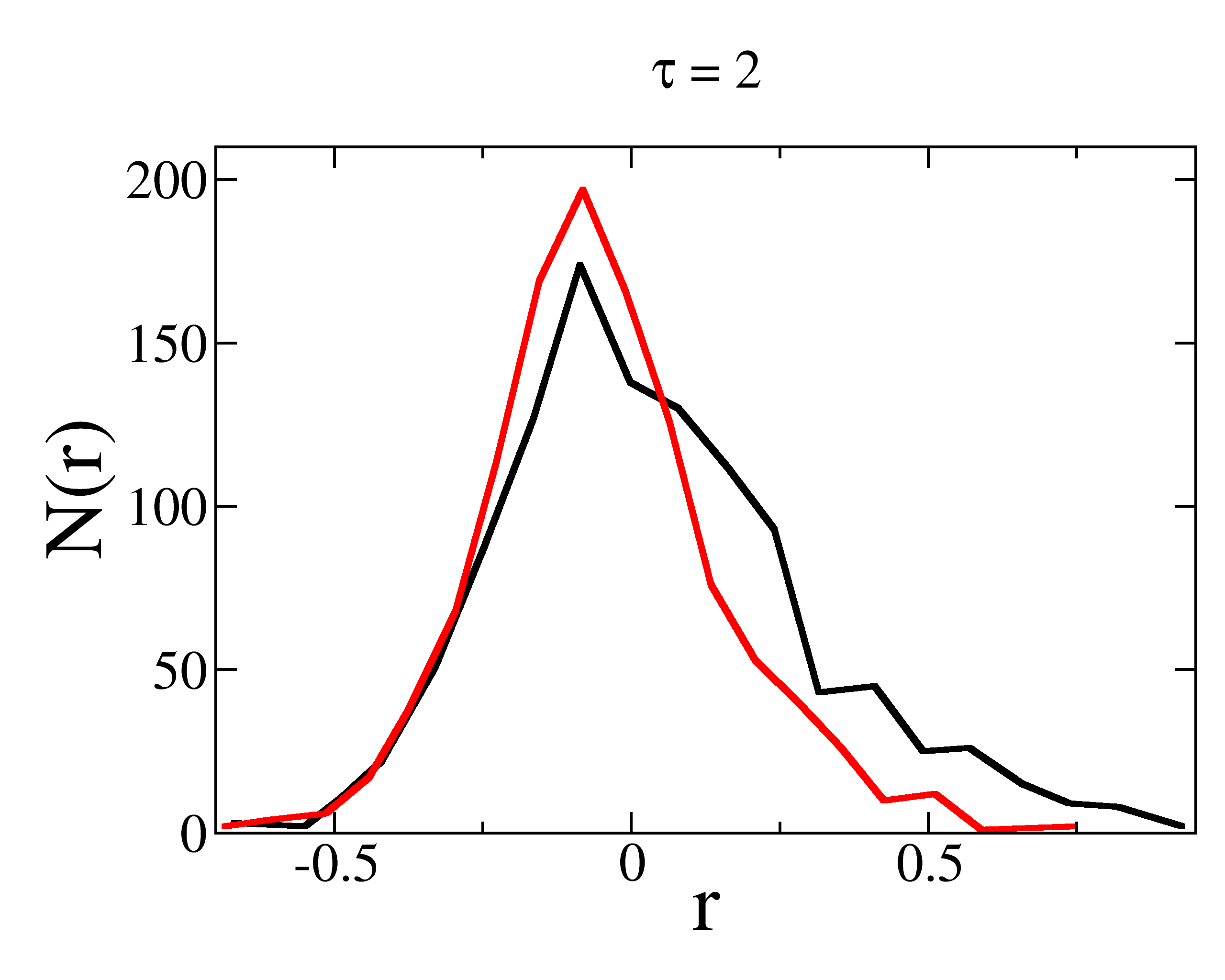}
\vskip .2cm
\includegraphics[width=7cm]{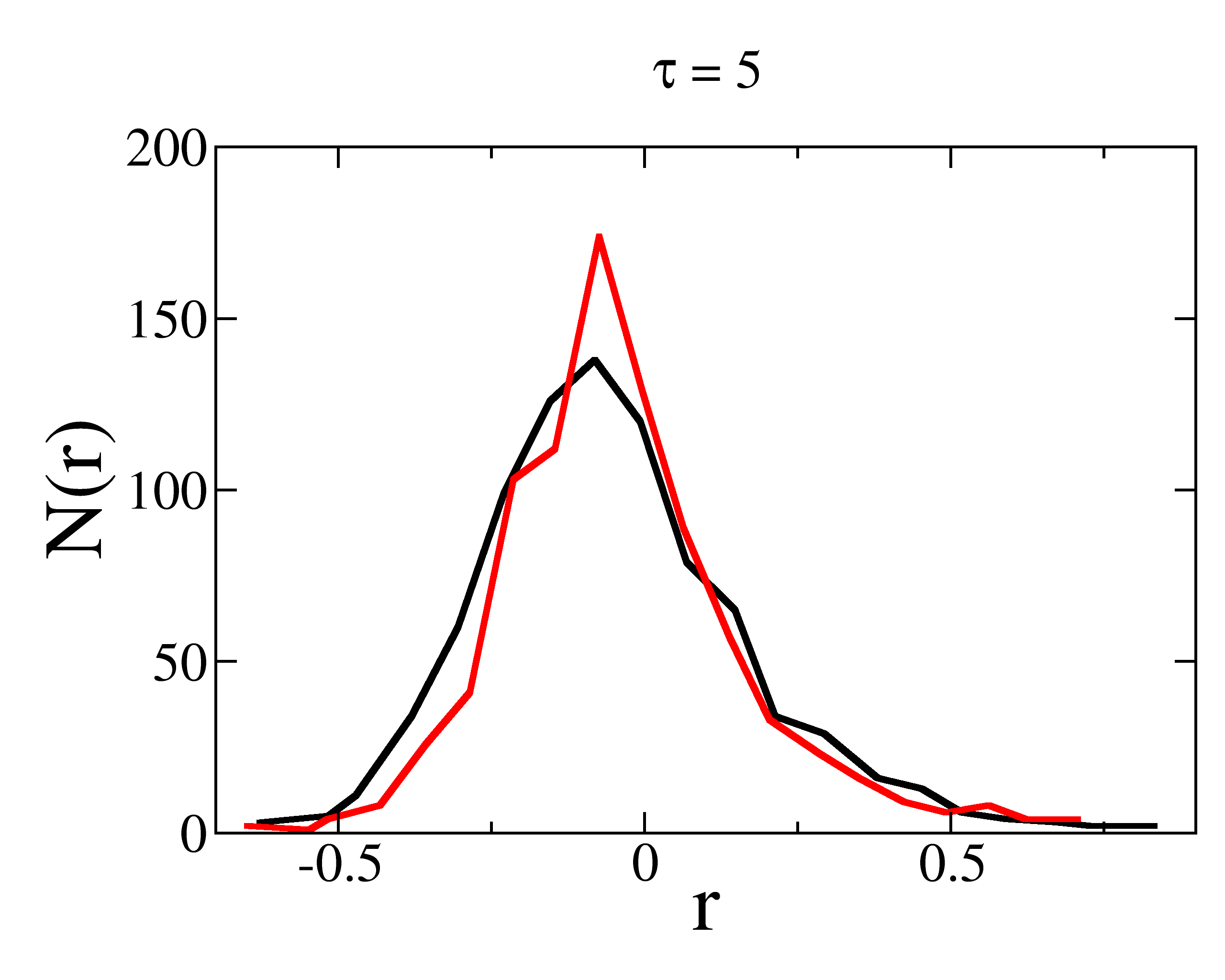}
\qquad
\includegraphics[width=7cm]{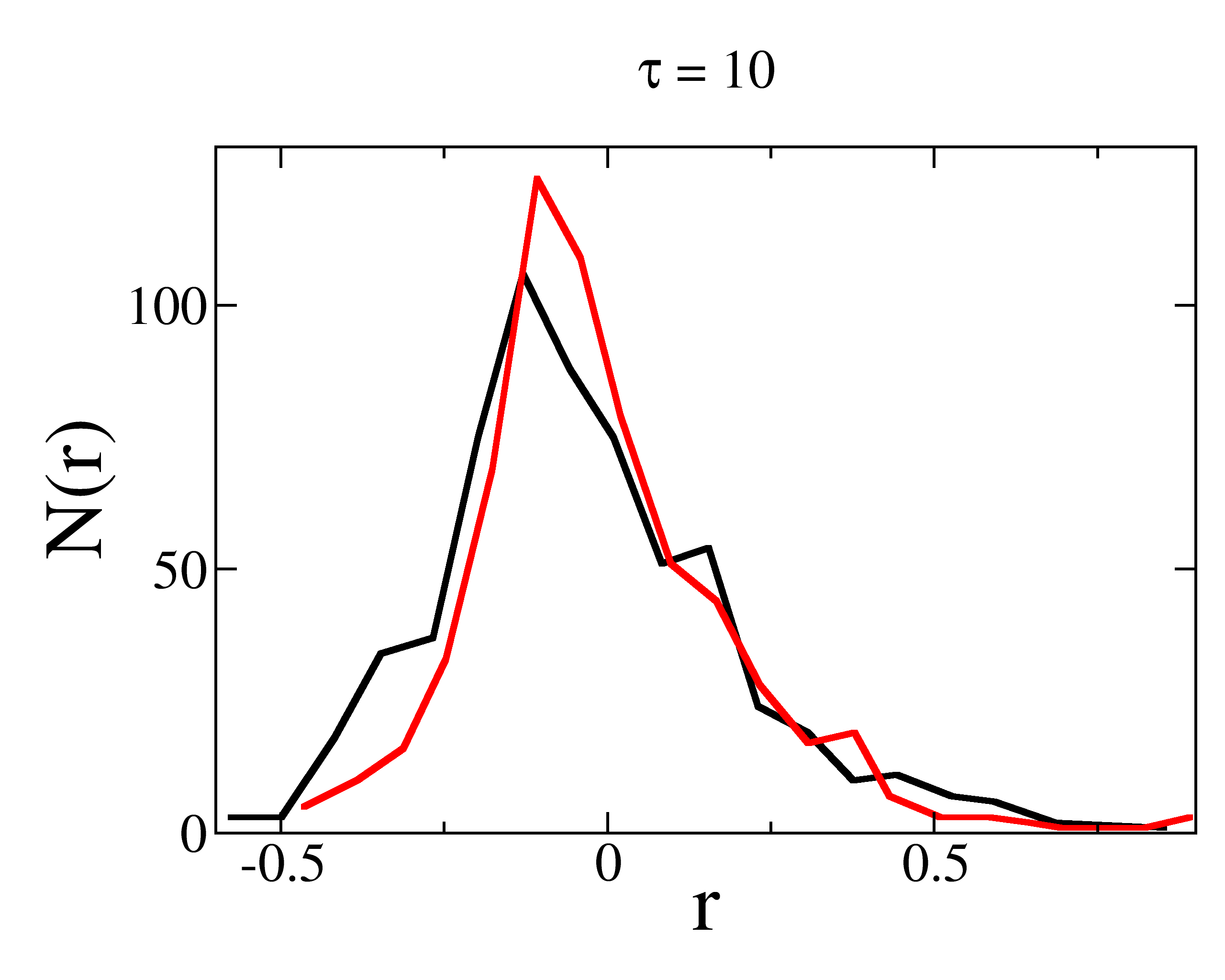}
\vskip .1cm
\caption{Same as those appearing
in Figure~\ref{fig:correlation} but for BM data set.}
\label{fig:correlation_bm}
\end{center}
\end{figure}

\noindent We further study the correlations between jumps.
Given an agent and an auction, we consider the list of all her
jumps $d_1, d_2, \ldots, d_{B-1}$ and calculate the Pearson's
correlation coefficient
\begin{equation}
r_\tau = \frac{\langle \left(d_t - \mu_t\right) \left(d_{t+\tau} - \mu_{t+\tau}\right)  \rangle}{\sigma_t \, \sigma_{t+\tau}} \;\; ,
\end{equation}
where $\langle \cdot \rangle$ stands for the average
over the entire time series (i.e., over all values of $t$ from $1$ to $T-1-\tau$). $\mu_t = \langle d_t \rangle$ and $\mu_{t+\tau} = \langle d_{t+\tau} \rangle$ are the average values of the bid gaps along the time series, while
$\sigma_t = \sqrt{\langle d_t^2 \rangle - \langle d_t \rangle}$ and
$\sigma_{t+\tau} = \sqrt{\langle d_{t+\tau}^2 \rangle - \langle d_{t+\tau} \rangle}$
are the respective standard deviations. We measure such coefficient
for every agent who has performed at least $10+\tau$ bids in the
same auction and show the number of agents $N\left(r\right)$ with given
value of $r$ in Figures~\ref{fig:correlation} and~\ref{fig:correlation_bm}. 
The same quantity
is also calculated for a randomized version of
the time series, where bid gaps are randomly reshuffled with
the only constraint that their partial sum cannot never be
smaller than one and larger than $M$. 
We consider several values of $\tau$. As one can clearly notice,
subsequent gaps (i.e., $\tau=1$) are slightly correlated. Such correlation,
becomes negligible when $\tau$ grows and already  for $\tau=2$,
$N\left(r\right)$ is negligible.
For $\tau=5$ and $\tau=10$, the curves corresponding to the
original time series and those obtained over randomly reshuffled
time series are almost identical.
\\
Such results show that agents perform almost uncorrelated
L\'evy flights. Once an agent makes a jump,
the length of this jump is slightly correlated
with the one of the jump made before. However, after 
few jumps there is not longer memory of what happened before.
In good approximation, the walk of the agent in the bid space 
can be therefore modeled as the one followed by
a random walker performing uncorrelated L\'evy flights.

\clearpage

\subsection{Testing the model}

\subsubsection{Maximum likelihood fit and Goodness of fit}
\noindent In this section, we compute the level of significance
of our model for the description of
real time series.
Suppose that 
a time series of $T$ bid values $b_1, b_2, \ldots, b_T$ 
describes a realization of our model. Fixed the exponent
$\alpha$, the bound $M$ of the lattice and the position $b_{t-1}$
at stage $t-1$, the probability
that the random walker jumps at $b_t$ at stage $t$ is
given by the transition matrix of Eq.~(6) of the main text. 
The probability or {\it likelihood} that the whole sequence
was extracted from our model is
\[
p\left(b_1, b_2, \ldots, b_T \left| \alpha \right.\right) =  \left(Q_\alpha\right)_{0,b_1} \; \left(Q_\alpha\right)_{b_1,b_2} \; \left(Q_\alpha\right)_{b_2,b_3} \; \cdots \; \left(Q_\alpha\right)_{b_{T-1},b_T} \;  .
\]
The value of $\alpha$ that maximizes the former equation represents
the most likely exponent of our model that could have generated
our particular sequence. In order to find its maximum, it is convenient 
to take the logarithm of both sides and write the log-likelihood
\begin{equation}
\mathcal{L}\left(b_1, b_2, \ldots, b_T \left| \alpha \right.\right) = \sum_{t=1}^T \; \ln{\left[  \left(Q_\alpha\right)_{b_{t-1},b_t} \right] } =
- \alpha \, \sum_{t=1}^T \ln{ \left| b_{t} - b_{t-1} \right| } -  \sum_{t=1}^T \ln{\left[ m_{b_{t-1}}\left(\alpha\right) \right]}  \;\;,
\label{eq:loglike}
\end{equation}
where we set $b_0=0$. The value $\alpha'$ 
at which the maximum of Eq.~(\ref{eq:loglike}) occurs 
can be estimated
numerically.

\noindent $\alpha'$ is the best exponent fitting  the data
in the hypothesis that they were produced
according to our model. The
significance level of the model for the description
of the data can be calculated by estimating the $p$-value associated
with our measurement. In this respect, we first compute the distance
between the theoretical distribution of the jump lengths 
\[
P_{\alpha'}\left(d \right) = \left[ \sum_{i,j} \left(Q_{\alpha'}\right)_{ij} \delta\left(d - \left|i - j \right| \right) \right] \; \left[ \sum_d \sum_{i,j} \left(Q_{\alpha'}\right)_{ij} \delta\left(d - \left|i - j \right| \right) \right]^{-1}
\]
and the one obtained
from our data $P\left(d\right)$ by calculating
\[
\eta_{data}= \max_{t} \left| P_{\alpha'}\left( \geq d_t \right) - P\left( \geq d_t \right)  \right|  \;\; ,
\]
where $P\left(\geq d\right) = \sum_{q \geq d} P\left(q\right)$ is the cumulative distribution of
the jump lengths. Notice that the distance between cumulative
distributions is the same
as the one adopted in the Kolmogorov-Smirnov test.
We then generate artificial time series of length $T$ 
from our model with exponent $\alpha'$
and compute their distance $\eta$ with respect to
the theoretical distribution. The $p$-value
is finally determined by the relative number of times in which 
we observe $\eta \geq \eta_{data}$.
\\
Synthetic time series generated according to our model
can be additionally used for the
determination of the error associated to the
estimation of $\alpha'$. The error associated to
$\alpha'$ is the standard deviation of the best exponents
estimated, with maximum likelihood, for the synthetic data sets.

\begin{figure}[!ht]
\begin{center}
\vskip .7cm
\includegraphics[width=7cm]{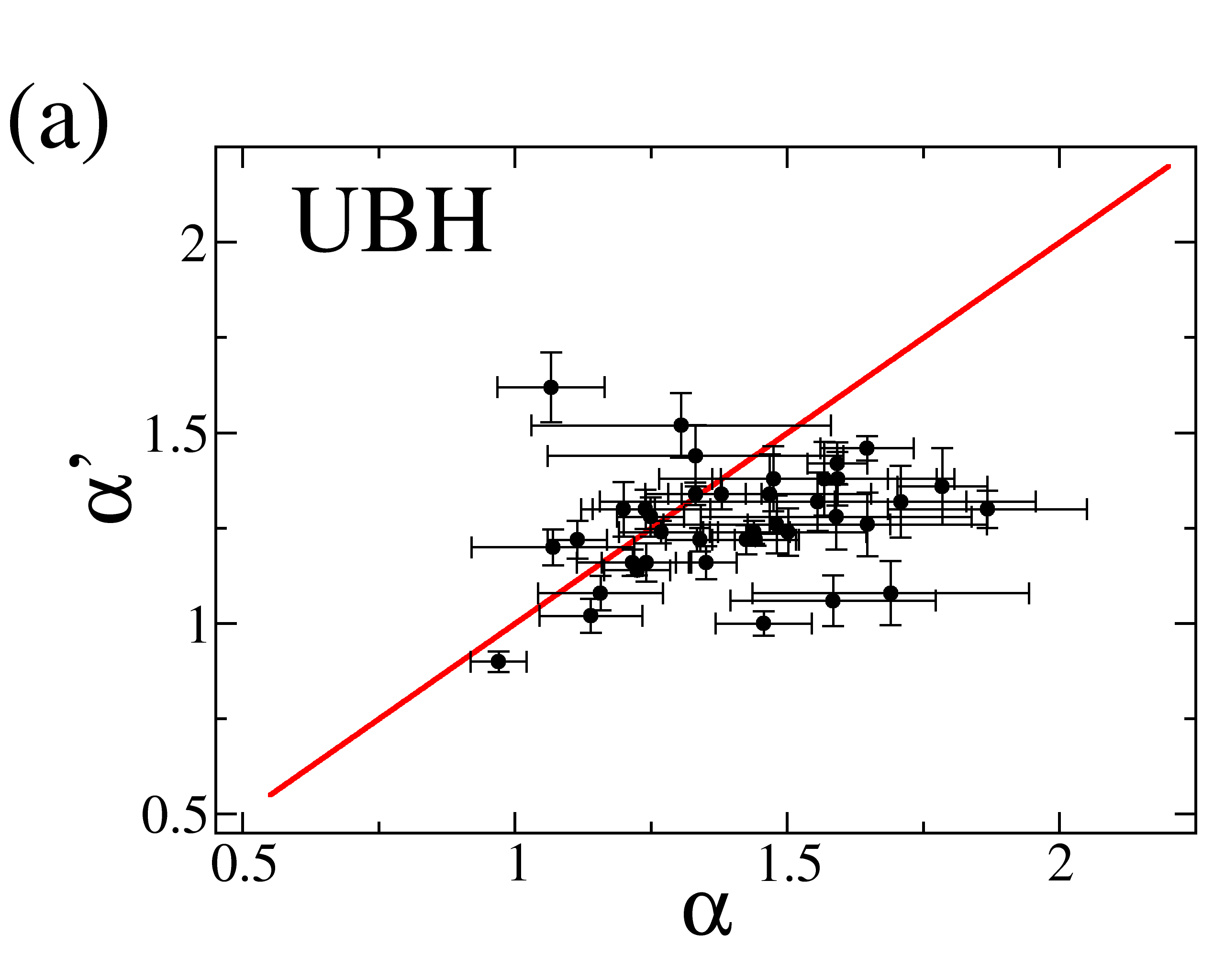}
\qquad
\includegraphics[width=7cm]{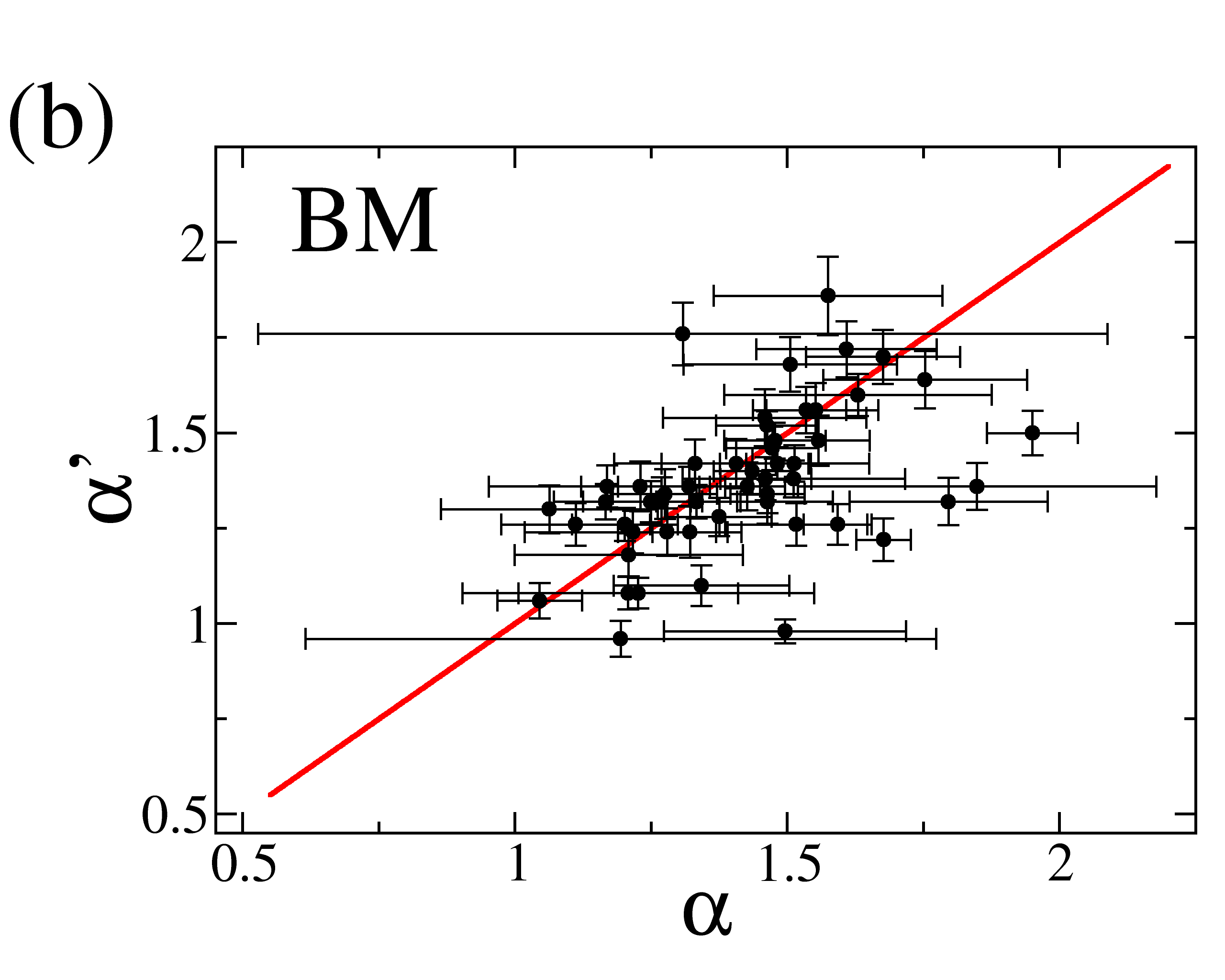}
\vskip .1cm
\caption{We consider only agents who have
performed at least $T=50$ bids in a single auction
in UBH data set and those with at least $T=100$ in the BM data set.
Our data set offers  $39$ agents that satisfy this constraint in the UBH data set
and $52$ in the BM data set.
Panels a (UBH) and b (BM) show
the scatter plot $\alpha'$ {\it versus} $\alpha$
for the best power-law exponents estimated by using
 maximum likelihood and least square methods, respectively.
The agreement between the two measurements is
good as demonstrated by the fact that
the majority of the points fall on the diagonal (red line).}
\label{fig:bayes}
\end{center}
\end{figure}

\noindent A graphical comparison between the exponents $\alpha$ (least square method) 
and $\alpha'$ (maximum likelihood method) is presented
in Figure~\ref{fig:bayes}. In general, the two methods
produce consistent results. For completeness, we list  the
results obtained in Tables~\ref{tab:ubh} and ~\ref{tab:bm}.

\begin{table}[!th]
\begin{center}

\begin{minipage}[!t]{0.4\textwidth}
\centering
\begin{tabular}{|c c c c c|}
 \hline
$a$ & $u$ & $\alpha$ & $\alpha'$ &   $p$
\\ \hline
\hline
$1^*$ & $23$ & $1.6(1)$ & $1.5(1)$ & $0.00$
 \\ \hline 
$1$ & $81$ & $1.2(1)$ & $1.3(1)$ & $0.17$
 \\ \hline 
$100$ & $1715$ & $1.6(2)$ & $1.1(1)$ & $0.01$
 \\ \hline 
$100^*$ & $81$ & $1.5(2)$ & $1.3(1)$ & $0.00$
 \\ \hline 
$104$ & $3093$ & $1.7(3)$ & $1.1(1)$ & $0.14$
 \\ \hline 
$108$ & $134$ & $1.7(1)$ & $1.3(1)$ & $0.02$
 \\ \hline 
$14$ & $134$ & $1.2(1)$ & $1.1(1)$ & $0.03$
 \\ \hline 
$14^*$ & $81$ & $1.9(2)$ & $1.3(1)$ & $0.00$
 \\ \hline 
$15$ & $423$ & $1.6(4)$ & $1.3(1)$ & $0.3$
 \\ \hline 
$179$ & $3663$ & $1.8(1)$ & $1.4(1)$ & $0.17$
 \\ \hline 
$19$ & $1$ & $1.3(1)$ & $1.3(1)$ & $0.28$
 \\ \hline 
$19$ & $1313$ & $1.2(1)$ & $1.2(1)$ & $0.14$
 \\ \hline 
$19^*$ & $134$ & $1.2(1)$ & $1.1(1)$ & $0.00$
 \\ \hline 
$19$ & $1433$ & $1.1(1)$ & $1.0(1)$ & $0.06$
 \\ \hline 
$19$ & $1448$ & $1.4(1)$ & $1.2(1)$ & $0.01$
 \\ \hline 
$19$ & $1558$ & $1.1(1)$ & $1.2(1)$ & $0.63$
 \\ \hline 
$19$ & $1576$ & $1.0(1)$ & $0.9(1)$ & $0.35$
 \\ \hline 
$19$ & $1601$ & $1.4(1)$ & $1.3(1)$ & $0.01$
 \\ \hline 
$19$ & $1632$ & $1.4(1)$ & $1.2(1)$ & $0.01$
 \\ \hline 
$19^*$ & $1640$ & $1.3(1)$ & $1.2(1)$ & $0.00$
 \\ \hline 
\end{tabular}
\end{minipage} 
\hspace{0.1cm}
\begin{minipage}[!t]{0.4\textwidth}
\centering
\begin{tabular}{|c c c c c|}
\hline
$a$ & $u$ & $\alpha$ & $\alpha'$ &   $p$
\\ \hline
\hline
$19$ & $1642$ & $1.3(1)$ & $1.2(1)$ & $0.13$
 \\ \hline 
$19^*$ & $1644$ & $1.4(1)$ & $1.2(1)$ & $0.00$
 \\ \hline 
$19^*$ & $1645$ & $1.4(1)$ & $1.2(1)$ & $0.00$
 \\ \hline 
$19$ & $3$ & $1.2(1)$ & $1.3(1)$ & $0.9$
 \\ \hline 
$19$ & $363$ & $1.2(1)$ & $1.1(1)$ & $0.02$
 \\ \hline 
$19$ & $434$ & $1.1(1)$ & $1.2(1)$ & $0.35$
 \\ \hline 
$19$ & $438$ & $1.5(1)$ & $1.0(1)$ & $0.02$
 \\ \hline 
$20$ & $617$ & $1.6(2)$ & $1.4(1)$ & $0.01$
 \\ \hline 
$22$ & $134$ & $1.2(1)$ & $1.3(1)$ & $0.95$
 \\ \hline 
$44$ & $433$ & $1.1(1)$ & $1.6(1)$ & $0.07$
 \\ \hline 
$46$ & $2003$ & $1.3(3)$ & $1.5(1)$ & $0.02$
 \\ \hline 
$5^*$ & $128$ & $1.6(1)$ & $1.4(1)$ & $0.00$
 \\ \hline 
$62$ & $2392$ & $1.3(3)$ & $1.4(1)$ & $0.43$
 \\ \hline 
$71^*$ & $324$ & $1.6(2)$ & $1.4(1)$ & $0.00$
 \\ \hline 
$73$ & $1640$ & $1.5(2)$ & $1.4(1)$ & $0.18$
 \\ \hline 
$73$ & $1715$ & $1.5(1)$ & $1.2(1)$ & $0.34$
 \\ \hline 
$79$ & $134$ & $1.6(2)$ & $1.3(1)$ & $0.11$
 \\ \hline 
$91$ & $1715$ & $1.5(2)$ & $1.3(1)$ & $0.32$
 \\ \hline 
$97$ & $1715$ & $1.6(2)$ & $1.3(1)$ & $0.09$
 \\ \hline 

\end{tabular}
\end{minipage} 

\end{center}
\caption{UBH data set. Each row corresponds to
one of the $39$ agents who have bid at least $50$ times in the same auction.
We report the id of the auction $a$, the id of the agent $u$, the
exponent $\alpha$ calculated with the least square method,
the exponent $\alpha'$  calculated with the maximum likelihood method and
the $p$-value. Entries with low $p$-values are marked with $^*$.
In the $77\%$
of the cases we find a $p$-value larger than $0$, which
indicates that our model well describe
the time series.}
\label{tab:ubh}
\end{table}

\begin{table}[!th]
\begin{center}

\begin{minipage}[!t]{0.4\textwidth}
\centering
\begin{tabular}{|c c c c c|}
 \hline
$a$ & $u$ & $\alpha$ & $\alpha'$ &   $p$
\\ \hline
\hline
$11$ & $28$ & $1.1(2)$ & $1.3(1)$ & $0.09$
 \\ \hline 
$13^*$ & $48$ & $1.5(1)$ & $1.3(1)$ & $0$
 \\ \hline 
$13$ & $28$ & $1.3(1)$ & $1.4(1)$ & $0.04$
 \\ \hline 
$15$ & $11$ & $1.8(3)$ & $1.4(1)$ & $0.05$
 \\ \hline 
$15$ & $36$ & $2.0(1)$ & $1.5(1)$ & $0.03$
 \\ \hline 
$28$ & $28$ & $1.4(1)$ & $1.4(1)$ & $0.58$
 \\ \hline 
$32^*$ & $150$ & $1.2(6)$ & $1.0(1)$ & $0$
 \\ \hline 
$39^*$ & $48$ & $1.3(1)$ & $1.3(1)$ & $0$
 \\ \hline 
$47$ & $28$ & $1.1(1)$ & $1.3(1)$ & $0.48$
 \\ \hline 
$50$ & $48$ & $1.5(1)$ & $1.3(1)$ & $0.27$
 \\ \hline 
$52$ & $28$ & $1.2(2)$ & $1.4(1)$ & $0.01$
 \\ \hline 
$55$ & $15$ & $1.3(1)$ & $1.3(1)$ & $0.38$
 \\ \hline 
$55^*$ & $150$ & $1.2(3)$ & $1.1(1)$ & $0$
 \\ \hline 
$61$ & $213$ & $1.3(1)$ & $1.2(1)$ & $0.62$
 \\ \hline 
$61$ & $36$ & $1.3(1)$ & $1.2(1)$ & $0.5$
 \\ \hline 
$62$ & $136$ & $1.6(1)$ & $1.5(1)$ & $0.2$
 \\ \hline 
$67$ & $98$ & $1.0(1)$ & $1.1(1)$ & $0.15$
 \\ \hline 
$69$ & $28$ & $1.5(1)$ & $1.4(1)$ & $0.39$
 \\ \hline 
$82^*$ & $36$ & $1.2(2)$ & $1.2(1)$ & $0$
 \\ \hline 
$89$ & $72$ & $1.4(1)$ & $1.3(1)$ & $0.15$
 \\ \hline 
$89$ & $48$ & $1.5(1)$ & $1.3(1)$ & $0.2$
 \\ \hline 
$89$ & $28$ & $1.2(1)$ & $1.3(1)$ & $0.16$
 \\ \hline 
$91$ & $15$ & $1.4(1)$ & $1.4(1)$ & $0.65$
 \\ \hline 
$92^*$ & $36$ & $1.8(2)$ & $1.3(1)$ & $0$
 \\ \hline 
$92$ & $28$ & $1.3(1)$ & $1.3(1)$ & $0.42$
 \\ \hline 
$94^*$ & $48$ & $1.6(1)$ & $1.3(1)$ & $0$
 \\ \hline 

\end{tabular}
\end{minipage} 
\hspace{0.1cm}
\begin{minipage}[!t]{0.4\textwidth}
\centering
\begin{tabular}{|c c c c c|}
\hline
$a$ & $u$ & $\alpha$ & $\alpha'$ &   $p$
\\ \hline
\hline
$94^*$ & $98$ & $1.5(2)$ & $1.0(1)$ & $0$
 \\ \hline 
$94$ & $15$ & $1.5(1)$ & $1.5(1)$ & $0.15$
 \\ \hline 
$105$ & $28$ & $1.5(1)$ & $1.5(1)$ & $0.45$
 \\ \hline 
$110$ & $28$ & $1.2(1)$ & $1.4(1)$ & $0.18$
 \\ \hline 
 $116$ & $15$ & $1.2(1)$ & $1.3(1)$ & $0.04$
 \\ \hline 
$122$ & $15$ & $1.5(1)$ & $1.4(1)$ & $0.78$
 \\ \hline 
$125$ & $28$ & $1.4(1)$ & $1.4(1)$ & $0.88$
 \\ \hline 
$133$ & $98$ & $1.7(1)$ & $1.2(1)$ & $0.02$
 \\ \hline 
$162$ & $15$ & $1.2(1)$ & $1.3(1)$ & $0.48$
 \\ \hline 
$181$ & $15$ & $1.2(2)$ & $1.1(1)$ & $0.1$
 \\ \hline 
$197$ & $15$ & $1.2(2)$ & $1.2(1)$ & $0.58$
 \\ \hline 
$201^*$ & $15$ & $1.3(2)$ & $1.1(1)$ & $0$
 \\ \hline 
 $262$ & $150$ & $1.3(8)$ & $1.8(1)$ & $0.04$
 \\ \hline 
$264$ & $1428$ & $1.5(2)$ & $1.7(1)$ & $0.12$
 \\ \hline 
\ $269$ & $122$ & $1.6(2)$ & $1.7(1)$ & $0.02$
 \\ \hline 
$279$ & $550$ & $1.5(2)$ & $1.5(1)$ & $0.22$
 \\ \hline 
$293^*$ & $3503$ & $1.5(1)$ & $1.5(1)$ & $0$
 \\ \hline 
$300$ & $550$ & $1.5(1)$ & $1.6(1)$ & $0.41$
 \\ \hline 
$300$ & $150$ & $1.7(1)$ & $1.7(1)$ & $0.17$
 \\ \hline 
$306^*$ & $150$ & $1.5(1)$ & $1.4(1)$ & $0$
 \\ \hline 
$317$ & $150$ & $1.8(2)$ & $1.6(1)$ & $0.22$
 \\ \hline 
 $318$ & $150$ & $1.5(2)$ & $1.4(1)$ & $0.04$
 \\ \hline 
$327$ & $1503$ & $1.3(1)$ & $1.4(1)$ & $0.19$
 \\ \hline 
$327$ & $150$ & $1.6(1)$ & $1.6(1)$ & $0.25$
 \\ \hline 
$331$ & $150$ & $1.6(2)$ & $1.6(1)$ & $0.29$
 \\ \hline 
$332$ & $1574$ & $1.6(2)$ & $1.9(1)$ & $0.15$
 \\ \hline 
\end{tabular}
\end{minipage} 

\end{center}
\caption{BM data set. Each row corresponds to
one of the $52$ agents who have bid at least $100$ times in the same auction.
We report the id of the auction $a$, the id of the agent $u$, the
exponent $\alpha$ calculated with the least square method,
the exponent $\alpha'$  calculated with the maximum likelihood method and
the $p$-value. Entries with low $p$-values are marked with $^*$. 
In the $79\%$
of the cases we find a $p$-value larger than $0$, which
indicates that our model well describe
the time series.}
\label{tab:bm}
\end{table}

\begin{figure}[!ht]
\begin{center}
\vskip .7cm
\includegraphics[width=7cm]{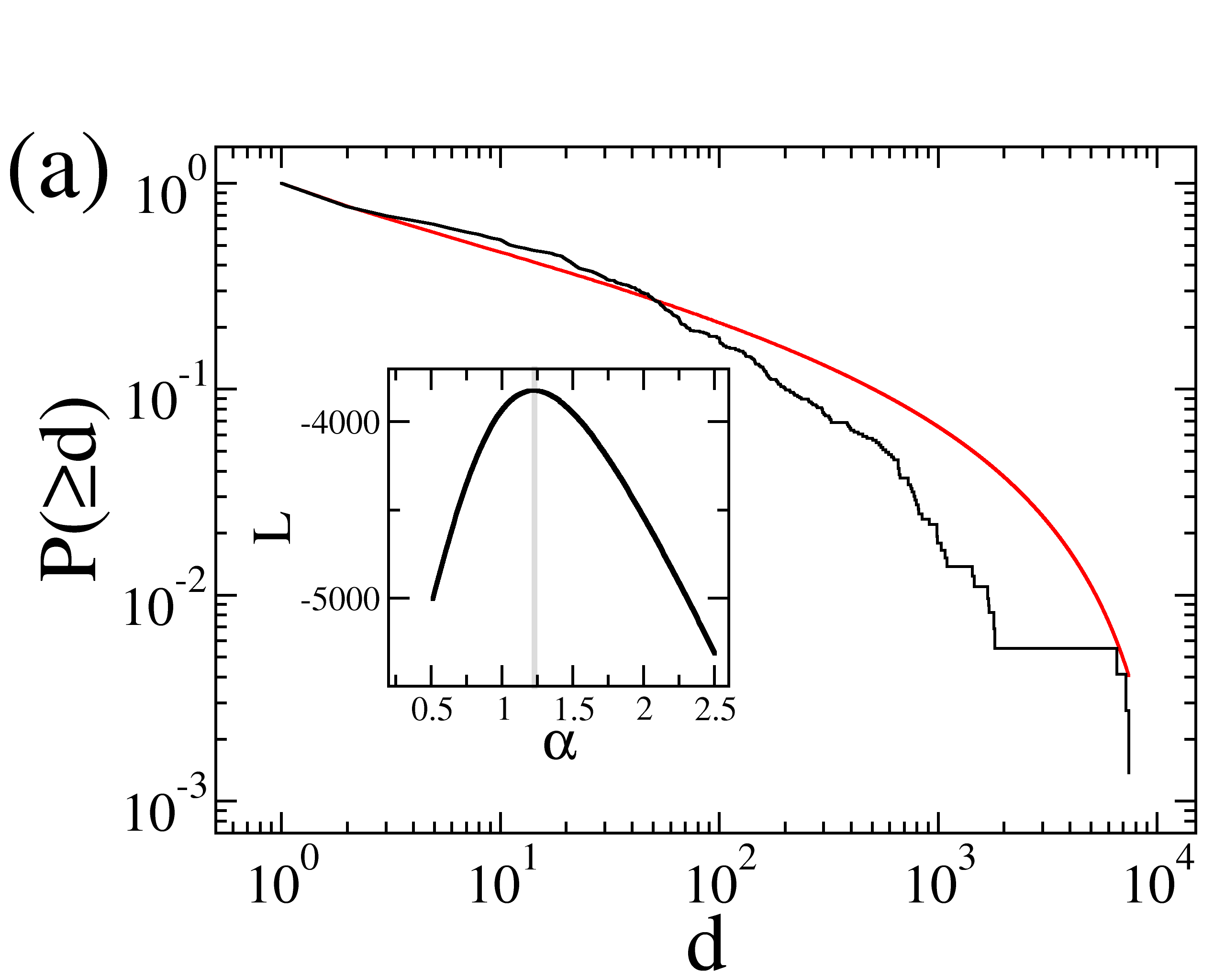}
\qquad
\includegraphics[width=7cm]{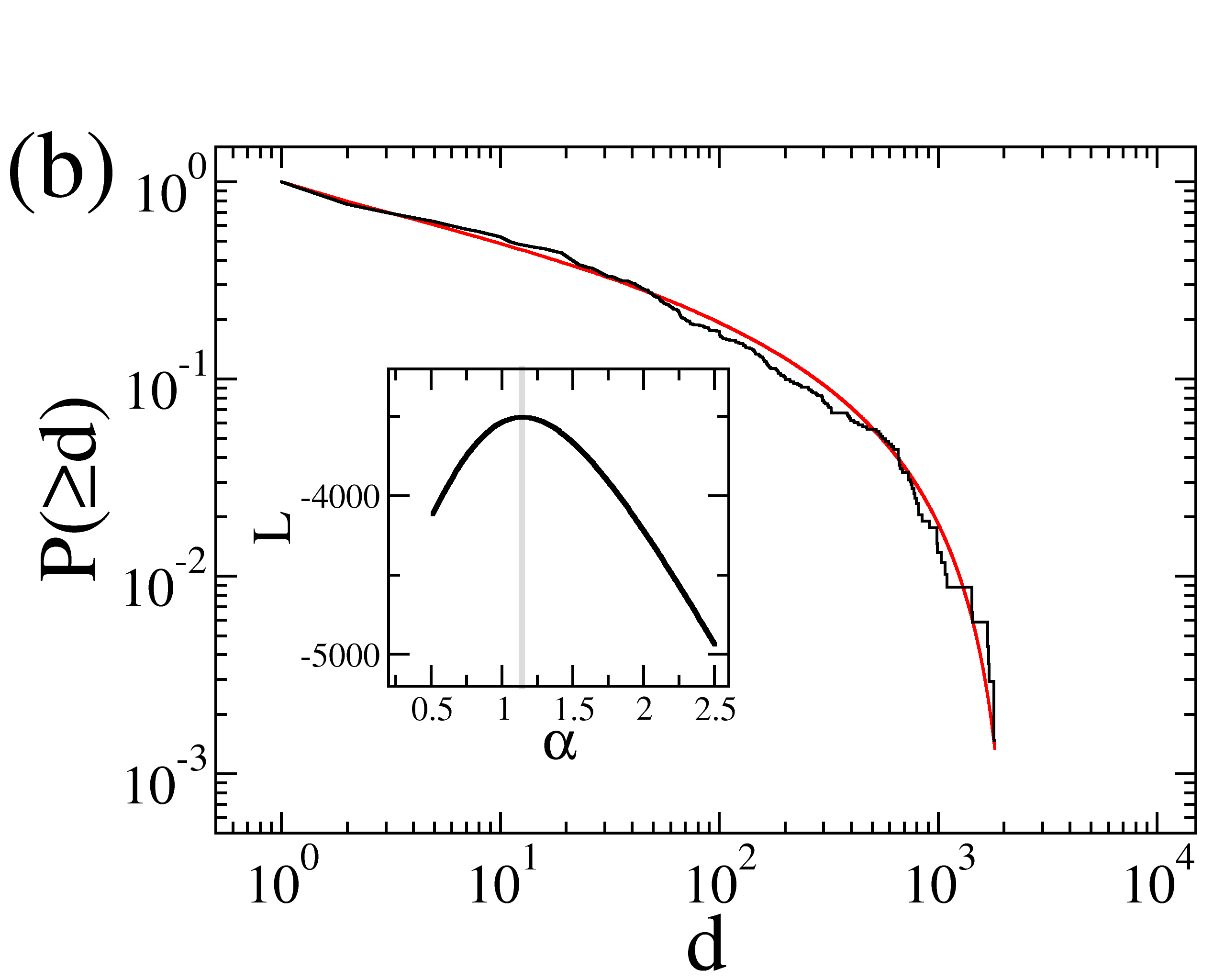}
\vskip .1cm
\caption{Example of the maximum likelihood method for the determination
of the exponent $\alpha'$ of the L\'evy flight. We consider
the time series of agent $1632$ in auction $19$ of the 
UBH data set (the same appearing in Figure~\ref{fig:example_user}).
In panel a, we set $M=13\,000$ which is the value of
the maximum bid amount that was allowed in the auction. The best exponent
$\alpha'=1.2(1)$ is obtained by looking at the maximum of the log-likelihood
as it is shown in the inset. 
The comparison between
the cumulative distribution
of the jump lengths expected from the model (red line)
and the one calculated over the time series (black)
do not well agree. It seems that the ``effective'' bound
is smaller than the real one.
 In panel b, we set $M=2\,200$ and 
consider only bid values smaller than this bound.
On the new time series, we perform a maximum likelihood
fit finding $\alpha'=1.1(1)$. The theoretical
expectation (red line) and the one obtained from the time series (black line)
are now very similar yielding a $p$-value equal to $0.1$.}
\label{fig:example_bayes}
\end{center}
\end{figure}

The $p$-values show also a general goodness
of our model for the description of the data.
Sometimes however, the value of $p$ is very small. This could be explained
in a simple manner. In Figure~\ref{fig:example_bayes} we plot for example
the statistical test performed over the same time series
appearing in Figure~\ref{fig:example_user}. In that auction, the maximum bid
amount was $M=13\,000$. However, this value of $M$ does not
correspond to the ``effective'' bound felt by the agent. This
bound seems to be around $M \sim 2\,000$ as the sudden drop
of $P\left(\geq d\right)$ would suggest. By setting $M=2\,200$ and running
again the statistical test, as we did in Figure~\ref{fig:example_bayes}b,
we see clearly that the curve predicted by our model
and the one measured on real data are in very good agreement.

\clearpage

\subsubsection{Another maximum likelihood fit}
\begin{figure}[!ht]
\begin{center}
\vskip .7cm
\includegraphics[width=7cm]{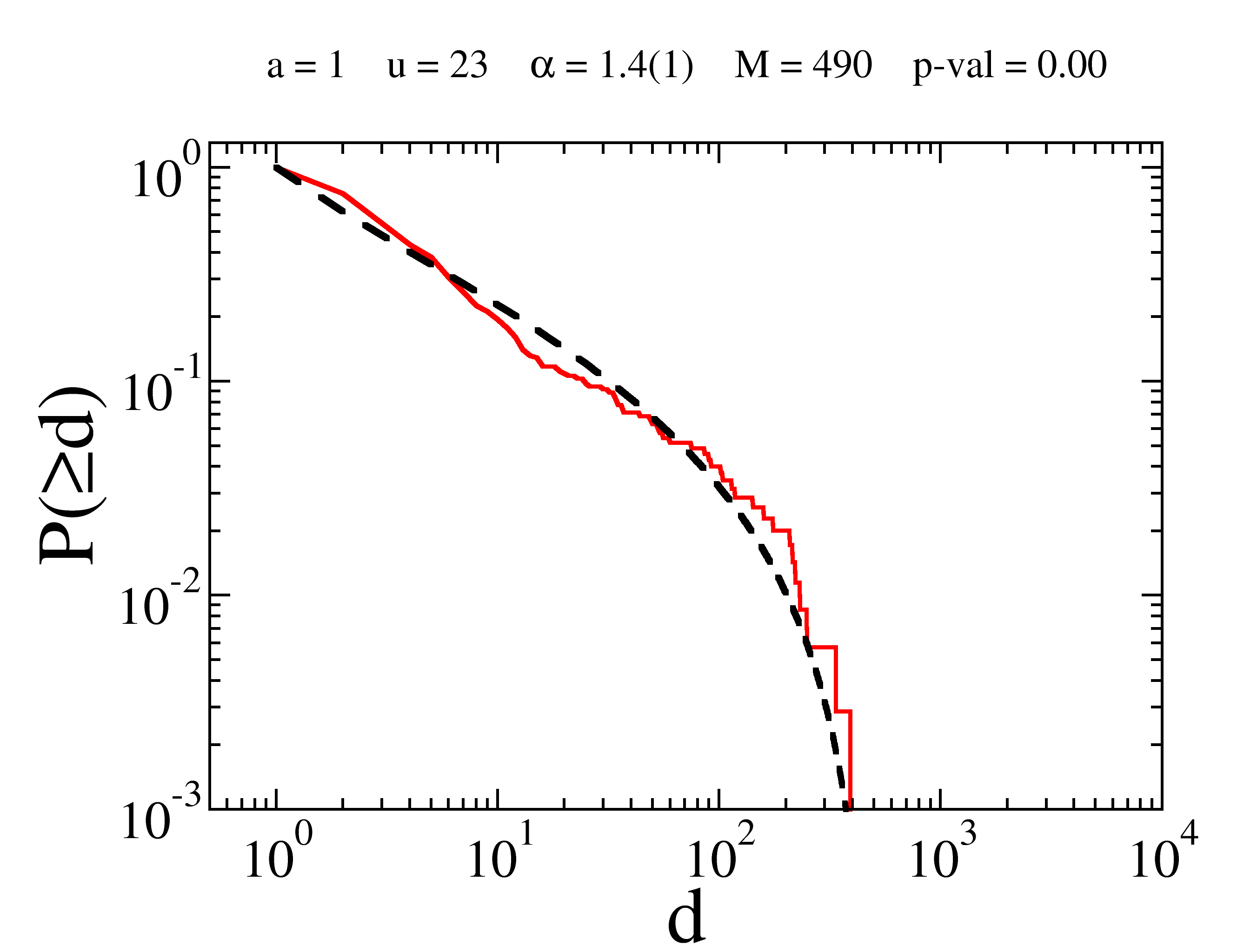}
\qquad
\includegraphics[width=7cm]{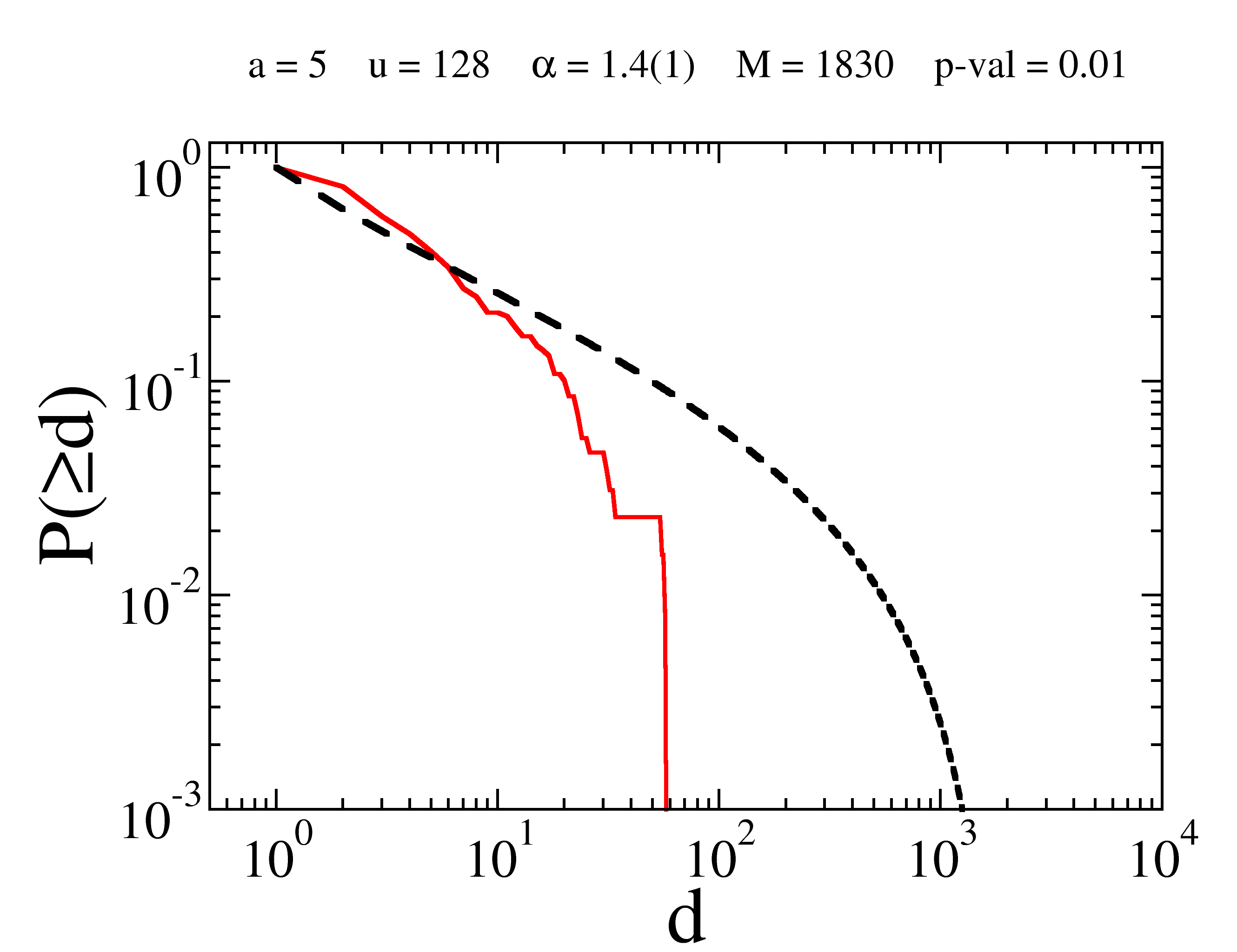}
\vskip .2cm
\includegraphics[width=7cm]{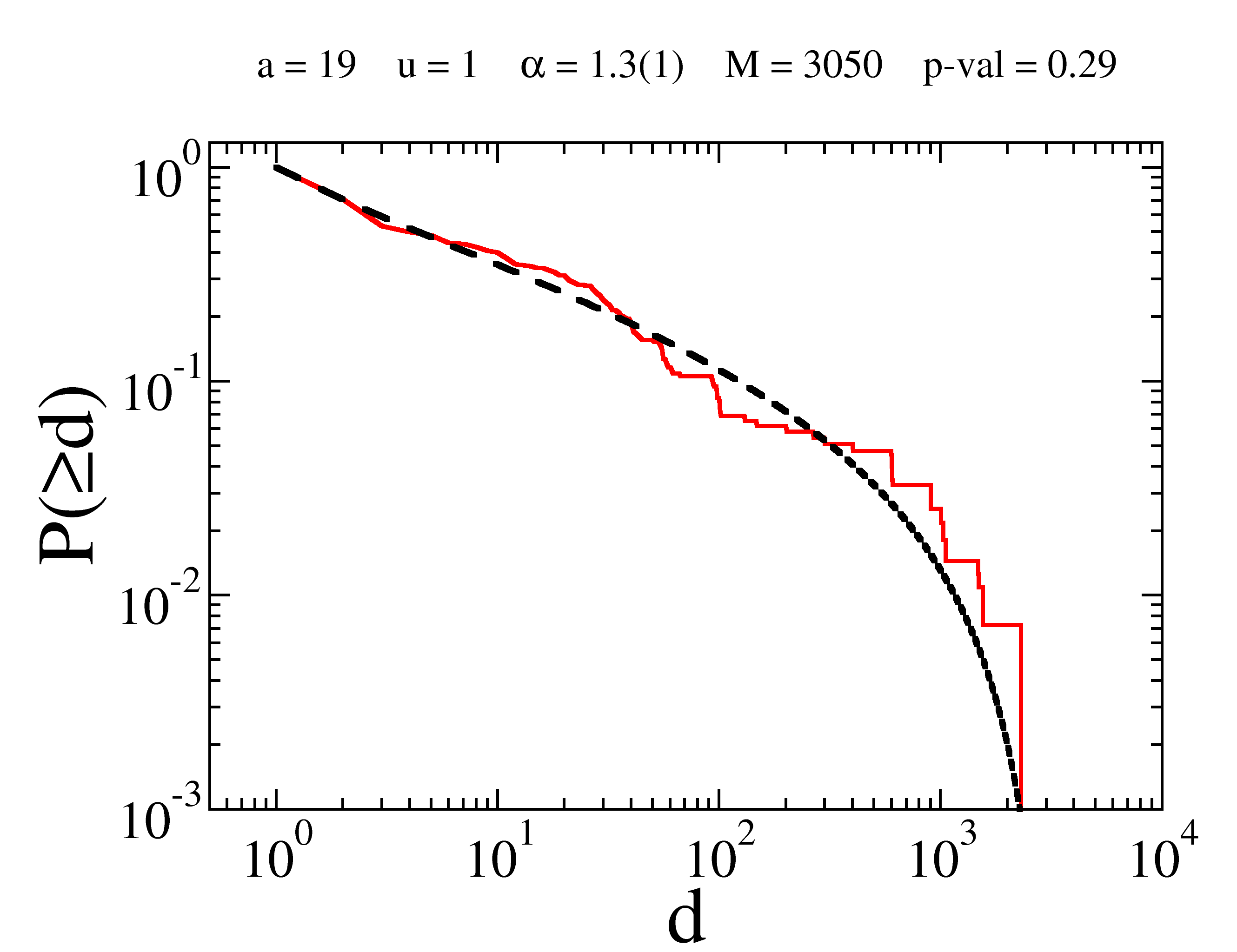}
\qquad
\includegraphics[width=7cm]{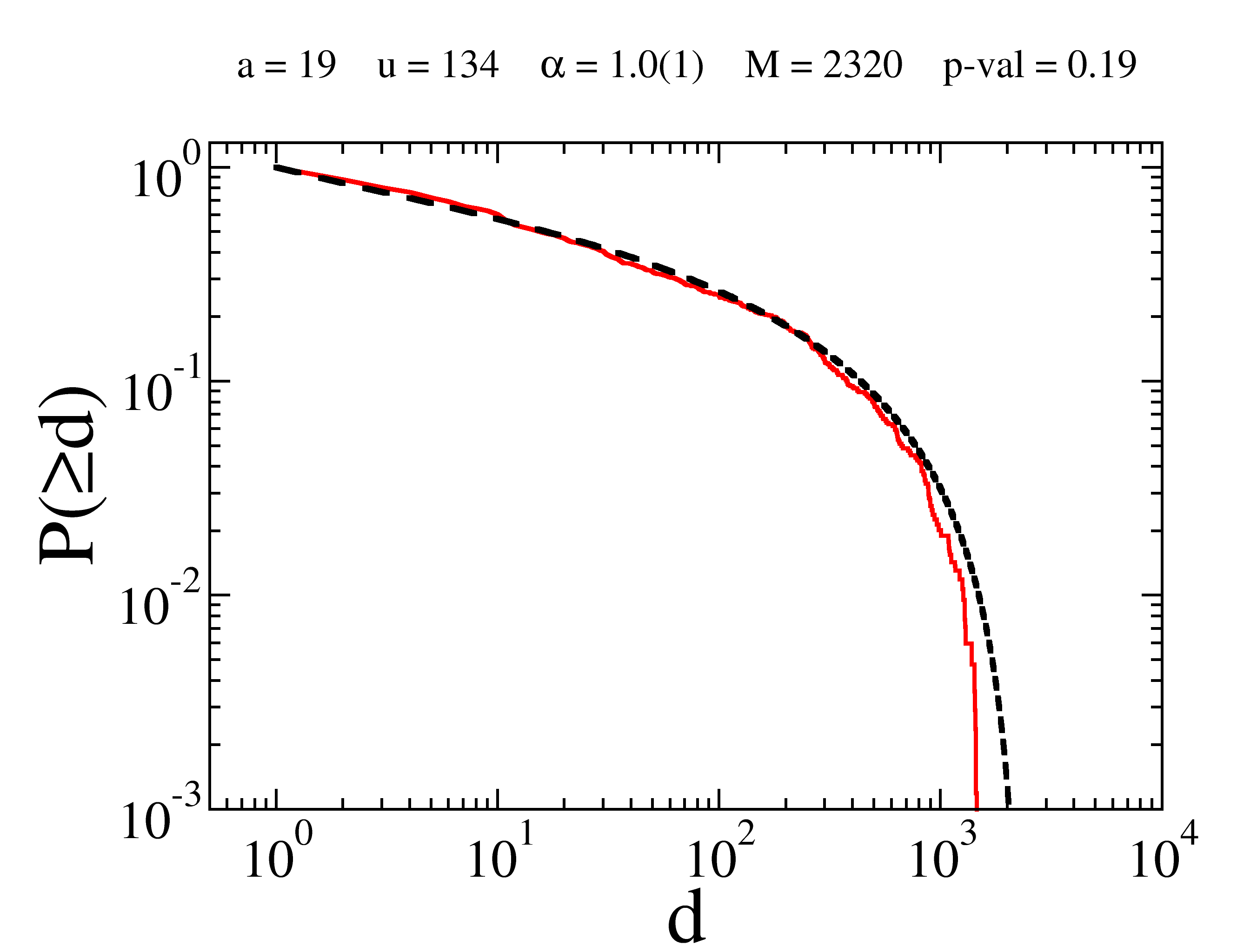}
\vskip .2cm
\includegraphics[width=7cm]{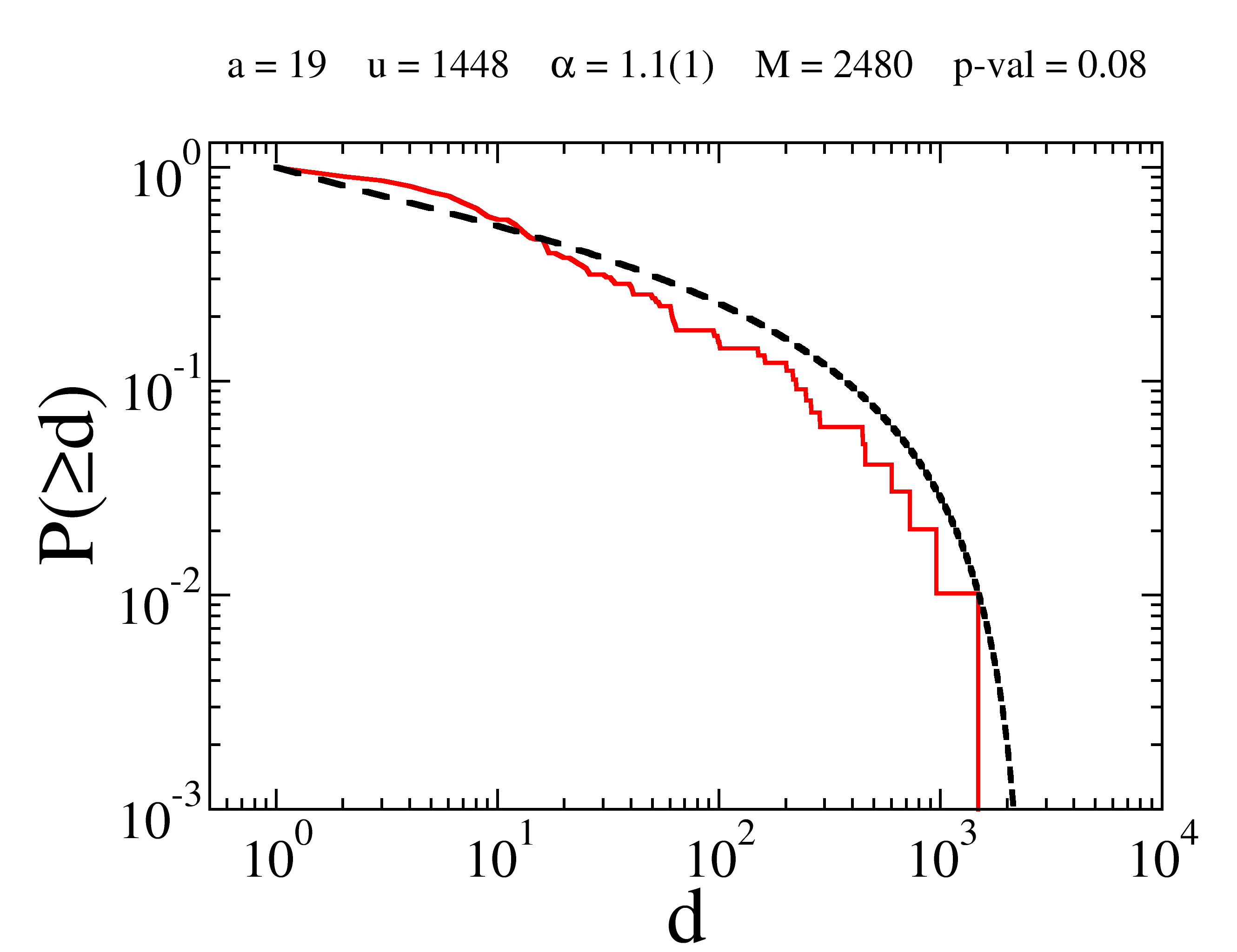}
\qquad
\includegraphics[width=7cm]{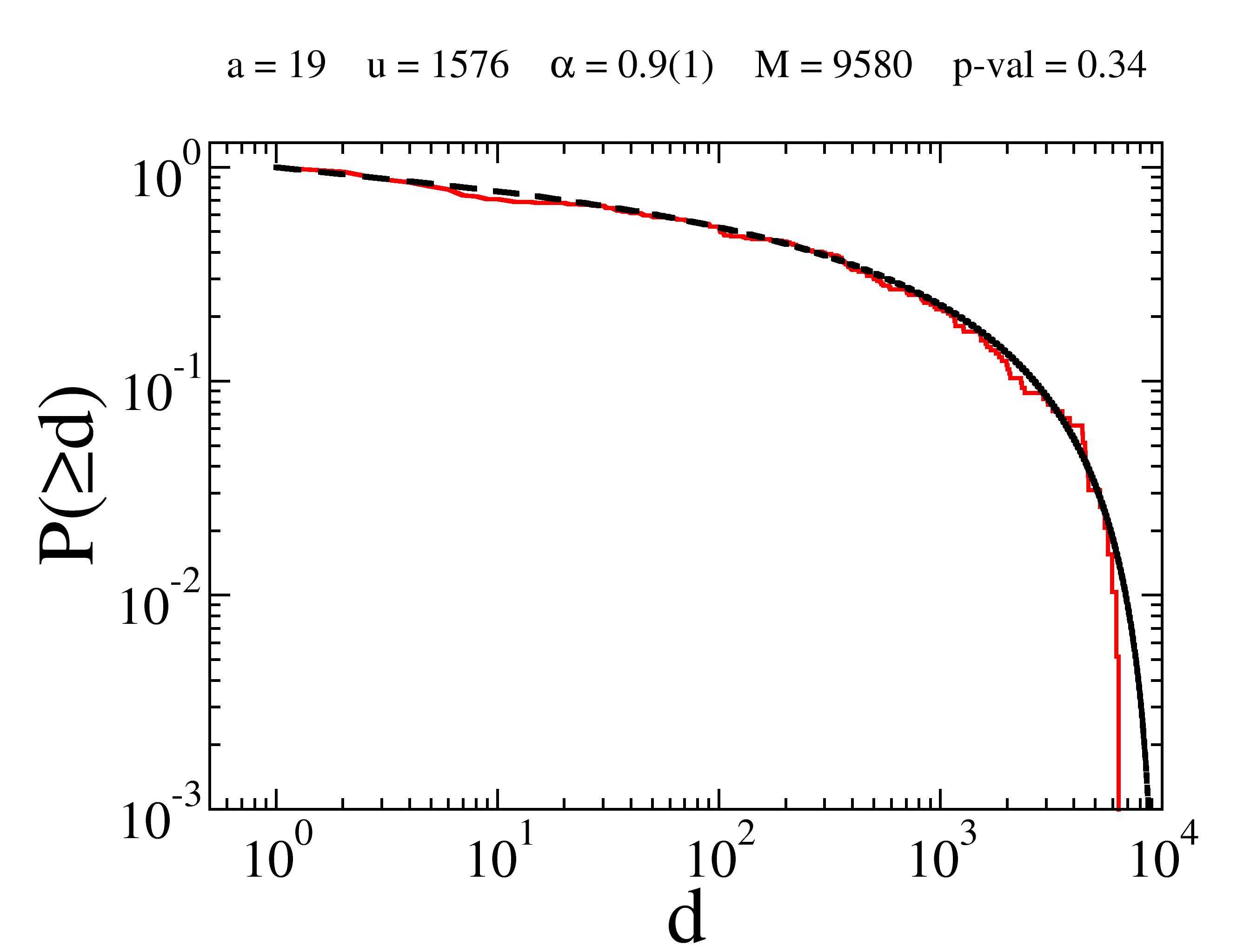}

\vskip .1cm
\caption{UBH data set. Cumulative distribution function $P\left(\geq d\right)$
measured for agent $u$ in auction $a$ (red full line)
compared with the theoretical distribution
(black dashed line). We show
several $P\left(\geq d\right)$s for different pairs $u$ and $a$.
We report also the best value of the upper bound $M$
and the $p$-value associated with our fit. }
\label{fig:exponents_single_like_1}
\end{center}
\end{figure}

\begin{figure}[!ht]
\begin{center}
\vskip .7cm
\includegraphics[width=7cm]{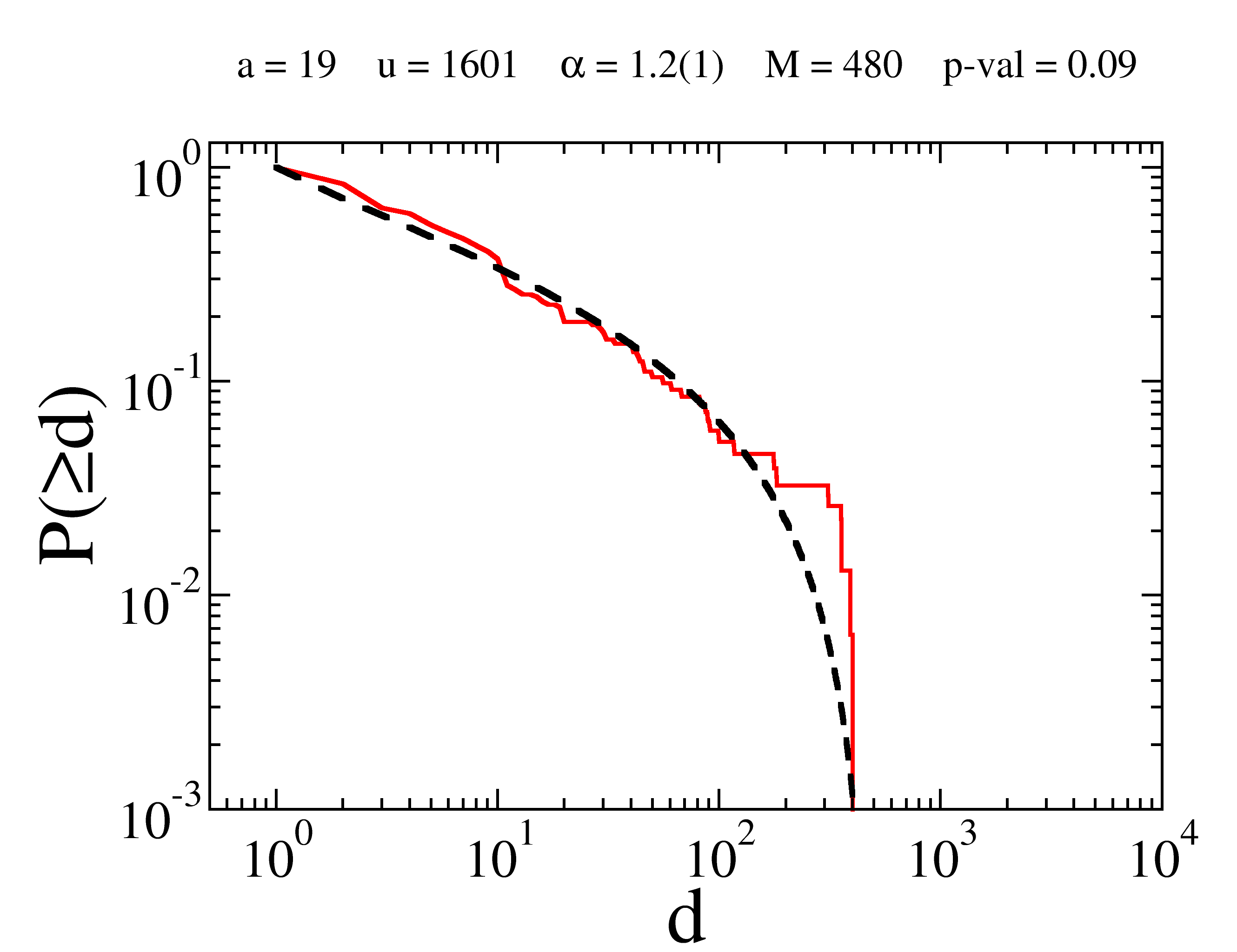}
\qquad
\includegraphics[width=7cm]{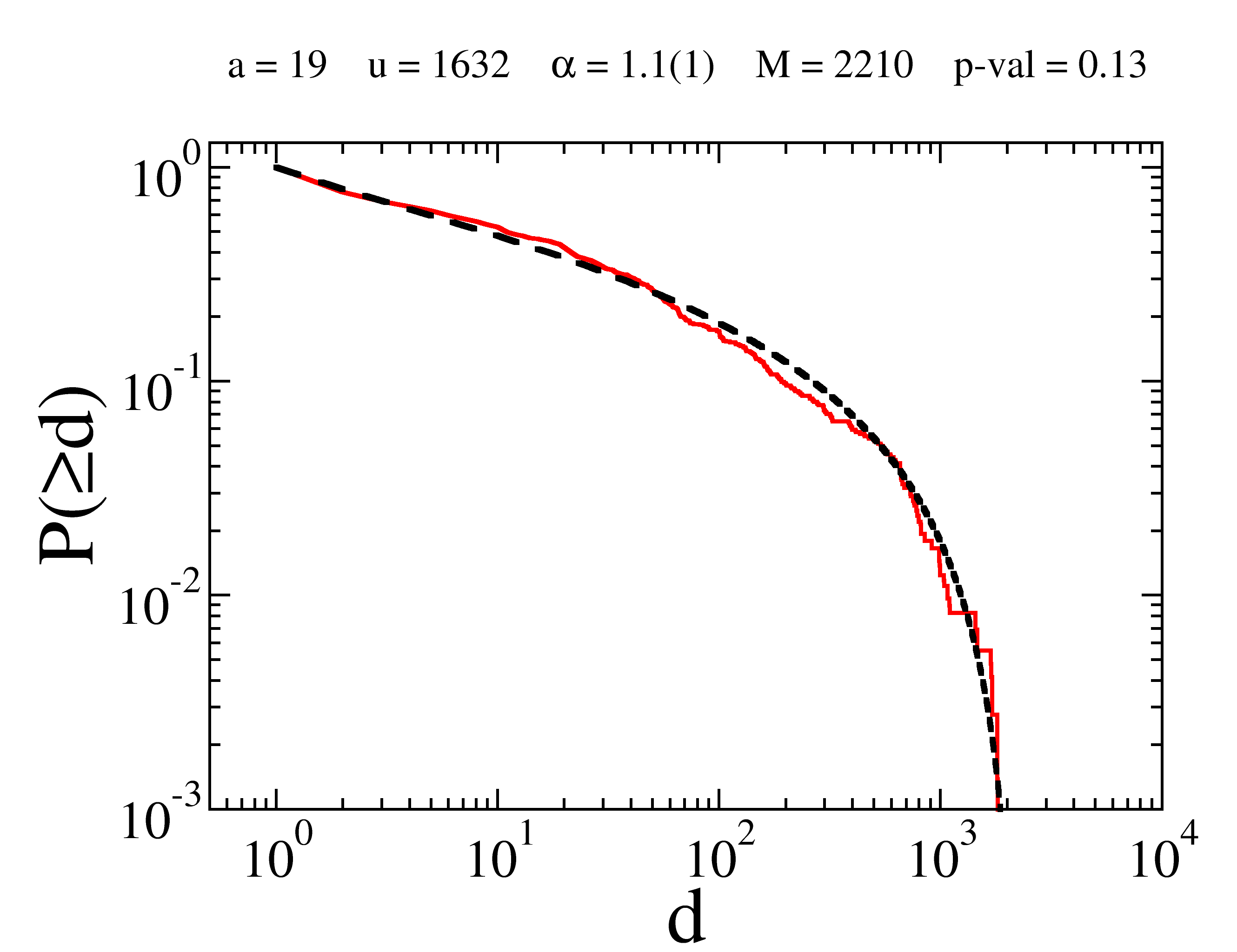}
\vskip .2cm
\includegraphics[width=7cm]{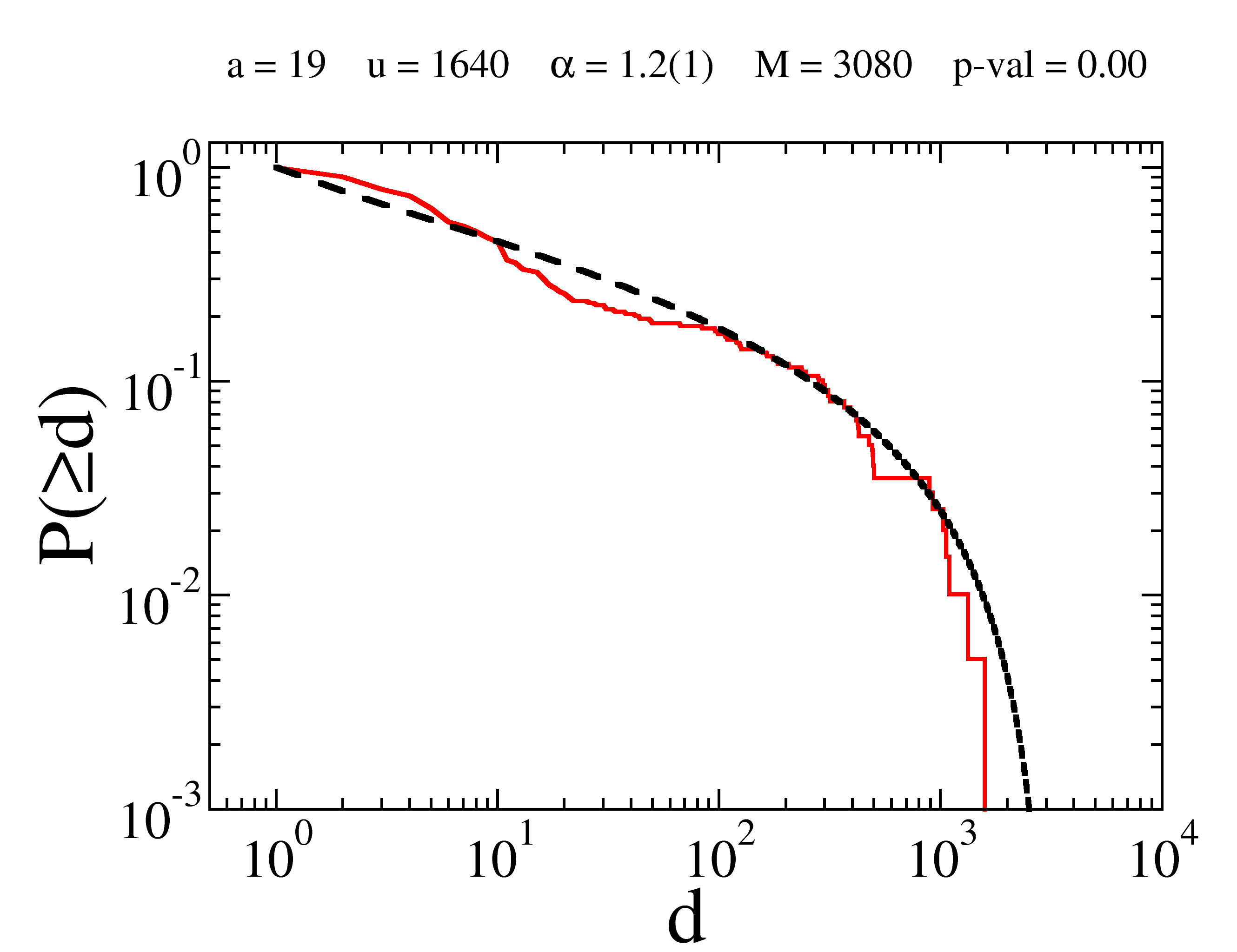}
\qquad
\includegraphics[width=7cm]{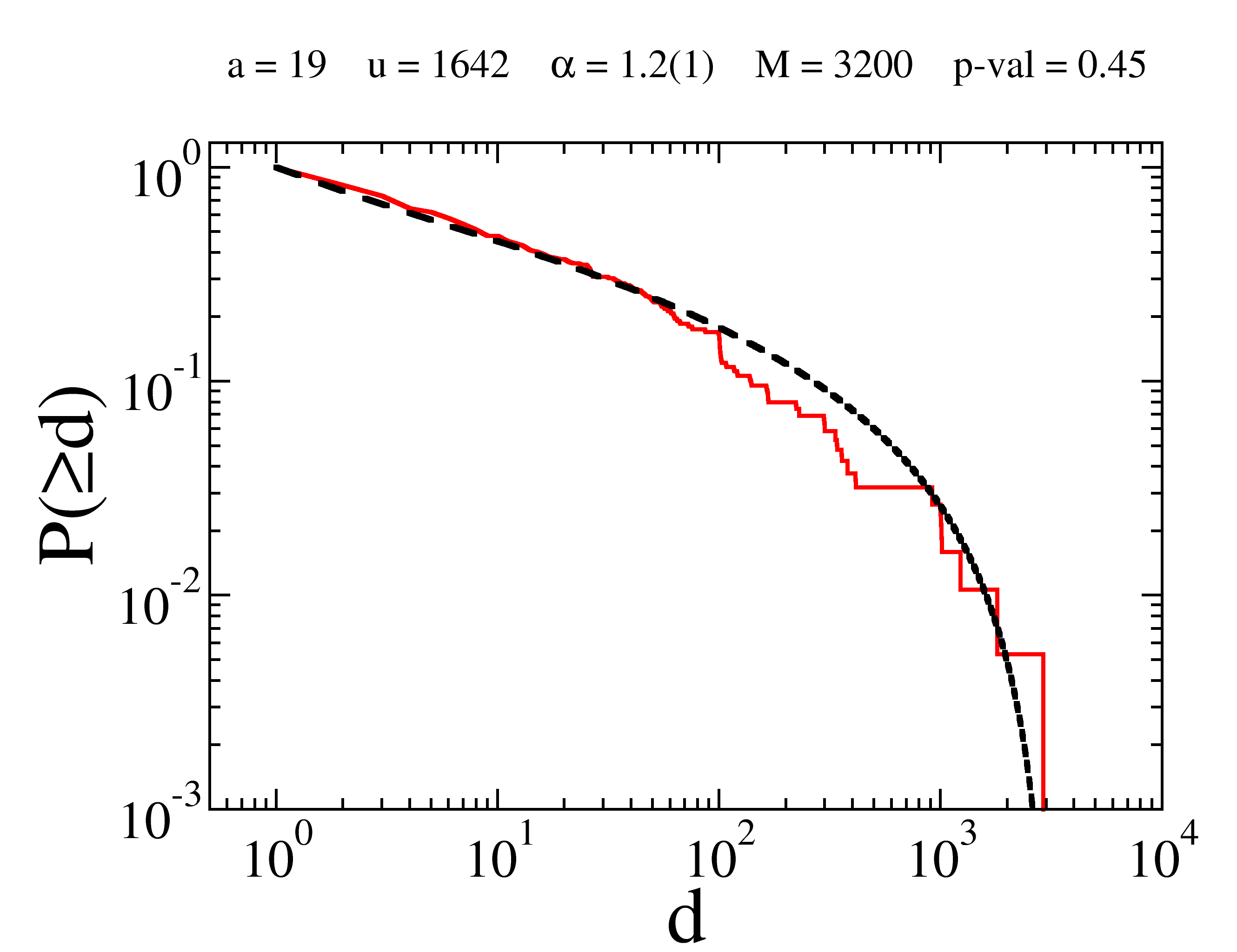}
\vskip .2cm
\includegraphics[width=7cm]{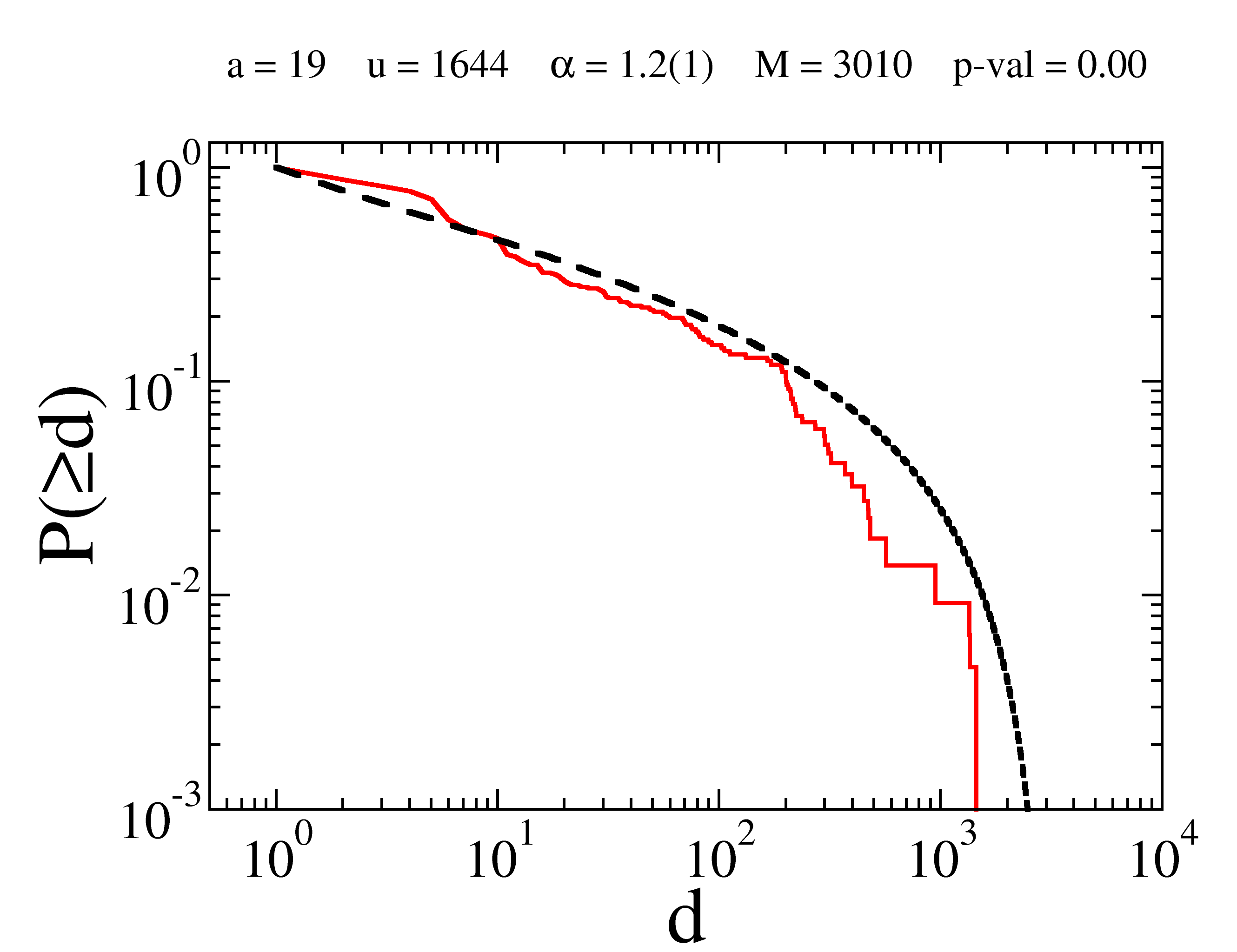}
\qquad
\includegraphics[width=7cm]{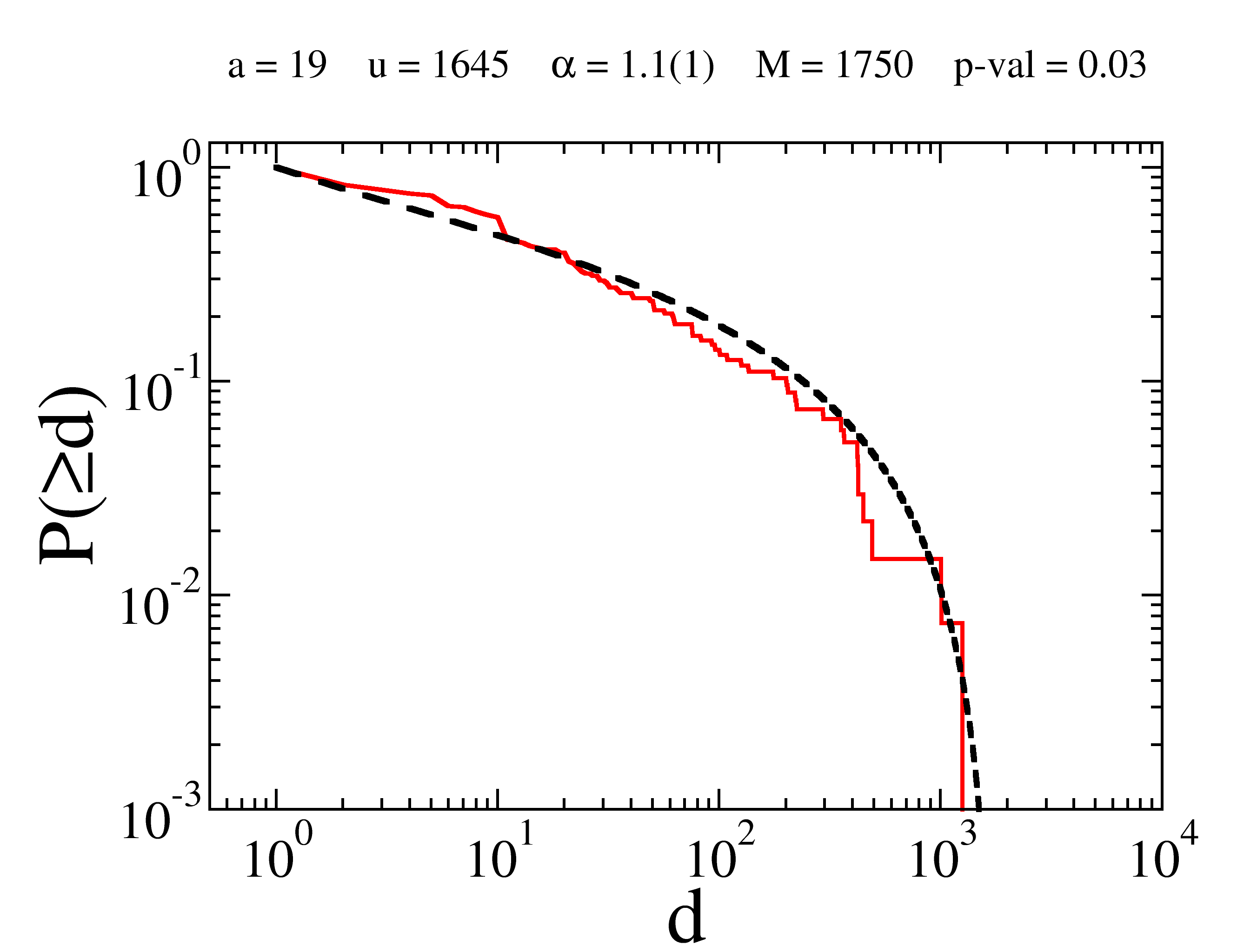}

\vskip .1cm
\caption{UBH data set. Same as Figure~\ref{fig:exponents_single_like_1}.}
\label{fig:exponents_single_like_2}
\end{center}
\end{figure}

\begin{figure}[!ht]
\begin{center}
\vskip .7cm
\includegraphics[width=7cm]{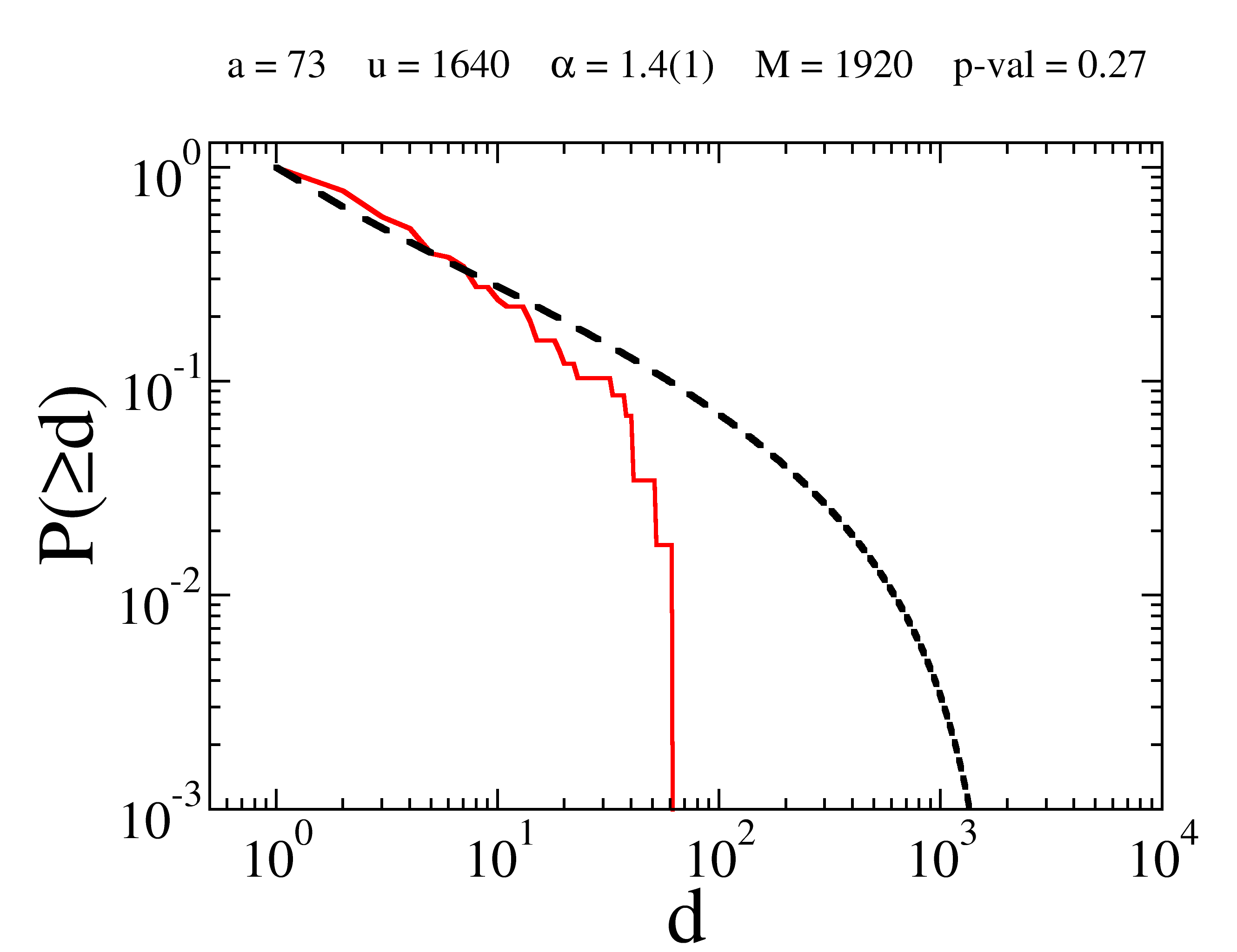}
\qquad
\includegraphics[width=7cm]{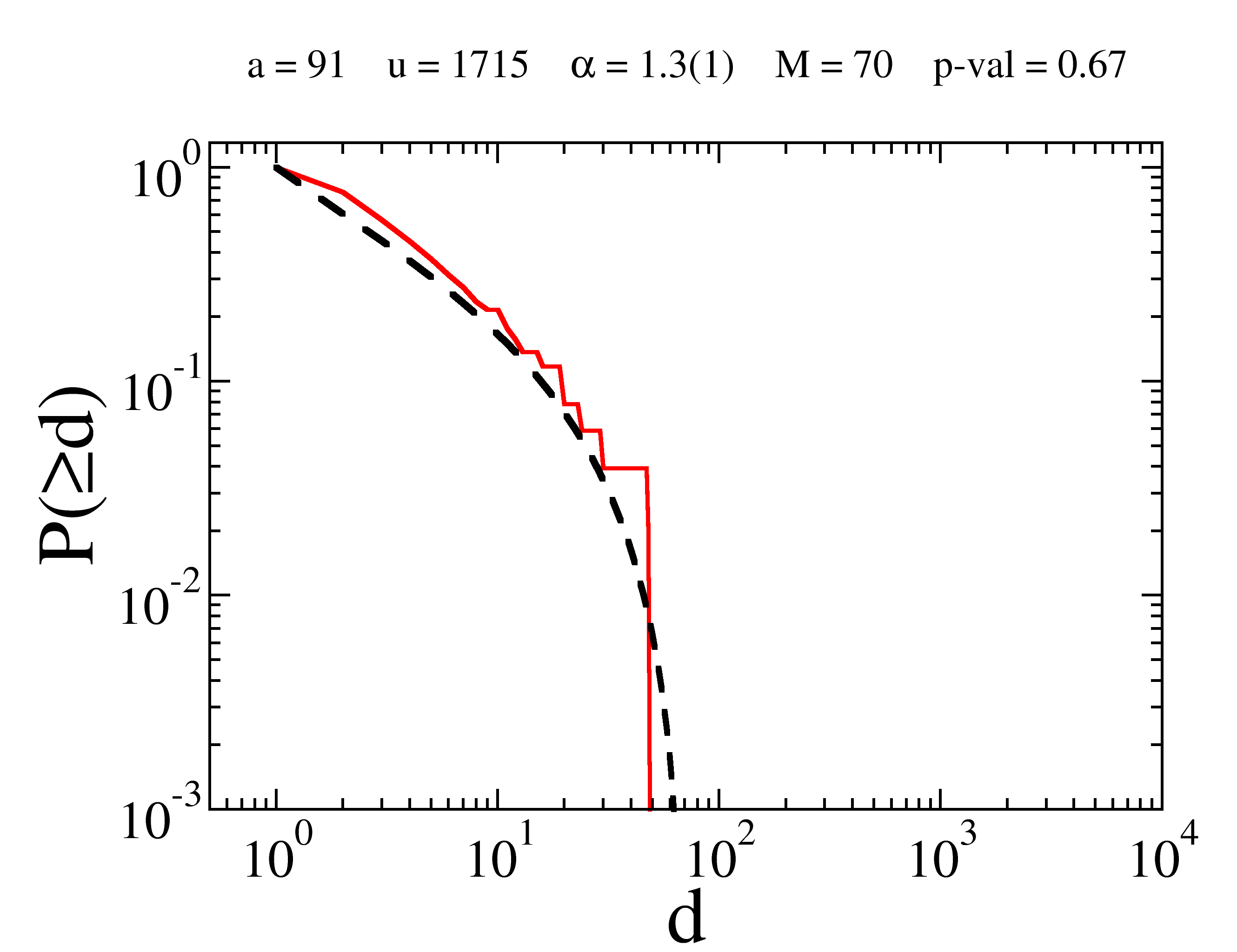}
\vskip .1cm
\caption{UBH data set. Same as Figure~\ref{fig:exponents_single_like_1} and 
~\ref{fig:exponents_single_like_2}.}
\label{fig:exponents_single_like_3}
\end{center}
\end{figure}

\begin{figure}[!ht]
\begin{center}
\vskip .7cm
\includegraphics[width=7cm]{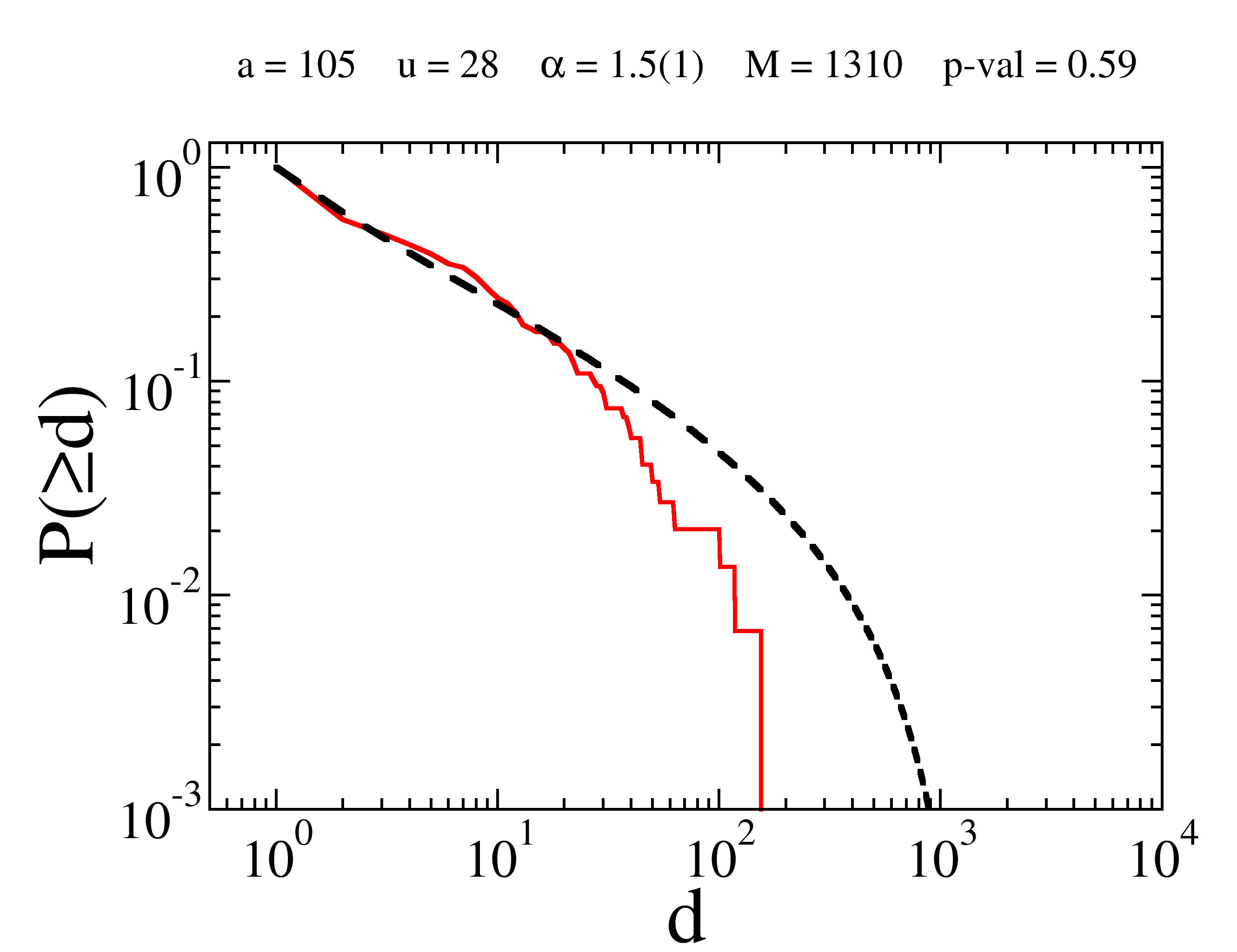}
\qquad
\includegraphics[width=7cm]{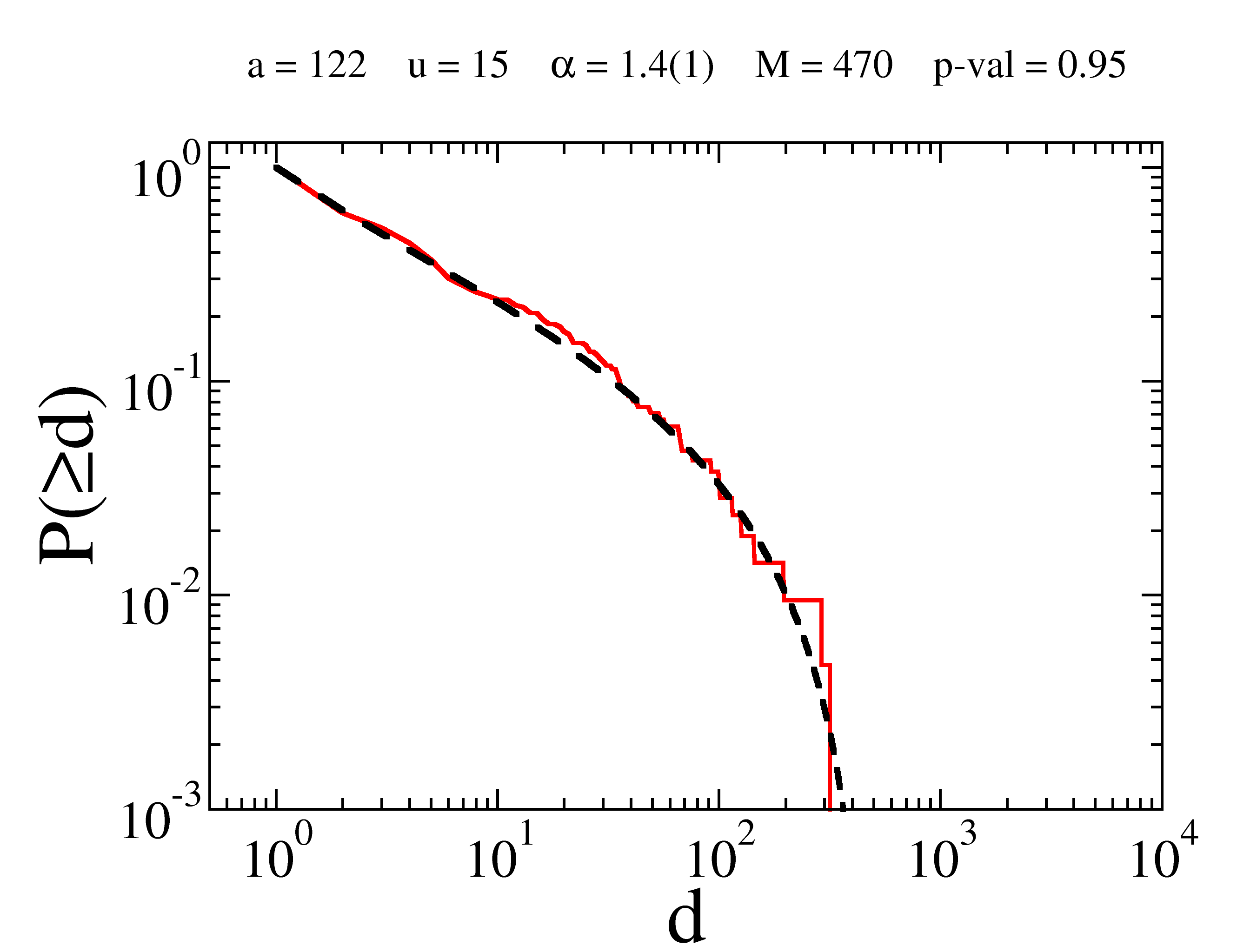}
\vskip .2cm
\includegraphics[width=7cm]{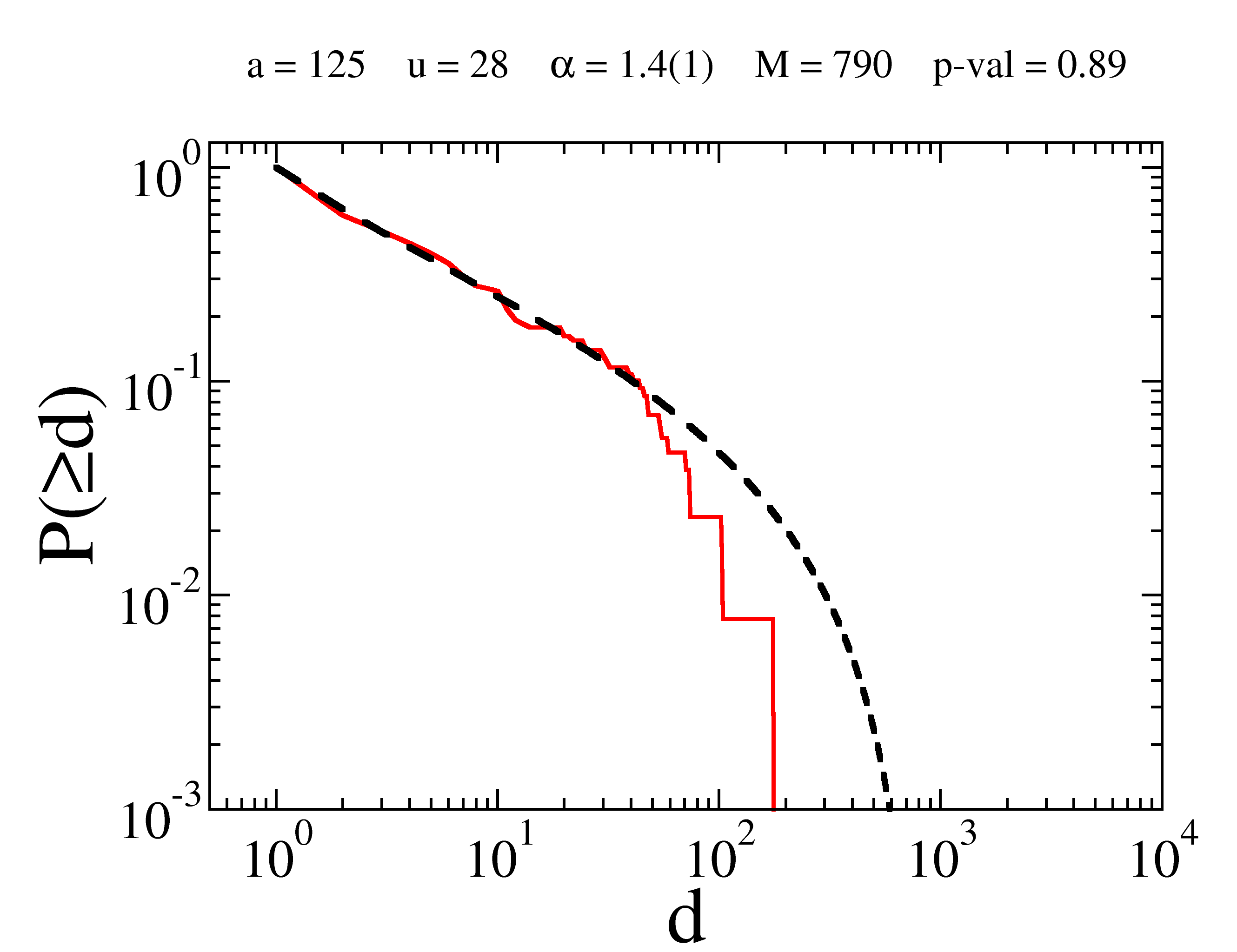}
\qquad
\includegraphics[width=7cm]{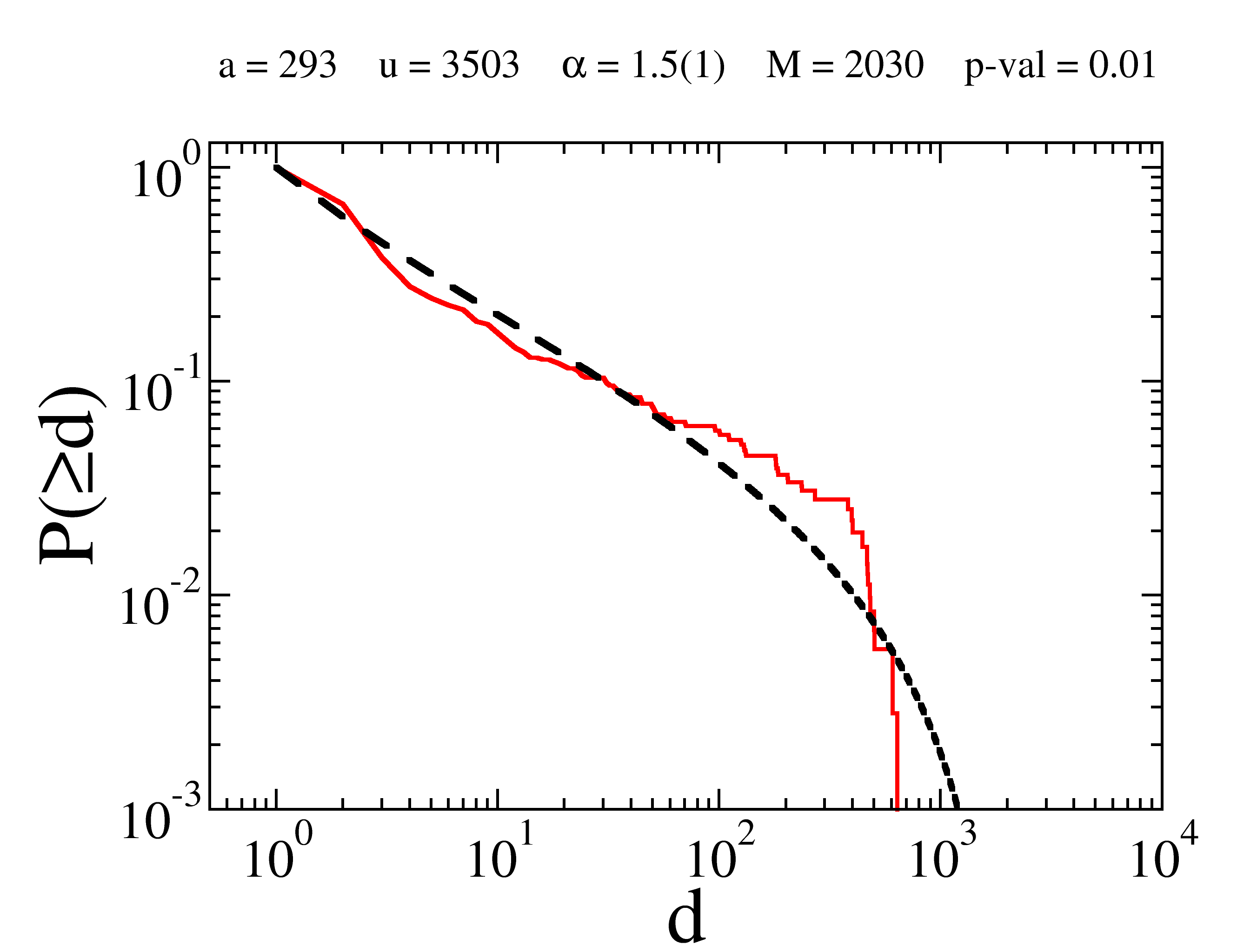}
\vskip .2cm
\includegraphics[width=7cm]{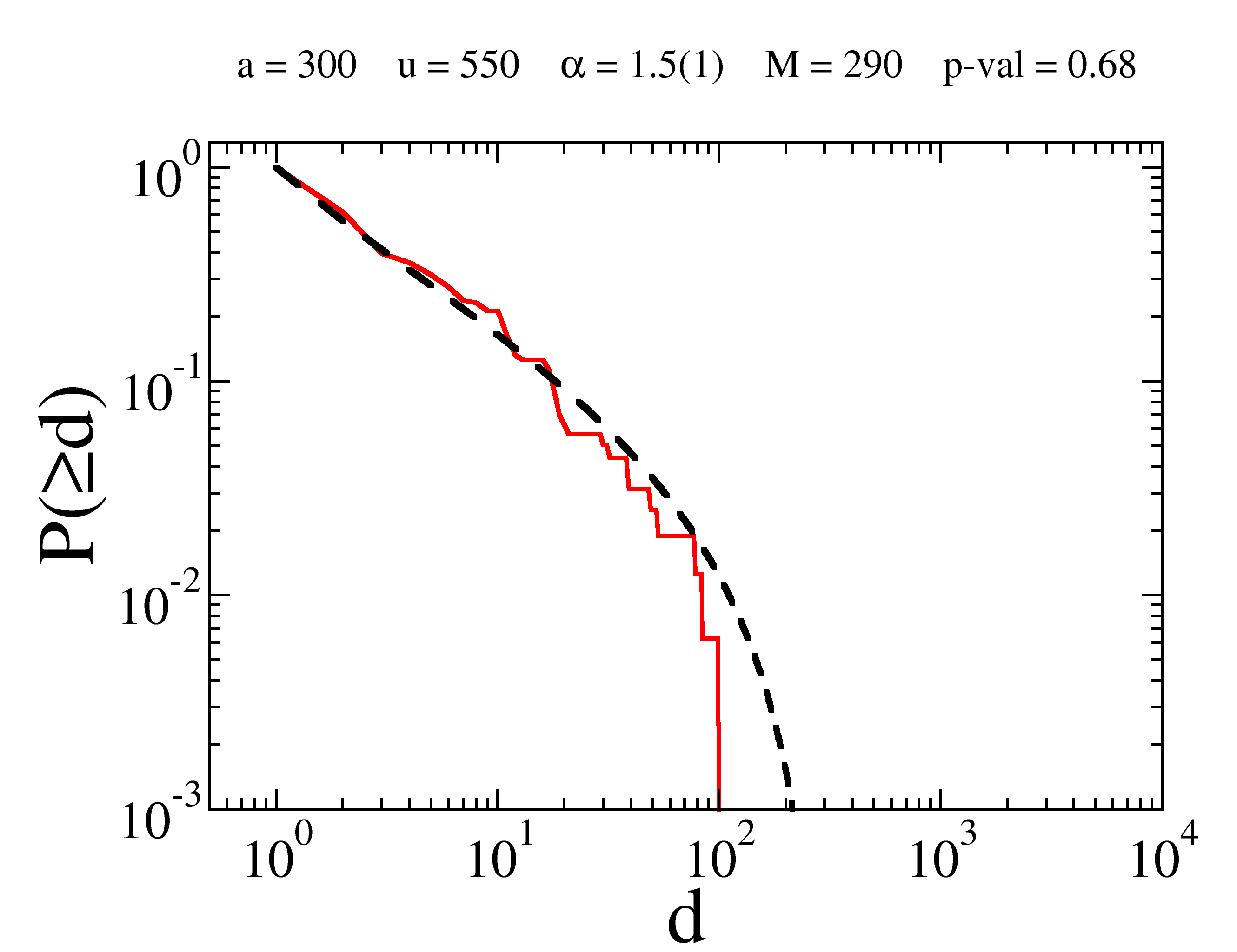}
\qquad
\includegraphics[width=7cm]{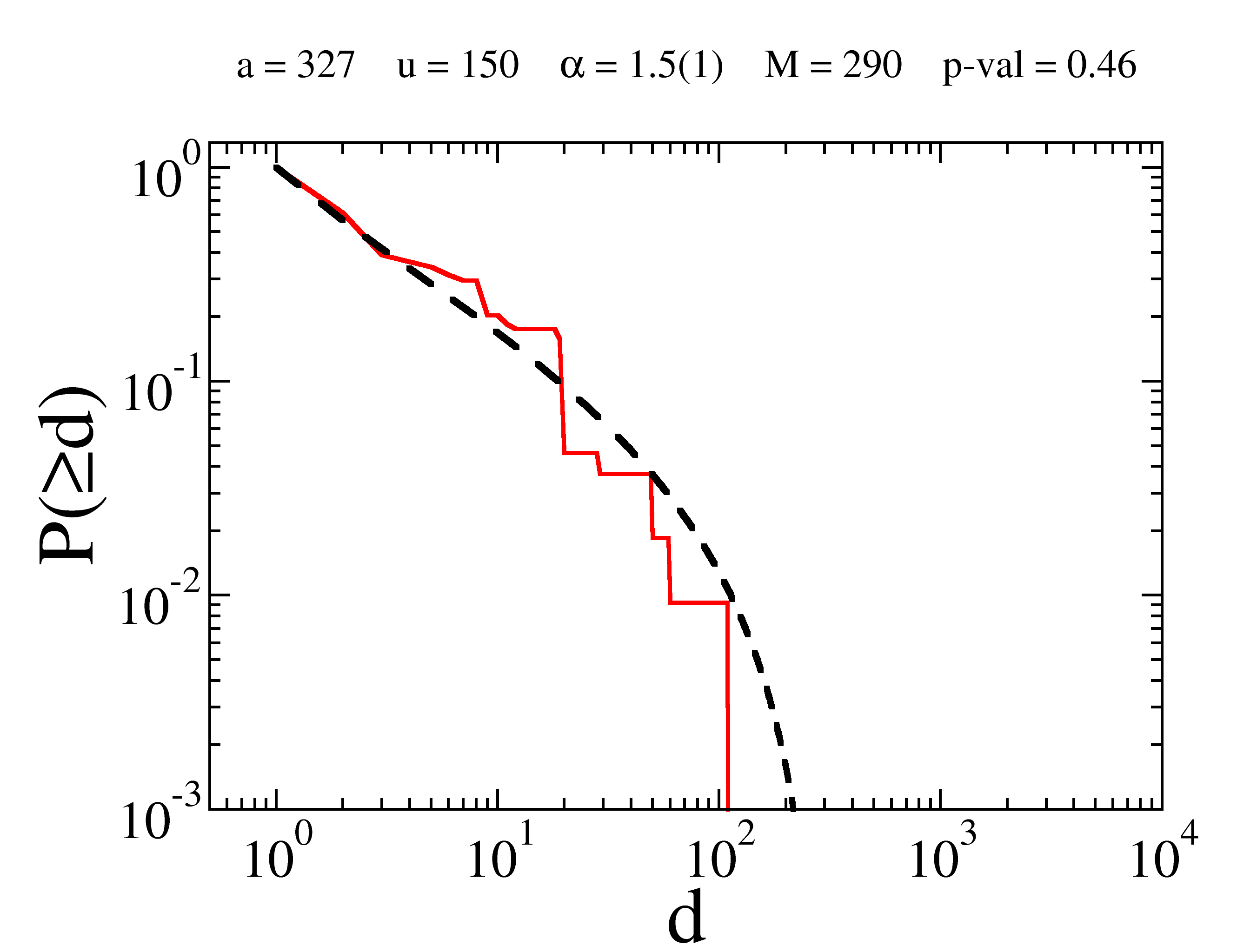}

\vskip .1cm
\caption{BM data set.
Cumulative distribution function $P\left(\geq d\right)$
measured for agent $u$ in auction $a$ (red full line)
compared with the theoretical distribution
(black dashed line). We show
several $P\left(\geq d\right)$s for different pairs $u$ and $a$.
We report also the best value of the upper bound $M$
and the $p$-value associated with our fit. }
\label{fig:bm_exponents_single_like_1}
\end{center}
\end{figure}

\begin{figure}[!ht]
\begin{center}
\vskip .7cm
\includegraphics[width=7cm]{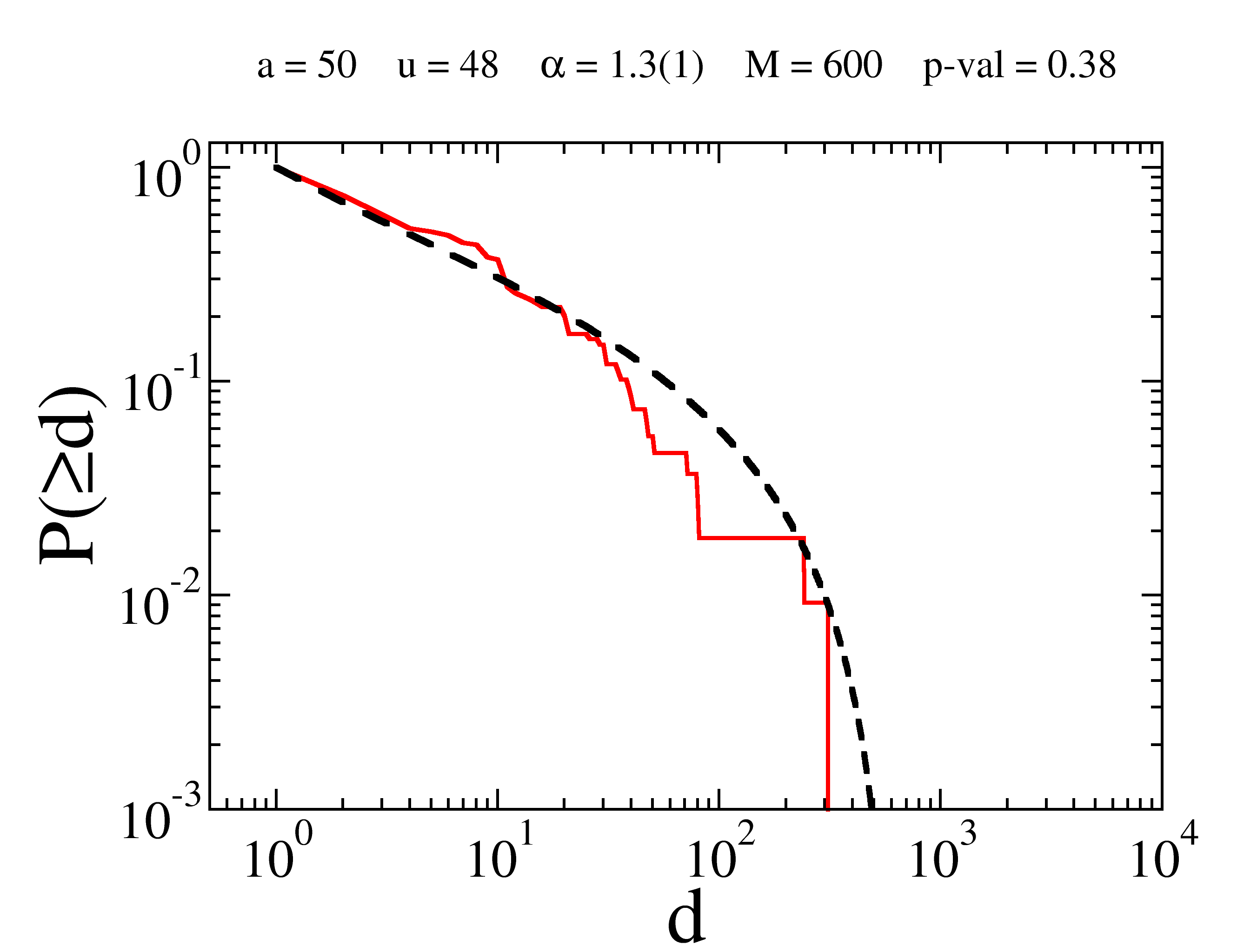}
\qquad
\includegraphics[width=7cm]{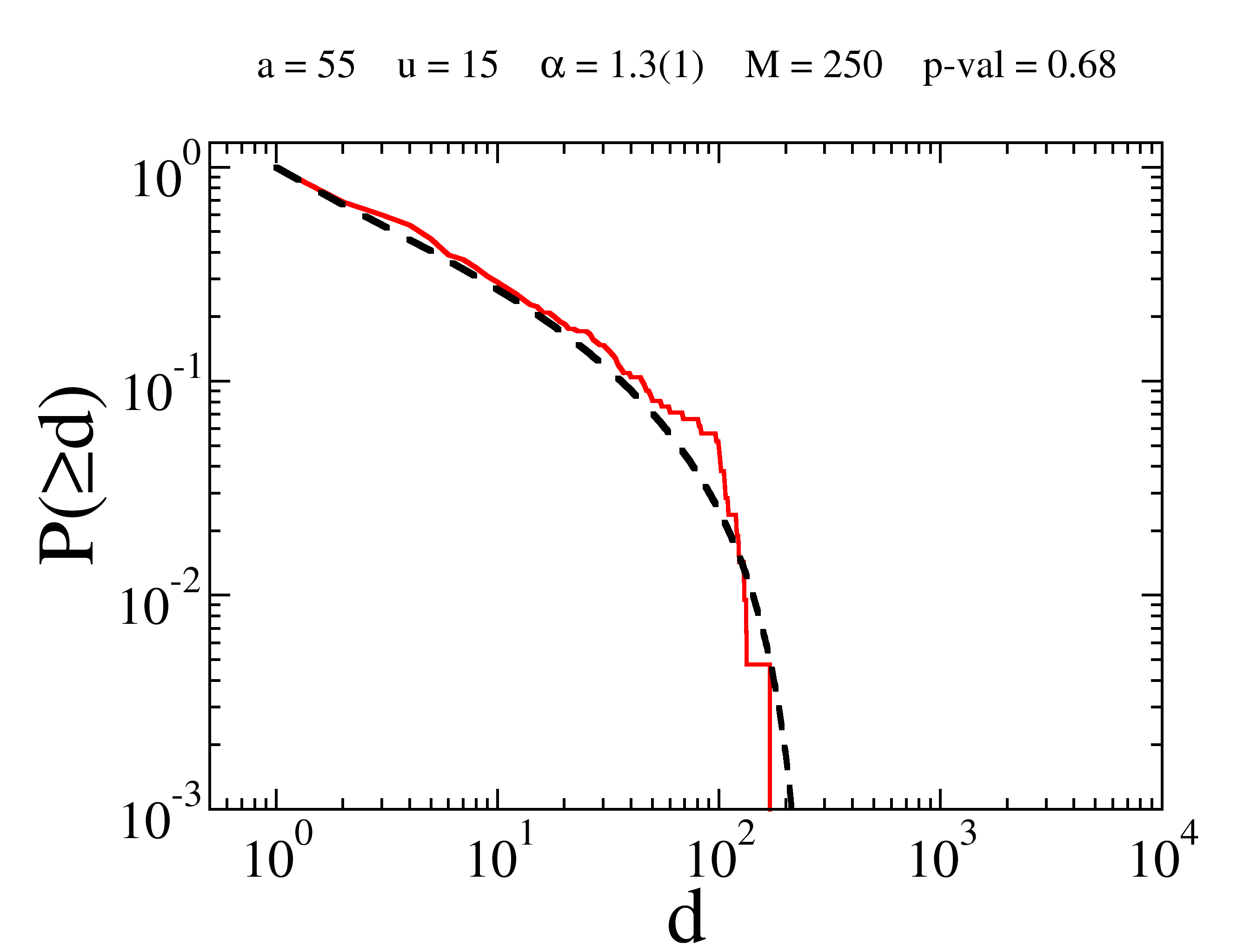}
\vskip .2cm
\includegraphics[width=7cm]{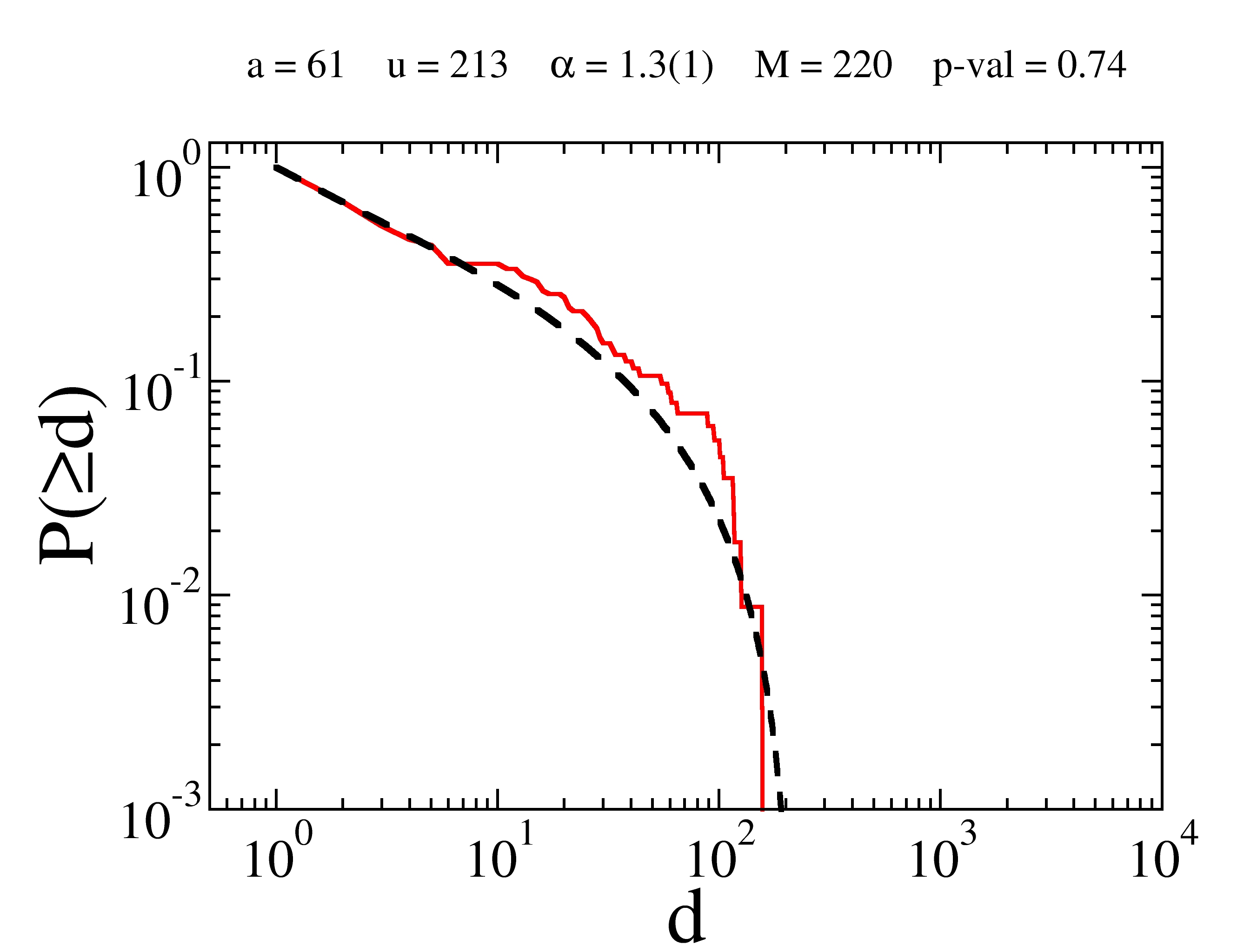}
\qquad
\includegraphics[width=7cm]{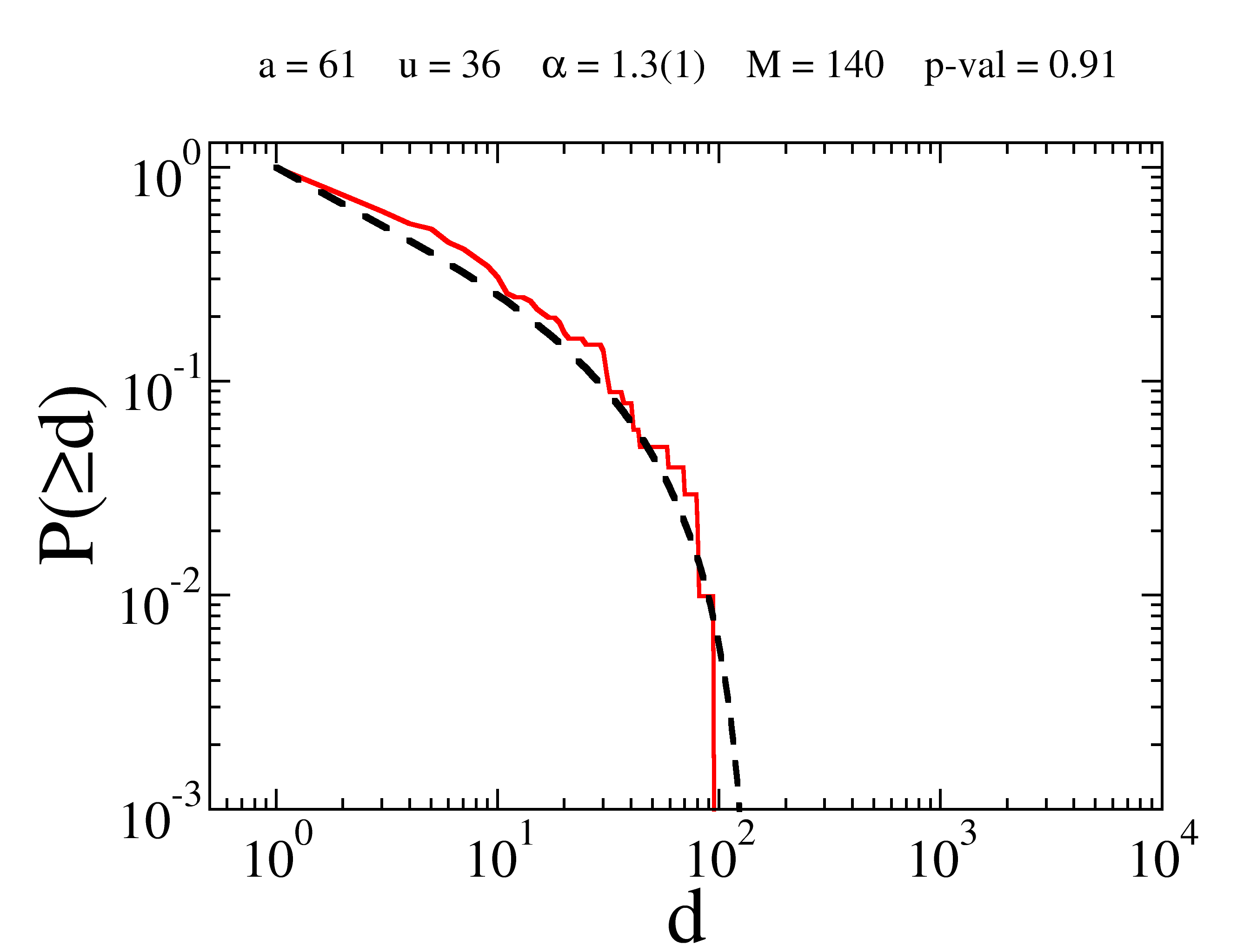}
\vskip .2cm
\includegraphics[width=7cm]{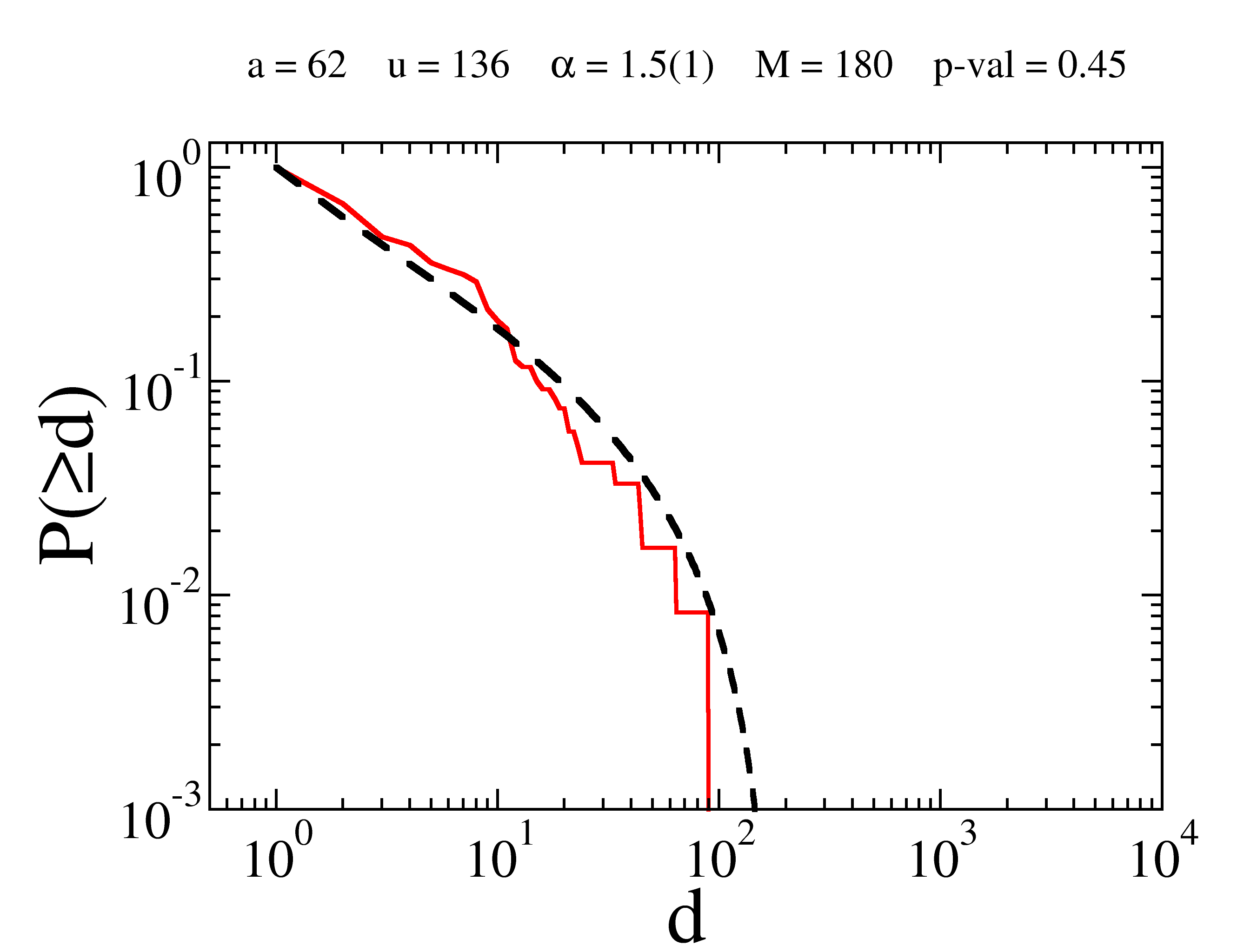}
\qquad
\includegraphics[width=7cm]{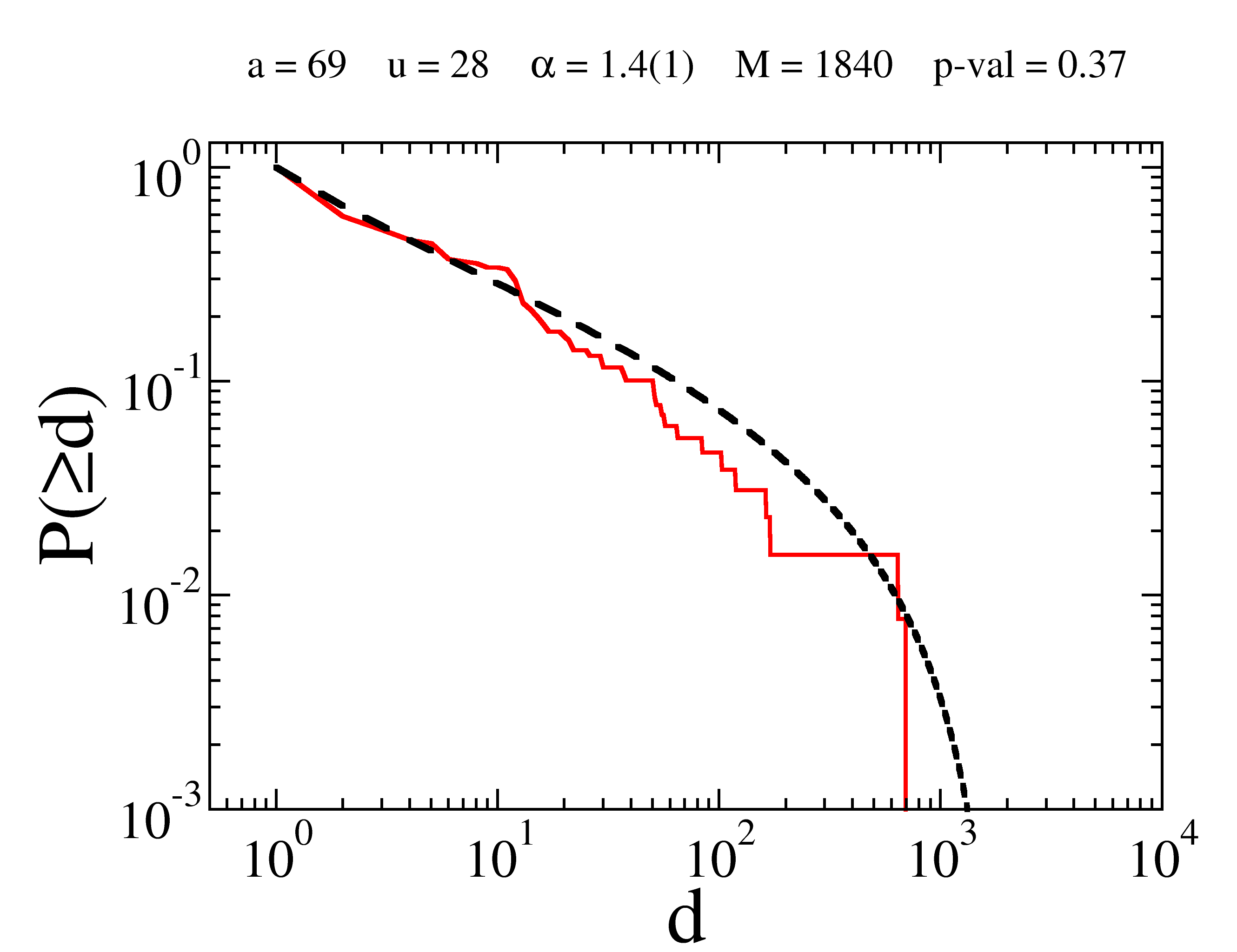}

\vskip .1cm
\caption{BM data set. Same as Figure~\ref{fig:bm_exponents_single_like_1}.}
\label{fig:bm_exponents_single_like_2}
\end{center}
\end{figure}

\begin{figure}[!ht]
\begin{center}
\vskip .7cm
\includegraphics[width=7cm]{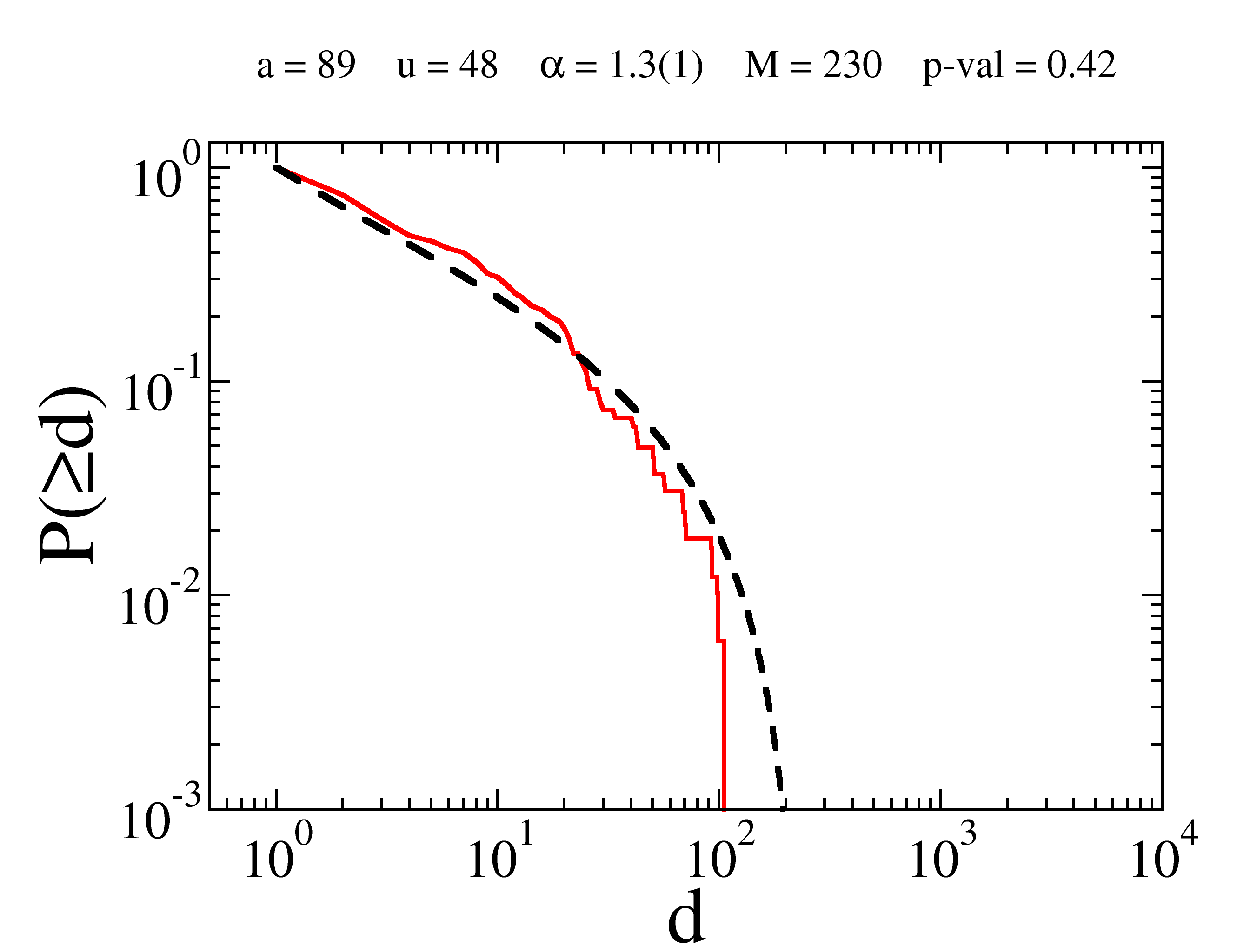}
\qquad
\includegraphics[width=7cm]{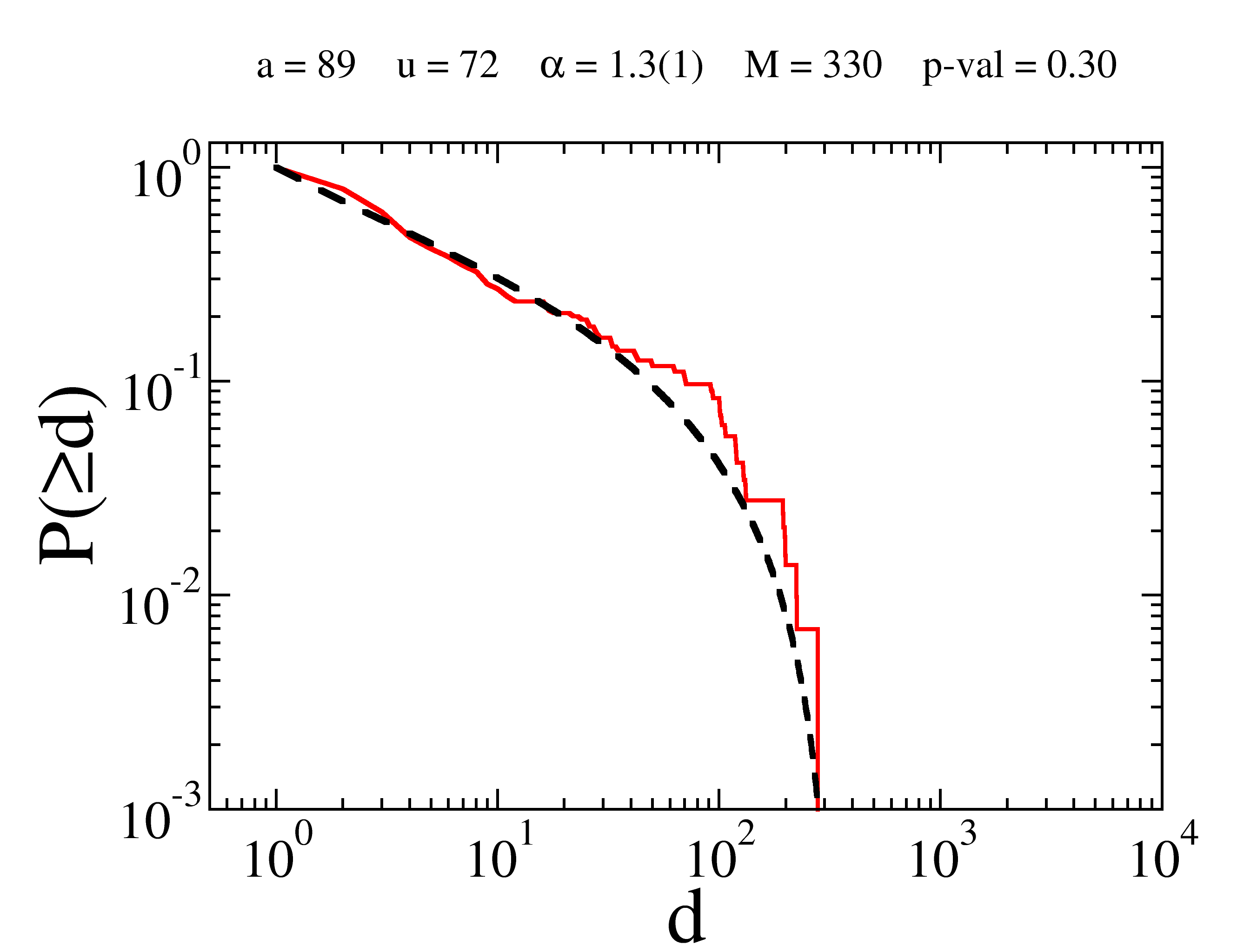}
\vskip .2cm
\includegraphics[width=7cm]{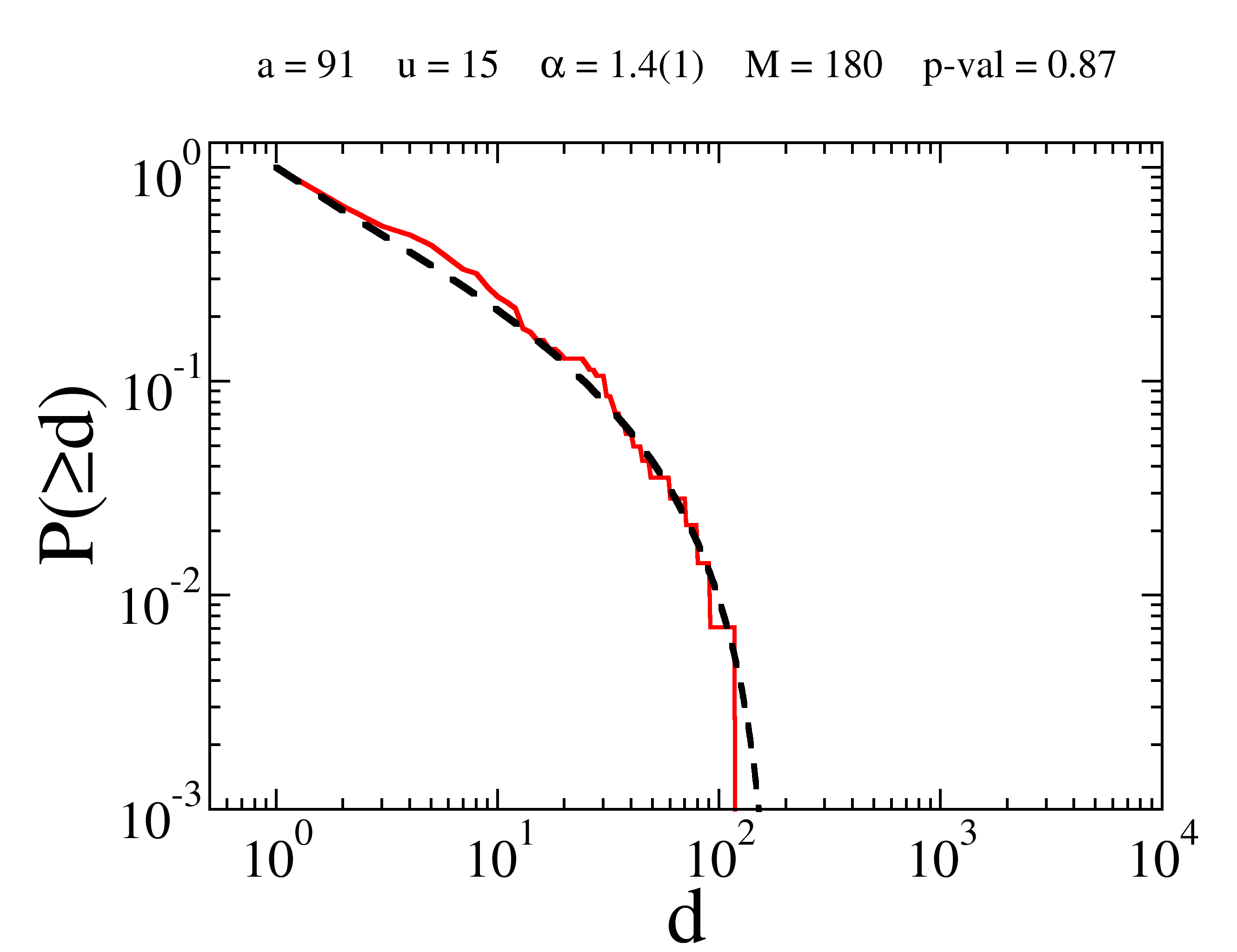}
\qquad
\includegraphics[width=7cm]{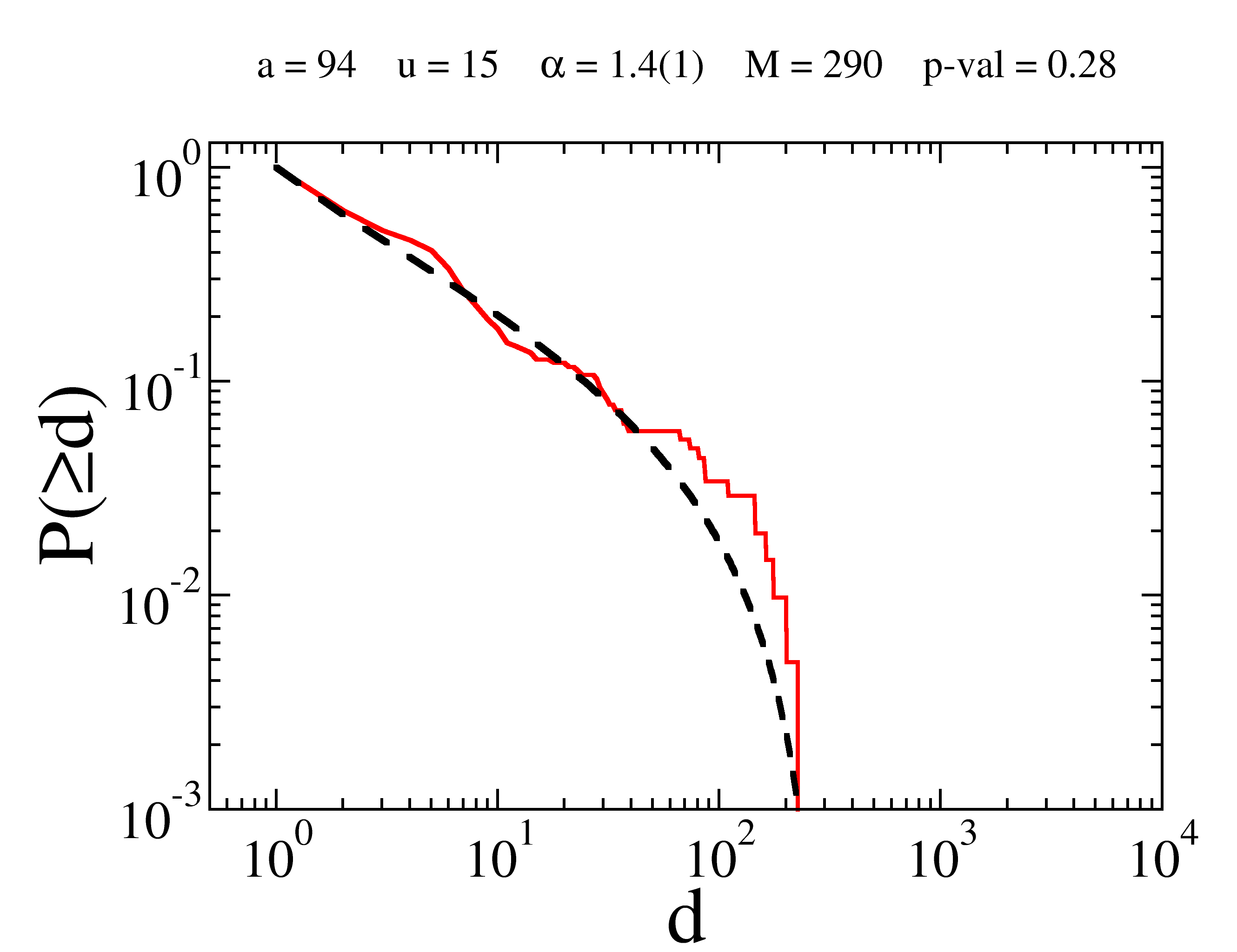}
\vskip .1cm
\caption{BM data set. Same as Figure~\ref{fig:bm_exponents_single_like_1} and 
~\ref{fig:bm_exponents_single_like_2}.}
\label{fig:bm_exponents_single_like_3}
\end{center}
\end{figure}

\noindent Since the upper-bound $M$ is agent dependent,
we perform an additional analysis where the upper bound $M$
is not directly taken from the data, but used as a parameter
for the fit. 
For each agent, we let the parameter $M$ vary
only in the range for which at 
least the $90\%$ of the bids values are below $M$.
Indicate with $\tilde{T}$ the number of bids below
the threshold $M$. We then find $\alpha'$ identifying the maximum 
of the likelihood function and calculate the $p$-value
as described so far. We consider the best value of
$M$ as the one which maximizes the product $\tilde{T} \times p$.  
Figures~\ref{fig:exponents_single_like_1},~\ref{fig:exponents_single_like_2} and~\ref{fig:exponents_single_like_3} report the
best fit for the same agents analyzed in Figures~\ref{fig:exponents_single_users_1},~\ref{fig:exponents_single_users_2} and~\ref{fig:exponents_single_users_3}
for UBH data set,
while Figures~\ref{fig:bm_exponents_single_like_1},~\ref{fig:bm_exponents_single_like_2} and~\ref{fig:bm_exponents_single_like_3} are
the analogous of Figures~\ref{fig:bm_exponents_single_users_1},~\ref{fig:bm_exponents_single_users_2} and~\ref{fig:bm_exponents_single_users_3}
for BM data set. For completeness, we report
in Tables~\ref{tab:ubh_like} and~\ref{tab:bm_like} the results
obtained with the maximum likelihood fit where $M$ is used as
parameter of the fit. The best exponents $\alpha'$ are still consistent
with those reported in Tables~\ref{tab:ubh} and~\ref{tab:bm}, but
the $p$-values result much increased.

\begin{table}[!th]
\begin{center}

\begin{minipage}[!t]{0.4\textwidth}
\centering
\begin{tabular}{|c c c c c c|}
 \hline
$a$ & $u$ & $\alpha$ & $\alpha'$ &  $M$ &   $p$
\\ \hline
\hline
$1^*$ & $23$ & $1.6(1)$ & $1.4(1)$ & $490$ & $0.00$ \\
\hline
$1$ & $81$ & $1.2(1)$ & $1.2(1)$ & $2220$ & $0.26$ \\
\hline
$100$ & $1715$ & $1.6(2)$ & $1.0(1)$ & $150$ & $0.15$ \\
\hline
$100$ & $81$ & $1.5(2)$ & $1.3(1)$ & $480$ & $0.01$ \\
\hline
$104$ & $3093$ & $1.7(3)$ & $1.0(1)$ & $70$ & $0.51$ \\
\hline
$108$ & $134$ & $1.7(1)$ & $1.3(1)$ & $70$ & $0.09$ \\
\hline
$14$ & $134$ & $1.2(1)$ & $1.1(1)$ & $530$ & $0.40$ \\
\hline
$14$ & $81$ & $1.9(2)$ & $1.3(1)$ & $870$ & $0.01$ \\
\hline
$15$ & $423$ & $1.6(4)$ & $1.2(1)$ & $120$ & $0.53$ \\
\hline
$179$ & $3663$ & $1.8(1)$ & $1.3(2)$ & $70$ & $0.35$ \\
\hline
$19$ & $1$ & $1.3(1)$ & $1.3(1)$ & $3050$ & $0.29$ \\
\hline
$19$ & $1313$ & $1.2(1)$ & $1.1(1)$ & $1650$ & $0.50$ \\
\hline
$19$ & $134$ & $1.2(1)$ & $1.0(1)$ & $2320$ & $0.19$ \\
\hline
$19$ & $1433$ & $1.1(1)$ & $0.9(1)$ & $1840$ & $0.99$ \\
\hline
$19$ & $1448$ & $1.4(1)$ & $1.1(1)$ & $2480$ & $0.08$ \\
\hline
$19$ & $1558$ & $1.1(1)$ & $1.2(1)$ & $2290$ & $0.93$ \\
\hline
$19$ & $1576$ & $1.0(1)$ & $0.9(1)$ & $9580$ & $0.34$ \\
\hline
$19$ & $1601$ & $1.4(1)$ & $1.2(1)$ & $480$ & $0.09$ \\
\hline
$19$ & $1632$ & $1.4(1)$ & $1.1(1)$ & $2210$ & $0.13$ \\
\hline
$19^*$ & $1640$ & $1.3(1)$ & $1.2(1)$ & $3080$ & $0.00$ \\
\hline 
\end{tabular}
\end{minipage} 
\hspace{0.1cm}
\begin{minipage}[!t]{0.4\textwidth}
\centering
\begin{tabular}{|c c c c c c|}
\hline
$a$ & $u$ & $\alpha$ & $\alpha'$ & $M$ &  $p$
\\ \hline
\hline
$19$ & $1642$ & $1.3(1)$ & $1.2(1)$ & $3200$ & $0.45$ \\
\hline
$19^*$ & $1644$ & $1.4(1)$ & $1.2(1)$ & $3010$ & $0.00$ \\
\hline
$19$ & $1645$ & $1.4(1)$ & $1.1(1)$ & $1750$ & $0.03$ \\
\hline
$19$ & $3$ & $1.2(1)$ & $1.3(1)$ & $3310$ & $0.91$ \\
\hline
$19$ & $363$ & $1.2(1)$ & $0.9(1)$ & $890$ & $0.20$ \\
\hline
$19$ & $434$ & $1.1(1)$ & $1.1(1)$ & $1750$ & $0.77$ \\
\hline
$19$ & $438$ & $1.5(1)$ & $0.9(1)$ & $2120$ & $0.43$ \\
\hline
$20$ & $617$ & $1.6(2)$ & $1.3(1)$ & $200$ & $0.09$ \\
\hline
$22$ & $134$ & $1.2(1)$ & $1.3(1)$ & $1000$ & $0.96$ \\
\hline
$44$ & $433$ & $1.1(1)$ & $1.6(1)$ & $3340$ & $0.08$ \\
\hline
$46$ & $2003$ & $1.3(3)$ & $1.4(1)$ & $110$ & $0.03$ \\
\hline
$5$ & $128$ & $1.6(1)$ & $1.4(1)$ & $1830$ & $0.01$ \\
\hline
$62$ & $2392$ & $1.3(3)$ & $1.3(1)$ & $130$ & $0.52$ \\
\hline
$71^*$ & $324$ & $1.6(2)$ & $1.4(1)$ & $860$ & $0.00$ \\
\hline
$73$ & $1640$ & $1.5(2)$ & $1.4(1)$ & $1920$ & $0.27$ \\
\hline
$73$ & $1715$ & $1.5(1)$ & $1.2(1)$ & $140$ & $0.68$ \\
\hline
$79$ & $134$ & $1.6(2)$ & $1.3(1)$ & $120$ & $0.21$ \\
\hline
$91$ & $1715$ & $1.5(2)$ & $1.3(1)$ & $70$ & $0.67$ \\
\hline
$97$ & $1715$ & $1.6(2)$ & $1.2(1)$ & $80$ & $0.19$ \\
\hline

\end{tabular}
\end{minipage} 

\end{center}
\caption{UBH data set. Each row corresponds to
one of the $39$ agents who have bid at least $50$ times in the same auction.
We report the id of the auction $a$, the id of the agent $u$, the
exponent $\alpha$ calculated with the least square method,
the exponent $\alpha'$  calculated with the maximum likelihood method,
the best value of the upper bound $M$ and
the $p$-value. Entries with low $p$-values are marked with $^*$.
In the $90\%$
of the cases we find a $p$-value larger than $0$, which
indicates that our model well describe
the time series.}
\label{tab:ubh_like}
\end{table}

\begin{table}[!th]
\begin{center}

\begin{minipage}[!t]{0.4\textwidth}
\centering
\begin{tabular}{|c c c c c c|}
 \hline
$a$ & $u$ & $\alpha$ & $\alpha'$ &  $M$ &  $p$
\\ \hline
\hline
$11$ & $28$ & $1.1(2)$ & $1.4(1)$ & $1290$ & $0.23$ \\
\hline
$13^*$ & $48$ & $1.5(1)$ & $1.2(1)$ & $570$ & $0.00$ \\
\hline
$13$ & $28$ & $1.3(1)$ & $1.3(1)$ & $310$ & $0.15$ \\
\hline
$15$ & $11$ & $1.8(3)$ & $1.3(1)$ & $180$ & $0.23$ \\
\hline
$15$ & $36$ & $2.0(1)$ & $1.5(1)$ & $200$ & $0.19$ \\
\hline
$28$ & $28$ & $1.4(1)$ & $1.4(1)$ & $530$ & $0.53$ \\
\hline
$32^*$ & $150$ & $1.2(6)$ & $1.0(1)$ & $160$ & $0.00$ \\
\hline
$39$ & $48$ & $1.3(1)$ & $1.3(1)$ & $170$ & $0.02$ \\
\hline
$47$ & $28$ & $1.1(1)$ & $1.2(1)$ & $320$ & $0.59$ \\
\hline
$50$ & $48$ & $1.5(1)$ & $1.3(1)$ & $600$ & $0.38$ \\
\hline
$52$ & $28$ & $1.2(2)$ & $1.4(1)$ & $2790$ & $0.11$ \\
\hline
$55$ & $15$ & $1.3(1)$ & $1.3(1)$ & $250$ & $0.68$ \\
\hline
$55^*$ & $150$ & $1.2(3)$ & $1.1(1)$ & $220$ & $0.00$ \\
\hline
$61$ & $213$ & $1.3(1)$ & $1.3(1)$ & $220$ & $0.74$ \\
\hline
$61$ & $36$ & $1.3(1)$ & $1.2(1)$ & $140$ & $0.91$ \\
\hline
$62$ & $136$ & $1.6(1)$ & $1.5(1)$ & $180$ & $0.45$ \\
\hline
$67$ & $98$ & $1.0(1)$ & $1.0(1)$ & $210$ & $0.83$ \\
\hline
$69$ & $28$ & $1.5(1)$ & $1.4(1)$ & $1840$ & $0.37$ \\
\hline
$82$ & $36$ & $1.2(2)$ & $1.2(1)$ & $360$ & $0.01$ \\
\hline
$89$ & $72$ & $1.4(1)$ & $1.3(1)$ & $330$ & $0.30$ \\
\hline
$89$ & $48$ & $1.5(1)$ & $1.3(1)$ & $230$ & $0.42$ \\
\hline
$89$ & $28$ & $1.2(1)$ & $1.3(1)$ & $1580$ & $0.17$ \\
\hline
$91$ & $15$ & $1.4(1)$ & $1.4(1)$ & $180$ & $0.87$ \\
\hline
$92$ & $36$ & $1.8(2)$ & $1.3(1)$ & $130$ & $0.04$ \\
\hline
$92$ & $28$ & $1.3(1)$ & $1.4(1)$ & $910$ & $0.46$ \\
\hline
$94$ & $48$ & $1.6(1)$ & $1.2(1)$ & $310$ & $0.18$ \\
\hline
\end{tabular}
\end{minipage} 
\hspace{0.1cm}
\begin{minipage}[!t]{0.4\textwidth}
\centering
\begin{tabular}{|c c c c c c|}
\hline
$a$ & $u$ & $\alpha$ & $\alpha'$ & $M$ &   $p$
\\ \hline
\hline

$94$ & $98$ & $1.5(2)$ & $0.9(1)$ & $280$ & $0.03$ \\
\hline
$94$ & $15$ & $1.5(1)$ & $1.4(1)$ & $290$ & $0.28$ \\
\hline
$105$ & $28$ & $1.5(1)$ & $1.5(1)$ & $1310$ & $0.59$ \\
\hline
$110$ & $28$ & $1.2(1)$ & $1.4(1)$ & $2870$ & $0.36$ \\
\hline
$116$ & $15$ & $1.2(1)$ & $1.3(1)$ & $3050$ & $0.13$ \\
\hline
$122$ & $15$ & $1.5(1)$ & $1.4(1)$ & $470$ & $0.95$ \\
\hline
$125$ & $28$ & $1.4(1)$ & $1.4(1)$ & $790$ & $0.89$ \\
\hline
$133$ & $98$ & $1.7(1)$ & $1.2(1)$ & $140$ & $0.15$ \\
\hline
$162$ & $15$ & $1.2(1)$ & $1.4(1)$ & $3270$ & $0.54$ \\
\hline
$181$ & $15$ & $1.2(2)$ & $1.0(1)$ & $250$ & $0.34$ \\
\hline
$197$ & $15$ & $1.2(2)$ & $1.2(1)$ & $180$ & $0.79$ \\
\hline
$201$ & $15$ & $1.3(2)$ & $1.1(1)$ & $110$ & $0.43$ \\
\hline
$262$ & $150$ & $1.3(8)$ & $1.8(1)$ & $1200$ & $0.08$ \\
\hline
$264$ & $1428$ & $1.5(2)$ & $1.7(1)$ & $1610$ & $0.16$ \\
\hline
$269$ & $122$ & $1.6(2)$ & $1.7(1)$ & $970$ & $0.02$ \\
\hline
$279$ & $550$ & $1.5(2)$ & $1.6(1)$ & $2600$ & $0.32$ \\
\hline
$293$ & $3503$ & $1.5(1)$ & $1.5(1)$ & $2030$ & $0.01$ \\
\hline
$300$ & $550$ & $1.5(1)$ & $1.5(1)$ & $290$ & $0.68$ \\
\hline
$300$ & $150$ & $1.7(1)$ & $1.7(1)$ & $1900$ & $0.26$ \\
\hline
$306^*$ & $150$ & $1.5(1)$ & $1.4(1)$ & $740$ & $0.00$ \\
\hline
$317$ & $150$ & $1.8(2)$ & $1.6(1)$ & $260$ & $0.38$ \\
\hline
$318$ & $150$ & $1.5(2)$ & $1.4(1)$ & $610$ & $0.05$ \\
\hline
$327$ & $1503$ & $1.3(1)$ & $1.4(1)$ & $230$ & $0.54$ \\
\hline
$327$ & $150$ & $1.6(1)$ & $1.5(1)$ & $290$ & $0.46$ \\
\hline
$331$ & $150$ & $1.6(2)$ & $1.6(1)$ & $1500$ & $0.42$ \\
\hline
$332$ & $1574$ & $1.6(2)$ & $1.9(1)$ & $1140$ & $0.15$ \\
\hline

\end{tabular}
\end{minipage} 

\end{center}
\caption{BM data set. Each row corresponds to
one of the $52$ agents who have bid at least $100$ times in the same auction.
We report the id of the auction $a$, the id of the agent $u$, the
exponent $\alpha$ calculated with the least square method,
the exponent $\alpha'$  calculated with the maximum likelihood method,
the best value of the upper bound $M$ and
the $p$-value. 
Entries with low $p$-values are marked with $^*$. 
In the $92\%$
of the cases we find a $p$-value larger than $0$, which
indicates that our model well describe
the time series.}
\label{tab:bm_like}
\end{table}

\clearpage

\subsubsection{Probability distribution of the L\'evy flight exponents}

\begin{figure}[!ht]
\begin{center}
\vskip .7cm
\includegraphics[width=7cm]{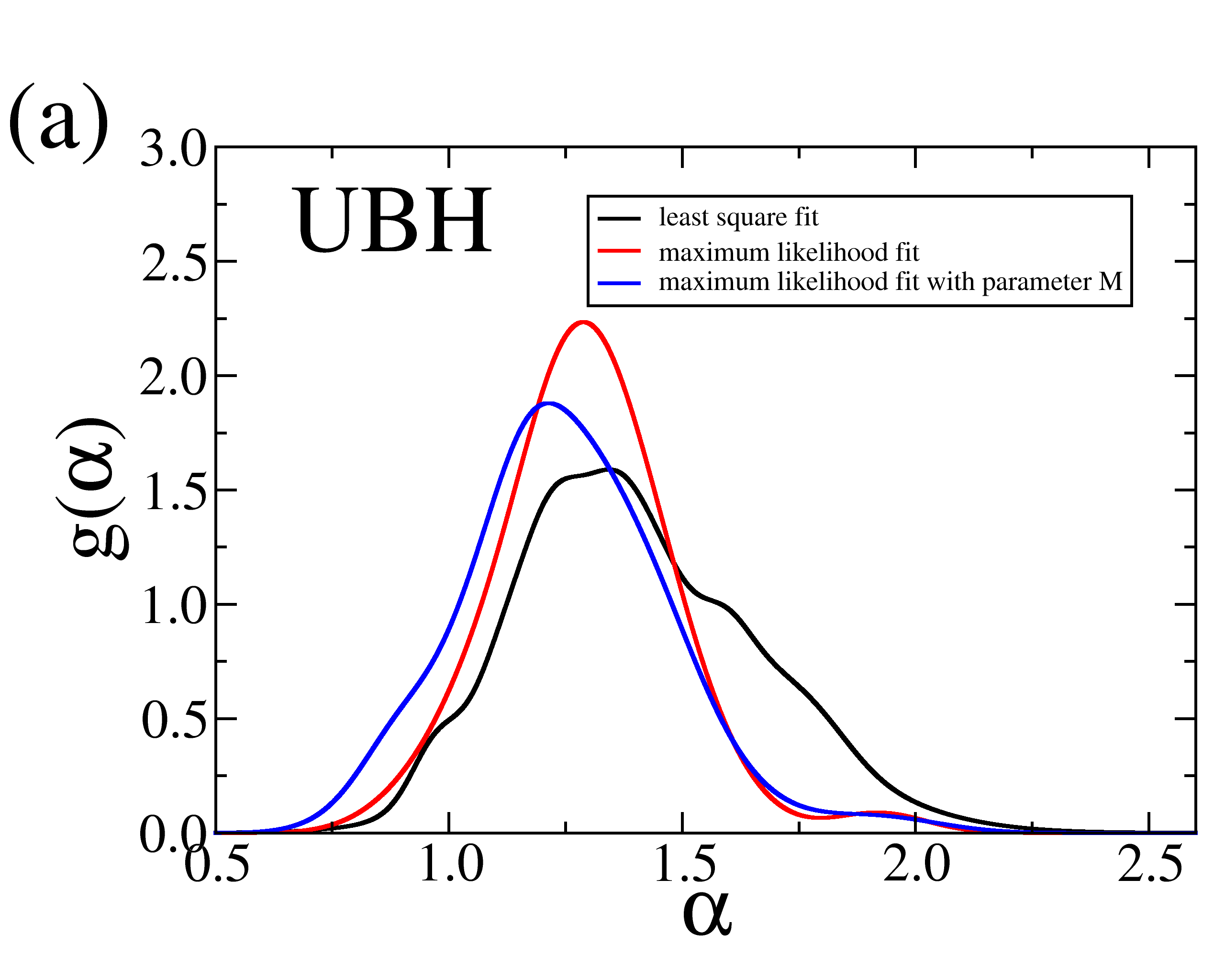}
\qquad
\includegraphics[width=7cm]{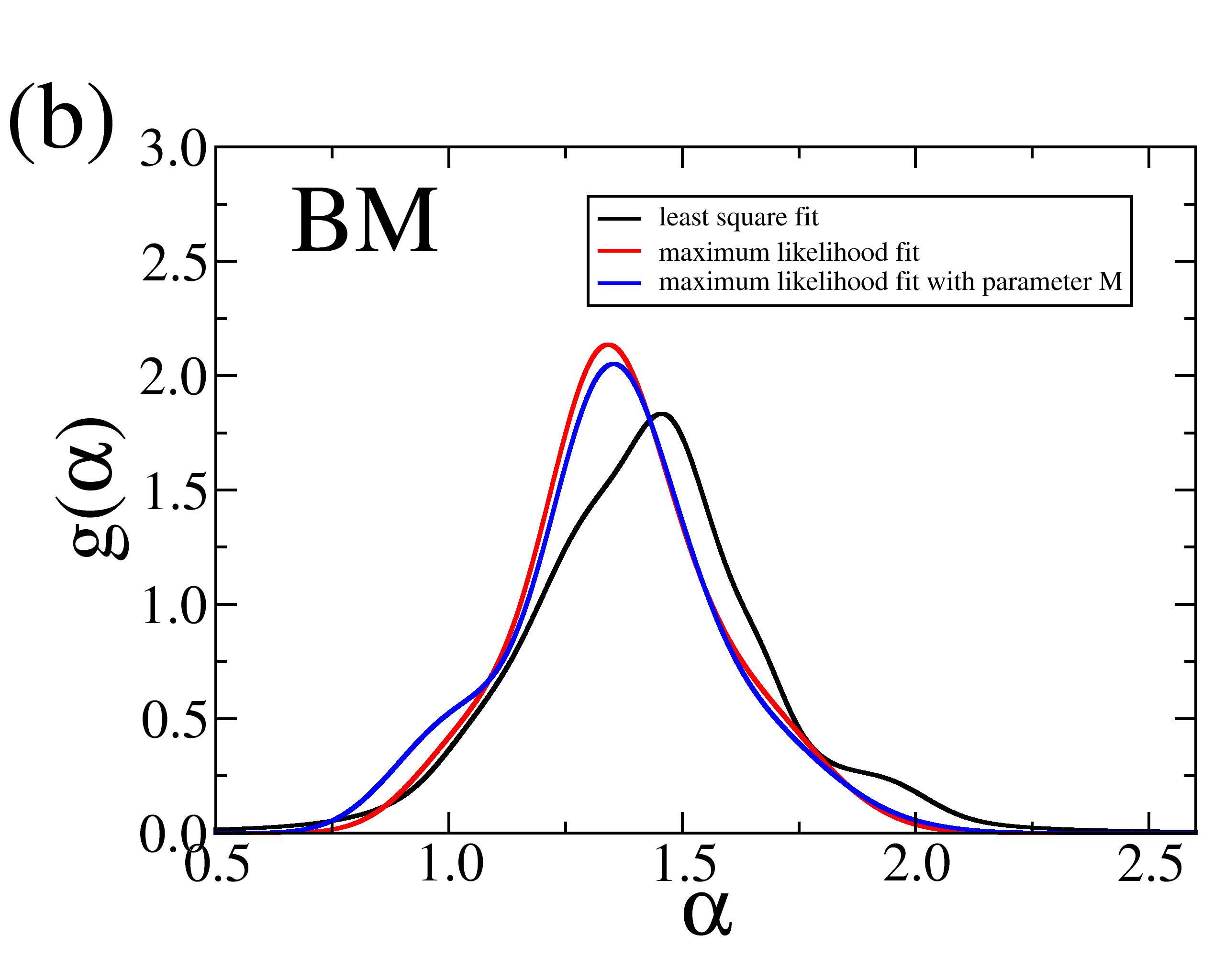}
\vskip .1cm
\caption{The pdf $g\left(\alpha\right)$ of the exponents 
calculated with the three fitting methods. 
The pdfs corresponding to the maximum likelihood
fit, where $M$ is a parameter of the fit,
are the same as those appearing in Figs.~2D
and~2E of the main text.
The distributions are characterized by the following
values of the mode $\alpha_b$, average $\langle \alpha\rangle$  and variance $\sigma$.
UBH data set:
for least square fit we have $\alpha_b = 1.34$, $\langle \alpha\rangle= 1.40$, $\sigma= 0.26$; for maximum likelihood fit we have $\alpha_b = 1.28$, $\langle \alpha\rangle= 1.29$, $\sigma= 0.20$; for maximum likelihood fit with
additional fitting parameter $M$ we have $\alpha_b = 1.21$, $\langle \alpha\rangle= 1.26$, $\sigma= 0.23$.
BM data set:
for least square fit we have $\alpha_b = 1.46$, $\langle \alpha\rangle= 1.42$, $\sigma= 0.27$; for maximum likelihood fit we have $\alpha_b = 1.34$, $\langle \alpha\rangle= 1.37$, $\sigma= 0.21$; for maximum likelihood fit with
additional fitting parameter $M$ we have $\alpha_b = 1.35$, $\langle \alpha\rangle= 1.36$, $\sigma= 0.23$.
}
\label{fig:bayes_h}
\end{center}
\end{figure}

\noindent As a final result, we compute the distribution
$g\left(\alpha\right)$ of the exponents measured
for single agents in single auctions.
We still consider only agents who have
performed at least $T=50$ bids in a single auction
in UBH data set and those with at least $T=100$ in the BM data set.
Assuming that the best estimation of the exponent
of agent $u$ is $\alpha_u$ and the associated error
of the measurement is $\Delta \alpha_u$, the empirical
distribution of the exponents is calculated as
\begin{equation}
g\left(\alpha\right) = C^{-1} \; \sum_u \;  \frac{1}{\Delta \alpha_u} \; e^{-\left(\alpha_u - \alpha\right)^2\, /\, \left[2\, \left(\Delta \alpha_u\right)^2\right]}\;\;,
\label{eq:kernel}
\end{equation} 
where $C= \int d\alpha \; \sum_u  \frac{1}{\Delta \alpha_u} \; e^{-\left(\alpha_u - \alpha\right)^2\, /\, \left[2\, \left(\Delta \alpha_u\right)^2\right]}$ is the proper
normalization constant. The resulting
pdfs are reported in Fig.~\ref{fig:bayes_h}.


\end{document}